\documentclass[sort&compress]{elsarticle}
\journal{IJHCS}

\def\plaintitle{Omnis Pr{\ae}dictio: Estimating the Full Spectrum of Human Performance with Stroke Gestures}

\def\plainkeywords{Stroke Gestures; Touch Gestures; Gesture user interfaces; Human Performance; Estimation; Prediction; Gesture Synthesis; Kinematic Theory}

\usepackage[utf8]{inputenc}
\usepackage[OT2,T1]{fontenc}
\usepackage{amsmath,amssymb,bm}
\usepackage{nicefrac}
\usepackage{xspace}
\usepackage{pgfplots}
\usepackage{booktabs,array}
\usepackage{graphicx,subfig}
\graphicspath{{figs/},{figs/gestures/},{figs/synth-samples/}}
\newcolumntype{C}[1]{>{\centering\arraybackslash}m{#1}}
\newcolumntype{R}[1]{>{\raggedright\arraybackslash}m{#1}}
\usepackage{balance}
\usepackage{fancyvrb}
\usepackage{listings}
\lstdefinelanguage{json}{
  comment=[l]{//},
}
\lstdefinelanguage{javascript}{
  morekeywords={let,const,for,of,function,return},
  morestring=[b]',
}

\usepackage{enumitem}
\setlist[enumerate,1]{leftmargin=1.5em, itemindent=0em, labelwidth=1em, labelsep=0em, align=left, style=multiline}
\setlist[enumerate,2]{leftmargin=1.5em, itemindent=1em, labelwidth=1em, labelsep=0em, align=left}

\usepackage[hidelinks,breaklinks]{hyperref}
\usepackage{cleveref}

\newcommand{\tth}[1]{\textbf{\shortstack{#1}}}

\def\omnis{\textsc{Omnis}\xspace}
\def\slm{$\Sigma\Lambda$\xspace}

\def\etal{et~al.\xspace}
\def\vs{vs.\xspace}

\def\ie{i.e.\xspace}
\def\eg{e.g.\xspace}

\def\mmStrokes{$\text{MM}_\text{S}$\xspace}
\def\mmFingers{$\text{MM}_\text{F}$\xspace}
\def\mmHands{$\text{MM}_\text{H}$\xspace}

\pgfplotsset{
  compat=newest,
  tick label style={font=\sf\large},
  label style={font=\bfseries\sffamily\large},
  x label style={yshift=0.5em},
  y label style={yshift=0.5em},
  title style={font=\bfseries\sffamily\large, yshift=0.5em},
  xtick pos=left,
  legend cell align=left,
  legend columns=5,
  legend style={
    font=\sf, at={(0.5,1.1)}, anchor=north,
    nodes={inner sep=0.25em},
    /tikz/every even column/.append style={column sep=1.5em},
    draw=none,
    fill=none,
  },
  every node near coord/.style={font=\large},
  /pgfplots-min/ybar legend/.style={
    /pgfplots-min/legend image code/.code={
    \draw[##1, /tikz/.cd, bar width=3pt, bar shift=0pt, yshift=-0.25em]

      plot coordinates { (0cm, 0.8em) (-\pgfplotbarwidth, 0.6em) (-\pgfplotbarwidth*2, 0.4em) };
    },
  },
}

\definecolor{nicered}{RGB}{166,206,227}
\definecolor{niceblue}{RGB}{31,120,180}
\definecolor{nicegreen}{RGB}{178,223,138}
\definecolor{niceturquoise}{RGB}{51,160,44}
\definecolor{niceorange}{RGB}{251,154,153}
\definecolor{niceyellow}{RGB}{227,26,28}
\definecolor{nicebrown}{RGB}{253,191,111}
\definecolor{nicegray}{RGB}{255,127,0}
\definecolor{nicepink}{RGB}{202,178,214}
\definecolor{nicepurple}{RGB}{106,61,154}

\definecolor{niceblackish}{RGB}{197,197,197}
\definecolor{niceblueish}{RGB}{76,147,195}
\definecolor{nicegreensish1}{RGB}{193,229,161}
\definecolor{nicegreensish2}{RGB}{131,203,65}
\definecolor{nicegreensish3}{RGB}{99,159,45}
\definecolor{nicegreensish4}{RGB}{54,86,24}

\pgfplotscreateplotcyclelist{nicebarsradu}{
  {niceblackish, fill=niceblackish},
  {niceblueish, fill=niceblueish},
  {nicegreensish1, fill=nicegreensish1},
  {nicegreensish2, fill=nicegreensish2},
  {nicegreensish3, fill=nicegreensish3},
  {nicegreensish4, fill=nicegreensish4},
}

\pgfplotscreateplotcyclelist{nicedotsradu}{
  {niceblackish, fill=none, draw=none, mark=*},
  {nicegreensish1, fill=none, draw=none, mark=square*},
  {nicegreensish2, fill=none, draw=none, mark=triangle*},
  {nicegreensish3, fill=none, draw=none, mark=diamond*},
  {nicegreensish4, fill=none, draw=none, mark=pentagon*},
}

\pgfplotscreateplotcyclelist{nicebars}{
  {nicered!80!black, fill=nicered!80!white},
  {niceblue!80!black, fill=niceblue!80!white},
  {nicegreen!80!black, fill=nicegreen!80!white},
  {niceturquoise!80!black, fill=niceturquoise!80!white},
  {niceorange!80!black, fill=niceorange!80!white},
  {niceyellow!80!black, fill=niceyellow!80!white},
  {nicebrown!80!black, fill=nicebrown!80!white},
  {nicegray!80!black, fill=nicegray!80!white},
  {nicepink!80!black, fill=nicepink!80!white},
  {nicepurple!80!black, fill=nicepurple!80!white},
}
\pgfplotscreateplotcyclelist{nicelines}{
  {nicered!80!black, fill=none, thick},
  {niceblue!80!black, fill=none, thick},
  {nicegreen!80!black, fill=none, thick},
  {niceturquoise!80!black, fill=none, thick},
  {niceorange!80!black, fill=none, thick},
  {niceyellow!80!black, fill=none, thick},
  {nicebrown!80!black, fill=none, thick},
  {nicegray!80!black, fill=none, thick},
  {nicepink!80!black, fill=none, thick},
  {nicepurple!80!black, fill=none, thick},
}
\pgfplotscreateplotcyclelist{nicedots}{
  {nicered!80!white, fill=none, draw=none, mark=*},
  {niceblue!80!white, fill=none, draw=none, mark=square*},
  {nicegreen!80!white, fill=none, draw=none, mark=triangle*},
  {niceturquoise!80!white, fill=none, draw=none, mark=diamond*},
  {niceorange!80!white, fill=none, draw=none, mark=pentagon*},
  {niceyellow!80!white, fill=none, draw=none, mark=o},
  {nicebrown!80!white, fill=none, draw=none, mark=square},
  {nicegray!80!white, fill=none, draw=none, mark=triangle},
  {nicepink!80!white, fill=none, draw=none, mark=diamond},
  {nicepurple!80!white, fill=none, draw=none, mark=pentagon},
}

\newcommand\storelabel[2]{\expandafter\xdef\csname label#1\endcsname{#2}}
\newcommand\getlabel[1]{\csname label#1\endcsname}

\newcommand{\stat}[1]{
\begingroup
\binoppenalty=\maxdimen
\relpenalty=\maxdimen
\setlength{\thinmuskip}{3mu}
\setlength{\medmuskip}{3mu}
\setlength{\thickmuskip}{3mu}
\ensuremath{#1}
\endgroup}

\newcommand{\statp}[1]{\ensuremath{\left(\stat{#1}\right)}}

\newcommand{\feat}[1]{{\small\texttt{\detokenize{#1}}}}
\newcommand{\featsmall}[1]{{\footnotesize\texttt{\detokenize{#1}}}}

\definecolor{ggplotRed}{HTML}{F8766D}
\definecolor{ggplotGreen}{HTML}{00BA38}
\definecolor{ggplotBlue}{HTML}{619CFF}

\begin{document}

\title{\textbf{\plaintitle}\tnoteref{t1}}
\tnotetext[t1]{\textbf{Int. J. Hum. Comput. Stud. 142 (2020)}
  DOI: \href{https://doi.org/10.1016/j.ijhcs.2020.102466}{10.1016/j.ijhcs.2020.102466}}

\author[1]{Luis A. Leiva\corref{cor}\fnref{sci}}
\ead{firstname.lastname@aalto.fi}

\author[2]{Radu-Daniel Vatavu\corref{cor}}
\ead{radu.vatavu@usm.ro}

\author[3]{Daniel Mart\'in-Albo\fnref{ind}}
\ead{dmas@wiris.com}

\author[4]{R\'ejean Plamondon}
\ead{rejean.plamondon@polymtl.ca}

\fntext[sci]{Work partially conducted while affiliated with Sciling,\,SL}
\fntext[ind]{Independent researcher}
\cortext[cor]{Corresponding author}

\address[1]{
  Aalto University,
  Finland
}
\address[2]{
  University ``\c{S}tefan cel Mare'' of Suceava,
  MintViz Lab | MANSiD Center,
  Romania
}
\address[3]{
  WIRIS Math,\,SL,
  Spain
}
\address[4]{
  Polytechnique Montréal,
  Laboratoire Scribens,
  Canada
}

\begin{abstract}
Designing effective, usable, and widely adoptable stroke gesture commands for graphical user interfaces is a challenging task that traditionally involves multiple iterative rounds of prototyping, implementation, and follow-up user studies and controlled experiments for evaluation, verification, and validation. An alternative approach is to employ theoretical models of human performance, which can deliver practitioners with insightful information right from the earliest stages of user interface design. However, very few aspects of the large spectrum of human performance with stroke gesture input have been investigated and modeled so far, leaving researchers and practitioners of gesture-based user interface design with a very narrow range of predictable measures of human performance, mostly focused on estimating production time, of which extremely few cases delivered accompanying software tools to assist modeling. We address this problem by introducing ``Omnis~Pr{\ae}dictio'' (\omnis for short), a generic technique and companion web tool that provides accurate user-independent estimations of \emph{any} numerical stroke gesture feature, including custom features specified in code. Our experimental results on three public datasets show that our model estimations correlate on average \stat{r_s > .9} with groundtruth data. \omnis also enables researchers and practitioners to understand human performance with stroke gestures on many levels and, consequently, raises the bar for human performance models and estimation techniques for stroke gesture input.

\end{abstract}


\maketitle

\section{Introduction}\label{sec:introduction}

Stroke gestures (also referred to as touch, pen, stylus, finger, or ink gestures)
represent two-dimensional pathlines that make up geometric symbols,
which are mapped by designers to specific application features, actions, and user interface commands.
For example, drawing the letter ``S'' on a smartphone's home screen could be used to query the list of contacts~\cite{Li10b}
or speed-dial a specific contact~\cite{oftSeen}.
Similarly, in the POW Studies video game ``Mr. Spiff's revenge''~\cite{POWStudios},
players can draw circles to create shields, arcs to launch fireballs, and hearts to drink life-saving potions.
More recently, the massive online game ``Harry Potter: Wizards Unite''
introduced a feature enabling players to draw stroke gestures for making spells against enemies~\cite{HPWU}.

Recent research has repeatedly demonstrated the practical convenience and utility of employing stroke gestures
as efficient shortcuts to access system functions~\cite{Li10b,Lu:2015:GEA} or specific applications~\cite{Zhang:2017:IAL, Zhang:2016:AMA}
on touchscreen devices, ranging from the tiniest gadgets and wearables~\cite{Funk:2014:UTW, Schneegass:2016:GUT}
to smartphones and tablets~\cite{Tu12} and to large touch and multitouch interactive displays~\cite{Bragdon:2011:GSA,Wobbrock09,Morris10}.
Stroke gestures also represent an effective input modality for users with low vision
to interact with smart devices~\cite{Vatavu:2018,Leiva17_chi} by providing a practical alternative to selecting touch targets,
challenging to see on small screens, or visual targets that cannot be easily acquired~\cite{Kane11,Kane08}.
Yet another direction of applications has demonstrated the viability of stroke gesture input
for user authentication~\cite{Liu:2017:USG, Yang:2016:FGA}.
For example, compared to traditional approaches, a gesture-based password enables users to authenticate faster on mobile devices
and, moreover, users are more willing to retry entering the password in case of erroneous input~\cite{Yang:2016:FGA}.
Ultimately, research on stroke gesture input has made possible new text entry techniques
for mobile devices~\cite{Rick:2010:POV, Burgbacher:2015:MHP},
such as shape-writing or gesture-typing~\cite{Kristensson04, Kristensson:2007:CSW},
widely available on today's smartphones and influencing research and development of future text entry methods
for wearable devices~\cite{Chen:2014:STE:2642918.2647354,Gordon:2016:WTG:2858036.2858242,Yu:2016:OHI:2858036.2858542}.

Besides heavily explored in research prototypes, stroke gestures have been used in commercial applications and products for decades,
and were featured by early interactive computing devices, such as the Apple Newton~\cite{AppleNewton} and Microsoft Tablet PCs~\cite{WindowsTablet}.
Examples of current applications that employ stroke gesture input include the Dolphin browser~\cite{Dolphin},
Quickify~\cite{Quickify},
oftSeen Gestures~\cite{oftSeen},
StrokeIt~\cite{StrokeIt},
or Lovely Charts~\cite{LovelyCharts}, to mention just a few.
The \$-family of stroke gesture recognizers~\cite{Wobbrock07,Vatavu12_p,Anthony10,Anthony12,Vatavu:2018:QSA}
has especially had a considerable impact in making stroke gesture input available on many platforms, devices, and programming languages,
with the net effect of stroke gestures being employed in the user interfaces of many applications, ranging from video games to wearables, virtual reality, and drone controlling~\cite{WobbrockDollarFamilyImpact}.
Alas, designing stroke gestures that represent a good fit to the functions they effect~\cite{Wobbrock09,Morris10,Kane11},
are easy to articulate~\cite{Vatavu11,Rekik14}, straightforward to recall by users~\cite{Nacenta13},
and recognized robustly by a computer~\cite{Leiva14_iwc,Long99,Rubine91,Blagojevic10,Vatavu12_p}
is a challenging task that involves user studies and experiments in an iterative design process
consisting of prototyping, implementation, verification, and validation steps.
Despite the practical benefits of involving actual users in this process,
conducting user studies and experiments to collect gesture training sets and data to inform design takes time, effort, and resources,
which unnecessarily delays the launch date of new products and applications.

Instead of recruiting participants for gesture user studies,
the alternative option for researchers and practitioners is to rely on theoretical models of human performance
with stroke gesture input~\cite{Cao07,Isokoski01,Leiva18_keytime}.
Such models and their associated prediction techniques can save precious time and provide insightful information
about suitable gesture commands right from the early stages of the design process.
However, despite considerable research on gesture recognition, analysis, elicitation,
and gesture-based interaction techniques in our community~\cite{Wobbrock09,Vatavu13_relacc,Anthony13gi,Zhai12,Morris10},
only few aspects of human performance with stroke gesture input have been investigated so far~\cite{Vatavu:2013:SML,Long00,Vatavu11}
with a considerable focus on the production time of stroke gestures~\cite{Cao07,Leiva18_keytime,Leiva18_gato}.
Unfortunately, when it comes to estimating other gesture features and measures of human performance
such as the curvature of gesture paths, the length and size of strokes, or the speed of gesture articulation,
no methods or tools are currently available.
We believe that this current state of affairs is an effect of the large emphasis that has been put on the production time of gesture input,
which represents an instantiation of the generic task time measure
heavily employed in the research and practice of Human-Computer Interaction (HCI)
to evaluate users' performance with interactive systems and tasks~\cite{Fitts54,Accot97,Zhai12,Isokoski01}.
However, human performance with stroke gesture input has been evaluated on many other dimensions
that are relevant to the user experience of gesture-based interaction on touchscreens.
For example, the perceived visual appearance of geometric shapes~\cite{Long00},
the scale of gesture input~\cite{Vatavu:2013:SML},
or the relative accuracy of gesture articulations with respect to canonical forms~\cite{Vatavu13_relacc}
are important dimensions on which to evaluate users' performance with stroke gesture articulations.
Despite the overall trend in the HCI community to focus on the equivalence \textit{performance} $\equiv$ \textit{time},
including for gesture input, designing stroke gestures that are not only fast to execute, but also smooth, short, or symmetric,
among other desirable geometric, kinematic, and articulation characteristics~\cite{Vatavu13_relacc,Vatavu14ghost},
represents an equally sensible design approach.
Next, we illustrate a few practical scenarios for which the ability to estimate such gesture
properties proves useful for user interface design.

\subsection{Motivating Examples}
Being able to estimate various characteristics of the stroke gestures that end users will produce represents a valuable asset for practitioners.
A few practical examples can demonstrate this point clearly:
\begin{enumerate}
	\itemsep 2pt
	\item
	Imagine a designer who prototypes a gesture user interface for a new text entry application on smartphones,
	where letters drawn by users are lowercased or uppercased automatically based on how large they are drawn,
	\eg, a small letter ``a'' will effect either a lowercase ``a'' or an uppercase ``A'' depending on its actual size in pixels on the screen.
	With such a design approach, the gesture recognizer will need to discriminate only between 26 gesture classes
	(\ie, the lowercase letters of the Roman alphabet) instead of 52 (both lowercase and uppercase letters),
	a reduction in the number of output classes that is expected to have a positive effect on classification error.
	The designer can use the bounding box area size of each articulated gesture path to estimate the user's intended scale of input,
	a gesture feature that was shown by prior work to be consistently produced by independent users~\cite{Vatavu:2013:SML}
	and, thus, the text boxes of the user interface will effect either the lowercase or uppercase letter modifiers automatically.
	However, the designer would have to run a gesture data collection experiment
	to understand how much variation is to be expected in terms of the \texttt{Area-Size} gesture feature for letters drawn by end-users.
	In this case, a predictive model or an equivalent estimator
	that could generate the distribution of the \texttt{Area-Size} gesture feature based on sound theoretical principles
	would save the designer considerable time.

	\item
	Consider a practitioner who has just implemented a statistical gesture classifier, similar to Rubine's popular recognizer~\cite{Rubine91},
	but perhaps using other gesture features~\cite{Blagojevic10} such as gesture \texttt{Curviness} or \texttt{Density},
	that they believe will improve classification accuracy significantly
	for a particular gesture set relevant for their particular application domain.
	The discriminatory power of a gesture feature is high when the variation of its values within gesture classes is small
	(\ie, the feature has a tight distribution centered on its mean, or low intra-class variation)
	and when the means of the distributions of the feature for different gesture classes are sufficiently different (\ie, high intra-class variation).
	Without any data or \emph{a priori} design knowledge, the practitioner has virtually no information
	regarding the empirical distributions of gesture features for various gesture types considered for the implement of the new recognizer.
	Having access to a theoretical model providing estimations of the distributions of those features
	would empower the practitioner with the ability to evaluate many potential gesture features for their gesture classifier with little effort,
	such as different definitions and variations of gesture \texttt{Density} or the ratio of \texttt{Curviness} and \texttt{Density}~\cite{Long00},
	if the practitioner believes that such new features can bring value to classification.
	Therefore, empowered with this information and knowledge, the practitioner could easily select the most easily recognizable gesture set
	to implement in their user interface and application, being confident in the high classification accuracy of their recognizer in practice.

	\item
	Consider now the example of a designer who is looking for an intuitive gesture set	for a touchscreen armband, skin, or cloth piece from the vicinity of the hand wrist to control music played by a smartwatch~\cite{Funk:2014:UTW,Han:2017:DTG,Schneegass:2016:GUT},
	such as the functions ``play song'', ``stop'', and ``repeat.''
	The designer concludes to associate the triangle shape with ``play,''
    rectangle with ``stop,'' and circle to ``repeat''
	to reflect well-known iconic symbols for these functions and, thus, to maximize the guessability and memorability of the gesture set for end users.
	However, the designer soon realizes that these gestures may look different in their geometric shapes,
	such as the aperture between the start and ending points,
	defined by gesture feature \texttt{F4} in the study of Tu \etal~\cite{Tu12},
	depending on where the gestures are produced, \eg, on the skin, the armband, or the piece of cloth.
	The designer wishes to know how large the variations in gesture feature \texttt{F4} are,
	which is important knowledge to discriminate between a valid closed-shape gesture and unintentional touches on the skin or clothes.
	In this case, an estimation technique based on theoretical models of human performance with gesture input
	that would present the designer with the user-independent distribution of the \texttt{F4} feature
	for the triangle, rectangle, and circle shapes would help inform a reliable rejection rule for unintentional touches.

	\item
	As a last illustrative example, consider a head-mounted display or smartglasses-based system
	designed for people with low vision that delivers enhanced representations of the visual reality,
	such as contrast enhancement, highlighted edges, or color correction, as demonstrated in~\citep{Zhao:2015:FCH:2700648.2809865,Tanuwidjaja:2014:CWA:2632048.2632091}.
	In that case, touch gestures could be captured on the side touch pad of the smartglasses~\cite{Grossman15,Yu:2016:OHI:2858036.2858542},
	a feature provided by most smart eyewear designs, to shortcut menu selections.
	For example, letter ``E'' could present users with an edge-enhanced view of the physical reality,
	and letter ``C'' would activate color enhancement.
	However, it is known that people with low vision produce stroke gestures that exhibit larger variations in terms of the \texttt{Length-Error}
	and \texttt{Bending-Error} features~\cite{Vatavu13_relacc} compared to people without visual impairments~\cite{Vatavu:2018},
	which reflect negatively in the accuracy rates of gesture recognizers.
	Our practitioner would like to know how large this variation is for the particular symbols ``E'' and ``C'' chosen for implementation in their user interface and wearable prototype.
	Again, an estimation method or tool that could generate the distributions of the values of these two features
	would help the practitioner form a good understanding of the variability expected in stroke gesture input
	for users with low vision and search for effective ways to deal with it.
\end{enumerate}

\subsection{Contributions}
\begin{figure*}[t]
	\centering
	\def\h{2cm}
	\includegraphics[height=\h]{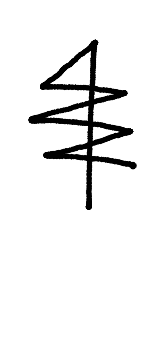}
	\includegraphics[height=\h]{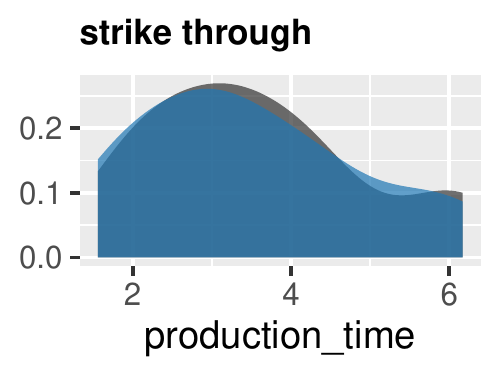}\hfill
	\includegraphics[height=\h]{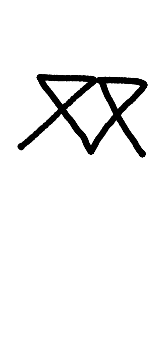}
	\includegraphics[height=\h]{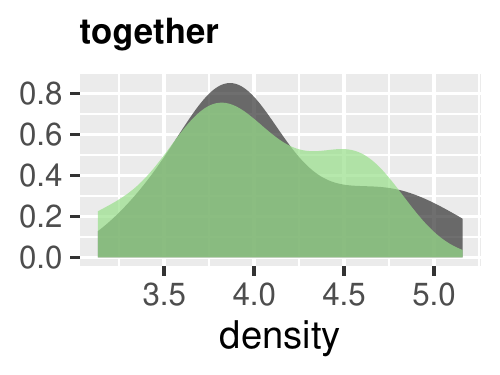}\hfill
	\includegraphics[height=\h]{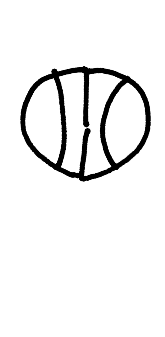}
	\includegraphics[height=\h]{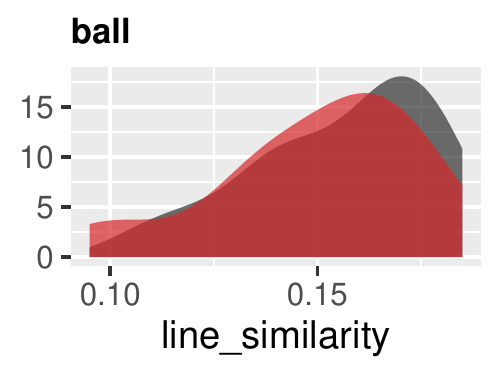}
	\caption{
		Distributions of three representative stroke gesture features
		examined in the scientific literature on gesture recognition and analysis. From left to right:
		production time~\cite{Leiva18_keytime},
		density~\cite{Long00},
		and line similarity~\cite{Anthony13gi}.
		Ground truth distributions are shown in dark gray (data from N=18 users)
		and distributions estimated by \omnis are superimposed in another color.
	}
	\label{fig:teaser}
\end{figure*}

The previous examples show that it is important for gesture user interface designers to estimate, as accurately as possible,
end users' performance with stroke gesture input on many dimensions.
Unfortunately, no models exist today to estimate gesture features beyond production time~\cite{Cao07,Isokoski01,Leiva18_gato,Leiva18_keytime}, which limit the toolbox available to designers to evaluate gesture sets, recognizers, and gesture features without running actual user studies and experiments, and render useful scenarios such as those presented above unattainable in practice.
In this work, we address this aspect by introducing ``Omnis~Pr{\ae}dictio'' (\omnis),
a generic and flexible technique that can accurately estimate \emph{any} numerical gesture feature
computed from the representation of a stroke gesture as a set of 2D points and associated timestamps.
\omnis synthesizes the distribution of the chosen gesture feature and generates relevant statistics, such as the mean, trimmed means, or variance.
\autoref{fig:teaser} shows examples of the estimated distributions for a few gesture features
that were reported by prior work as key to inform gesture set design~\cite{Rekik14},
evaluate user performance with stroke gesture input~\cite{Cao07,Isokoski01},
and improve gesture recognition accuracy~\cite{Rubine91,Blagojevic10}.
In sum, the contributions of this article are as follows:
\begin{enumerate}
  \itemsep 1pt
  \item
  We introduce \omnis, a generic human performance estimation technique
  informed by the core principles of the Kinematic Theory~\cite{Plamondon95a}
  that estimates the values of \emph{any} numerical gesture feature
  that can be computed from the representation of a stroke gesture as a set of 2D points.
  \omnis reports user-independent distributions of the feature values, as illustrated in \autoref{fig:teaser}.
  Moreover, \omnis accepts stroke gestures of all kinds, \eg, unistrokes, multistrokes, and single- and multi-touch gestures.

  \item
  We evaluate \omnis on three public datasets
  (consisting of $7,200$ samples of $30$ distinct gesture types collected from $18$ participants)
  and on a set of $18$ representative gesture features
  frequently employed in the scientific literature on gesture recognition and analysis.
  On average, the estimations delivered by \omnis for these features correlated \stat{r_s > .90} with groundtruth data
  and as high as \stat{r_s =.98} for some of the features that we evaluated; see \Cref{tbl:results1}.

  \item
  We release a companion web application and RESTful API that practitioners can readily use
  to compute accurate, user-independent estimates of a wide palette of gesture features,
  such as the gesture features common in the gesture literature that we evaluated, which are implemented by default in \omnis.
  Moreover, the application flow of \omnis enables practitioners to define their own custom gesture features
  via simple code-based definitions of those features, and also to integrate the \omnis API with third party applications;
  see \Cref{fig:api1,fig:api2} from the Appendix for some examples.
\end{enumerate}

Aside from these contributions, this article attempts to raise awareness in the HCI community interested in gesture input and gesture user interfaces that other measures of human performance besides the production time of stroke gestures
are relevant and important for an effective user experience with stroke gesture input
and, thus, recommendable to explore and inform the design of gesture sets for interactive products and applications.
Also, by being able to estimate any numerical gesture feature,
\omnis raises the bar for human performance models and estimation techniques for stroke gesture input~\cite{Cao07,Leiva18_keytime,Leiva18_gato}
with unprecedented flexibility and accuracy.
We hope that our theoretical developments, together with our empirical results, web application, and API
will help researchers to attain new discoveries and advance knowledge in gesture input and practitioners to design gesture sets
better suited to end users' articulation abilities and preferences.

\section{Related Work}\label{sec:relwork}

In this work, we employ the concepts and principles of the Kinematic Theory~\cite{Plamondon95a,Plamondon95b} to devise a theoretical model, from which to estimate various statistics and measures of stroke gesture features. While the Kinematic Theory has been used to synthesize and analyze human handwriting successfully in many applied contexts,
practical applications in HCI have been primarily directed at generating gesture training data~\cite{Leiva16_g3,Leiva17_chi,Taranta16}
and estimating gesture production times~\cite{Leiva18_keytime,Leiva18_gato,Ungurean18mobilehci}.
For example, Leiva \etal~\cite{Leiva18_keytime} used the \slm model of the Kinematic Theory to develop KeyTime,
a very accurate technique for estimating the production times of unistroke gestures. In a follow-up work,
KeyTime was superseded by GATO~\cite{Leiva18_gato},
able to estimate the production times of multistroke and multitouch gestures as well.
In this work, we build on top of these recent advances and tools made available in our community
to advance the state of the art in estimation methods for stroke gesture input, as follows:
\begin{enumerate}
	\itemsep 1pt
	\item
	Both KeyTime~\cite{Leiva18_keytime} and GATO~\cite{Leiva18_gato} are limited to estimating just the \emph{production time} of stroke gestures, and they were explicitly formulated to synthesize timestamps only~\cite[Eq.~2]{Leiva18_gato}. On the contrary, \omnis can estimate any numerical feature, including new gesture features that researchers and practitioners still have to invent. The new features can be defined and submitted via our online web application and API.
	\item
	Going beyond previous methods and tools, \omnis estimates \emph{feature distributions}, from which it is possible to compute any descriptive statistics related to location (\eg, mean, median, trimmed means) and dispersion (\eg, standard error and variance) to estimate end-user performance with stroke gesture input.\footnote{While our implementation provides a series of built-in statistics, developers can compute new statistical measures in the same way they can define new gesture features, since our API also provides the raw list of estimated values; see the \nameref{apx:api} from the Appendix}.
\end{enumerate}
From this perspective, \omnis substantially advances the state of the art
and represents the only general-purpose, comprehensive, and flexible estimation model currently available for stroke gesture input. In addition, \omnis also creates important theoretical knowledge. For example, the Kinematic Theory
models the velocity profile of human movements (see \autoref{sec:system}) and, therefore,
it is expected to perform well for estimating measures
that are derived directly from velocity profiles,
such as production time~\cite{Leiva18_gato,Leiva18_keytime}.
However, it is unclear how the \slm model will scale to estimating more sophisticated measures
related to the shape and geometry of stroke gestures,
such as \texttt{Area-Size} or \texttt{Bending-Error} from the examples from the \nameref{sec:introduction} section.
To the best of our knowledge, the HCI community has not been offered any estimator of human performance with stroke gesture input
beyond production time, which is unfortunate as many gesture features are key to inform
gesture set design~\cite{Long00, Anthony13gi, Rekik14}
or recognizers~\cite{Rubine91, Blagojevic10}.
In the following sections, we review prior work that examined user performance with stroke gesture input.
We also discuss applications of the Kinematic Theory to stroke gesture synthesis and analysis.

\subsection{Stroke Gesture Input}
Several studies have pointed out the benefits of stroke gestures as command shortcuts.
By leveraging the rich capabilities of the human motor system,
stroke gestures enable efficient interaction with touchscreen devices~\cite{Anderson:2013:LPG}.
Stroke gestures also provide rich perceptual cues to users by
creating an association between symbolic shapes and the meaning of specific application functions~\cite{Appert09, Zheng:2018:MGM}. Moreover,
compared to traditional interactions implemented by selecting options from menus and pointing in graphical user interfaces, stroke gestures have
the potential to lower cognitive load and the need for visual attention~\cite{Zhai12}.

Simple forms of stroke-based input, such as pointing and menu selection,
have been studied using Fitts' law~\cite{Fitts54} and its variations~\cite{Bi13,Wobbrock08}:
the steering law~\cite{Accot97} or the Keystroke-Level Model~\cite{Card80}.
However, more complex stroke input, such as handwriting, shape-writing, or free-form gesture paths,
requires more sophisticated models to characterize human performance effectively.
Comprehensive surveys in this direction are provided by Quinn and Zhai~\cite{Quinn16} and M\"uller \etal~\cite{Muller17}.
Interestingly, in a cross-cultural study involving forty people from nine countries,
Mauney~\cite{Mauney2010} noted that the most common gesture for the ``Print'' function was represented by letters,
although the cultural background of the participants had a strong influence on their choices of gestures overall, \eg, Chinese participants employed symbolic gestures more frequently than participants from other countries. In a large-scale, in-the-wild study with 388 participants, Poppinga \etal~\cite{Poppinga14}
reported that mobile users preferred symbolic and letter-shaped gestures to launch applications on their mobile devices.

There are many ways to produce a stroke gesture
depending on the number of strokes or the number of fingers and hands touching the screen.
For example, common touchscreen technology found in commodity smartphones and tablets
can detect multiple discrete touch points at once,
which enables practitioners to access a rich design space of unistrokes, multistrokes, single- and multi-touch, and bimanual gestures for their user interfaces and interactive applications.
In addition, expert gesture input design often involves the use of multiple fingers~\cite{Bailly12,Azenkot12,Ghomi13,Luo15},
various finger parts~\cite{Harrison11}, or even the entire hand for expressive input~\cite{Matulic17}.
At the same time, users are known for exhibiting variations in articulating multistroke and multitouch gestures
in terms of the number of strokes and fingers touching the screen~\cite{Anthony13gi,Rekik14},
especially when there are no constraints imposed~\cite{Hinrichs11}.
Therefore, an ideal human performance estimation technique should be able to handle
all sorts of measurable variation in stroke gesture input,
reflected by appropriate gesture features. From this perspective,
\omnis is a significant step forward by enabling estimation of any numerical gesture feature computable from a set of two-dimensional touch points.

\subsection{Features for Stroke Gesture Recognition}
Besides evaluating user performance with stroke gesture input,
the community has devised reliable techniques to recognize stroke gestures effectively~\cite{Wobbrock07,Kristensson04,Li10a,Taranta17}.
For example, the \$-family of gesture recognizers consists of several easy-to-implement practical approaches
based on the Nearest-Neighbor classification algorithm
that enable practitioners to implement gesture recognition on virtually any platform, device, and programming language.
For example, the \$1 recognizer~\cite{Wobbrock07} is a simple and effective technique
for classifying unistroke gestures with high accuracy rates using the Euclidean distance
between the candidate gesture and templates from a training set;
\$N~\cite{Anthony10} is an extension of \$1 to gestures composed of multiple strokes;
and \$P~\cite{Vatavu12_p} is an articulation-invariant gesture recognizer
that classifies stroke gestures regardless of how they are produced by users
(\ie, \$P ignores the effects of the number of strokes, stroke orientation, and stroke direction during the articulation of the gesture).
Other gesture recognition approaches, such as Protractor~\cite{Li10a}, \$N-Protractor\cite{Anthony12}, PennyPincher~\cite{Taranta:2016:SAG}, and \$Q~\cite{Vatavu:2018:QSA} were introduced to make stroke gesture recognition fast for platforms with little computing resources~\cite{Li10a,Vatavu:2018:QSA} or for time budget scenarios~\cite{Taranta:2016:SAG,Taranta:2015:PPB};
\eg, \$Q is a significant speed-up of the \$P point-cloud gesture recognizer designed for mobile, wearable, and embedded devices.
Yet more recent approaches have employed vector-based algebra to implement effective recognition of stroke gestures,
such as the !FTL method~\cite{Vanderdonckt:2018:FAS}
using a local shape distance between vectors defined on the gesture path.

Other researches have looked at the problem and challenge of gesture recognition from a global perspective,
and introduced recognizers for multiple modalities of gesture input,
\ie, recognizers that are modality-agnostic.
For example, Jackknife~\cite{Taranta17} is a gesture recognition technique
that employs the Dynamic Time Warping algorithm for touch, accelerated motion, free-hand, and whole-body gestures.
Recent work has also focused on online, partially-entered stroke gestures, such as the G-Gene technique~\cite{Carcangiu:2018:GGA}.
Regarding multitouch input, several gesture representation, recognition, and design tools were introduced, such as Gesture Studio~\cite{Lu13} and Gesture Coder~\cite{Lu12}, that enabled designers to readily implement multitouch interaction in their applications.
Yet another category of tools represented by Proton~\cite{Kin12a} and Proton++~\cite{Kin12b}
introduced declarative approaches to describe multitouch gestures as regular expressions of touch events.

\subsection{Evaluating User Performance with Stroke Gesture Input}
Researchers have employed a variety of measures to characterize user performance with stroke gesture input.
For example, Blagojevic \etal~\cite{Blagojevic10} examined 114 distinct gesture features
to inform the design of feature-based statistical classifiers.
Other researchers looked for representative features to depict various aspects of user performance.
For example, Anthony \etal~\cite{Anthony13gi} evaluated gesture articulation consistency,
and reported high levels of consistency within users and
lower consistency for gestures produced with more strokes.

Cao and Zhai's CLC (Curves, Lines, Corners) model~\cite{Cao07} was specifically designed
to quantify the production time of \emph{unistroke} gestures.
The model operates by dividing the gesture shape into curves, straight lines, and corners,
for which production times are estimated individually.
The total production time for a particular gesture is computed as the sum of the individual production times
needed to articulate each of the gesture's elementary sub-parts represented by curves, lines, and corners.
The CLC model works very well as a first-order estimator,
i.e., to estimate the relative ranking of gestures according to their production times.
However, it can only provide a single estimation value,
which is insufficient to characterize the variation in gesture articulation
within and between users~\cite{Anthony13gi,Vatavu12_p,Vatavu13_relacc} (low flexibility).
Also, CLC is known to overestimate the actual magnitudes of production times~\cite{Cao07,Castellucci08,Vatavu11} (low accuracy),
presumably because it doesn't compensate for users' articulation skills~\cite{Cao07}.
To address these issues, Leiva \etal~\cite{Leiva18_keytime} introduced KeyTime, a technique for unistroke gestures that accepts free-form drawing as input,
and GATO~\cite{Leiva18_gato}, an extension of KeyTime for multistroke and multitouch gestures.
However, these techniques can only estimate production times,
which is just one dimension of the user experience of stroke gesture-based interaction.

Gesture features and measures have also been used to inform the design of effective gesture sets and commands.
For example, Long \etal~\cite{Long00} were interested in gesture shapes that would be easy for users to learn and recall.
They found that user perception of gestures' visual similarity correlated with several features, such as length, area, or various angles,
and derived a model for perceived gesture similarity based on their observations.
Vatavu \etal~\cite{Vatavu:2013:SML} found that gesture size,
implemented with the area of the gesture's bounding box,
was a good estimator for the user-perceived scale of gesture input.
Researchers have employed other gesture measures to understand differences in performance between users or between input conditions.
For example, Vatavu \etal~\cite{Vatavu13_relacc} used relative accuracy measures
to quantify deviations from ``ideal'' gesture shapes or templates from a training set.
Kane \etal~\cite{Kane11} and Tu \etal~\cite{Tu12} examined specific gesture features,
such as ``line steadiness'' or ``axial symmetry''
to understand the differences between stroke gestures produced
with the pen or the finger~\cite{Tu12}, or by users with and without visual impairments~\cite{Kane11}.

Such gesture measures have proven useful
to characterize various aspects of stroke gesture input as well as to inform gesture-based user interface design.
However, another line of work has focused on a more fundamental understanding of human movements during stroke gesture production by relating to key aspects from the motor control theory. We discuss this work in the following section.

\subsection{Models of Human Movement Applied to Gesture Research}
Djioua and Plamondon~\cite{Djioua09} demonstrated
that the Kinematic Theory~\cite{Plamondon95a}
and its associated Sigma-Lognormal~(\slm) model~\cite{Plamondon06}
represent a compelling theory and model for handwriting production analysis.
Since then, the \slm model has proved to be a very accurate descriptor of human movement
in a wide range of application scenarios, such as
verifying human signatures~\cite{Plamondon14},
reproducing wrist movements and eye saccades~\cite{Plamondon95a},
and, more recently, stroke gestures~\cite{Almaksour11}.
Furthermore, researchers have evaluated the characteristics of stroke gestures synthesized with the \slm model
from the perspective of both classification performance~\cite{Leiva16_g3} and similarity to gesture shapes articulated by real users~\cite{Leiva17_iwc}.
Furthermore, it has been shown that synthetic stroke gestures are on par with their human counterparts~\cite{Leiva17_dis}
in terms of articulation speed and geometric characteristics~\cite{Leiva16_g3,MartinAlbo16_g3}, and that gesture synthesis is successful for various user groups, such as users with low vision~\cite{Leiva17_chi} or with motor impairments~\cite{Ungurean18chi,Ungurean18mobilehci}.

Many models have been proposed to study human movement production; \eg,
models relying on neural networks~\cite{Bullock88},
behavioral models~\cite{Thomassen83},
and models exploiting minimization principles~\cite{Flash85}.
Among these, the Kinematic Theory has provided a well-established and solid framework
for the practical study of the production of human movement and applications~\cite{Plamondon93}.
This framework takes into account different psychophysiological features,
such as the neuromuscular response time, and has been shown to outperform many other approaches~\cite{Djioua09,Plamondon93}.
The \slm model~\cite{Plamondon06} is the latest instantiation of the framework,
which has been adopted and repurposed recently for stroke gesture analysis.

Viviani \etal~\cite{Viviani95,Viviani82} were among the first researchers to investigate the fundamental aspects of human handwriting and drawing behavior.
Since then, a fruitful line of research has been the application of minimization principles to motor control,
such as Flash and Hogan's Minimum-Jerk Theory~\cite{Flash85}.
Further investigations showed that lognormal-based models,
such as those postulated by the Kinematic Theory~\cite{Plamondon95a,Plamondon93},
are arguably the most accurate descriptors of human movements that are known in the scientific literature,
compared to which \textit{``other models can be considered as successive approximations''}~\cite{Djioua09}.
Actually, it has been shown that the concepts postulated by the Minimum-Jerk Theory
and the Kinematic Theory are linked and describe, with different arguments, a model of velocity profiles~\cite{Djioua10}.

Other models of human movements have addressed specific application domains for stroke gesture input, such as text entry.
For example, Quinn and Zhai~\cite{Quinn16} developed a model of gesture production
that could estimate realistic gesture paths for arbitrary shape-writing tasks.
The model employed ``statistical via-points'' located in each key traveled by the user's finger
with distributions that reflected the sensorimotor noise and speed-accuracy trade-off while typing.
However, this model does not predict movement time (rather, it is assumed a fixed parameter)
and assumes an interaction task involving a particular keyboard layout.
Thus, it would need adaptation before it could be applied to free-form stroke gesture input for other practical applications.

\section{Kinematic Theory Overview}\label{sec:kt}

\begin{figure}[!h]
  \centering
  \includegraphics[width=0.9\linewidth]{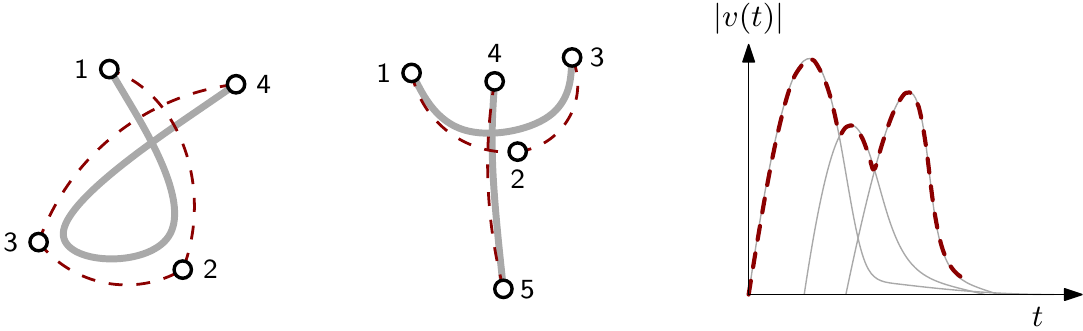}
  \caption{
    A handwritten stroke gesture (solid lines) is described by the temporal overlap of a series of primitives (dashed arcs)
    connecting a sequence of virtual targets (numbered circles).
    Each primitive is modeled and described by a lognormal velocity profile (right image).
    The velocity of the overall gesture path is described by the vectorial summation
    of all the primitives involved in the production of each stroke.
  }
  \label{fig:theory-example}
\end{figure}

For purposes of self-containment, we provide a brief description
of the core principles of the Kinematic Theory and their mathematical formulation. In broad terms, the Kinematic Theory is a general framework for studying human movement from the perspective of movement production.
The latest instantiation of this framework is the \slm model~\cite{Plamondon06}.
Under the \slm framework, it is assumed that a complex handwritten trace,
such as a character, word, signature, or stroke gesture, is composed of a series of \emph{primitives}\footnote{
  The Kinematic Theory uses the term ``stroke'' to denote what we call a ``primitive'' in this article.
  In HCI, a stroke is the points sequence between two consecutive touch-down and touch-up events.
} in the form of arcs connecting a sequence of \emph{virtual targets}, such as those illustrated in \autoref{fig:theory-example}.
The virtual targets correspond to near-zero velocity peaks
and are automatically computed by the \slm model~\cite{MartinAlbo16_g3}.
For our application domain, these primitives form the action plan of the user for a specific gesture that,
by means of the neuromuscular network, will produce a two-dimensional path.

Each gesture primitive is modeled according to a lognormal function of their velocity profile defined by a set of \emph{central} ($D, t_0, \theta$) and \emph{peripheral} ($\mu, \sigma$) parameters~\cite{Plamondon95a}:
\begin{align}\label{eq:lognormal}
\|\vec{v}_i(t)\|
  &= D_i \Lambda(t; t_{0_i}, \mu_i, \sigma_i^2) \notag\\
  &= \frac{D_i}{\sigma_i \sqrt{2\pi} (t - t_{0_i})} \exp{\left( \frac{-[\ln (t - t_{0_i}) - \mu_i]^2}{2 \sigma_i^2} \right)}
\end{align}
\begin{equation}\label{eq:sum-lognormals}
  \vec{v}(t) = \sum_{i=1}^{N} \vec{v}_i(t) = \sum_{i=1}^{N} \left[
    \begin{matrix}
    \cos \phi_i(t) \\
    \sin \phi_i(t)
    \end{matrix}
    \right] D_i \Lambda(t; t_{0_i}, \mu_i, \sigma_i^2)
\end{equation}
\begin{equation}\label{eq:ang}
  \phi_i(t) = \theta_{s_i} + \frac{\theta_{e_i} - \theta_{s_i}}{2} \left[
    1 + \text{erf} \left(\frac{\ln (t - t_{0_i}) - \mu_i}{\sigma_i \sqrt{2}}\right)
  \right]
\end{equation}

Then, a \slm extractor~\cite{MartinAlbo15} computes the values of the model parameters that best fit to the observed velocity profiles.
The resulting model can become \emph{generative},
in the sense that it is possible to add perturbations to the model parameters
and produce different gesture variations as a result~\cite{Leiva17_iwc}, as follows:
\begin{equation}\label{eq:noise}
  \begin{aligned}
  p^*_i &= p_i + n_{p_i}
  \end{aligned}
\end{equation}
where $p_i=\{\,\mu_i, \sigma_i, D_i, \theta_{i}\,\}$ denotes the set of \slm parameters that are modified for this purpose.

In this work, the noise $n_{p_i} = \mathcal{U}(-n_i, n_i)$ applied to each primitive
follows a uniform distribution (\ie, a rectangular distribution with constant probability)
centered around the expected human variability ranges calculated and reported in the motor control theory literature~\cite{Galbally12b,Leiva17_iwc}:
$n_\mu = n_\sigma = 0.1$, $n_D = 0.15$, and $n_{\theta} = 0.06$.
Concretely, these values were estimated from $6,400$ signature samples collected from $400$ users~\cite{Galbally12b}
from five different geographic locations under controlled conditions.\footnote{
  Each signature example was annotated both at the stroke and at the ``primitive'' level (see \protect\autoref{fig:theory-example}).
}
Furthermore, these samples were captured over four sessions distributed over six months,
which makes this data robust enough to reflect both inter- and intra-session variability.

Note that other distributions and different noise values may be needed for different user categories, such as gestures articulated by users with visual impairments~\cite{Leiva17_chi} or by users with motor impairments~\cite{Ungurean18chi}.
We also should note that no perturbations are added to the $t_0$ parameter in this work,
since $t_0$ is very sensitive even to small fluctuations~\cite{Djioua09, Leiva16_g3}.
Nevertheless, perturbations in $t_0$ have been suggested to reflect changes in the sequence of command instantiation,
\eg, due to a decrease in attention or neuromotor fatigue~\cite{Djioua09}.
Therefore, further analysis of the $t_0$ parameter is left as an opportunity for future work.

\section{Omnis Pr{\ae}dictio}\label{sec:system}
\label{sec:OmnisPraedictio}

\omnis builds on the principles and concepts of the Kinematic Theory
and stroke gesture synthesis~\cite{Leiva17_chi,Leiva17_iwc,Leiva17_dis,Leiva16_g3}
to produce estimates of the user-independent distribution of features of interest for stroke gesture input.
In the following, we describe the fundamentals of our technique.

\subsection{A Note on Estimation vs. Prediction}
Often, the terms ``estimation'' and ``prediction'' are used interchangeably
to denote ``an approximation of a result,''
although there is a subtle distinction between them.
In particular, an \emph{estimation} usually refers to the creation of a mathematical model
that explains some observation, process, or natural phenomena,
whereas a \emph{prediction} refers to the use of an already built mathematical model
to calculate new values for unlabeled data.
In this article we use the term ``estimation'' to refer to the distribution results delivered by \omnis,
as they represent an approximate model for the unseen population of values for a given gesture feature,
\eg, the distributions illustrated in \autoref{fig:teaser} represent models of the variation expected to observe in stroke gesture input
in terms of three features.
Based on the estimation of frequency distributions,
\omnis predicts useful statistics, such as the mean or variance of various features and gesture types.
For reasons of consistent treatment and discussion,
we use the term ``estimation'' in the remainder of this article.

\subsection{Technical Details}
Let $g = \left\{ \left(x_i, y_i, t_i, p_i \right) \,|\, i=1..n \right\}$ be a stroke gesture
defined as a series of 2D points with associated timestamps and identifiers
reported by a touch-sensitive surface in response to finger or stylus movements.
Let $\mathcal{G}$ be the set of all gestures, and $\phi$ a real-valued function
defined over $\mathcal{G}$, \eg, $\phi(g)$ may be the path length of gesture $g$, the area size of its bounding box, etc.
Our goal for \omnis is to estimate the user-independent distribution of the values attainable by $\phi$ for gestures of the same type
with minimum effort from the designer's part, such as by starting from a single example, $g_0$,\footnote{
  We use the index $0$ for the gesture seed
  and indices $1$ to $N$ for the synthesized data.
} articulated by the designers themselves.

\begin{figure}[t]
  \centering
  \def\w{1cm}
  \includegraphics[width=\w]{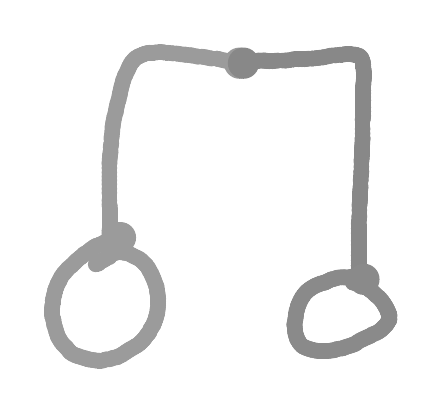}
  \includegraphics[width=\w]{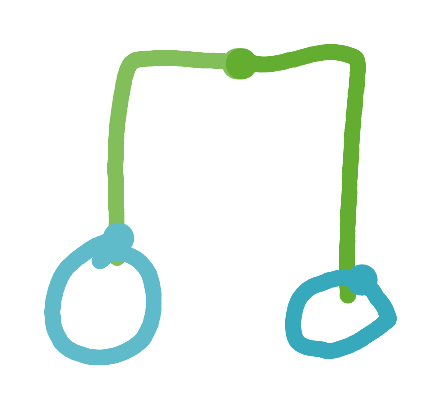}
  \includegraphics[width=\w]{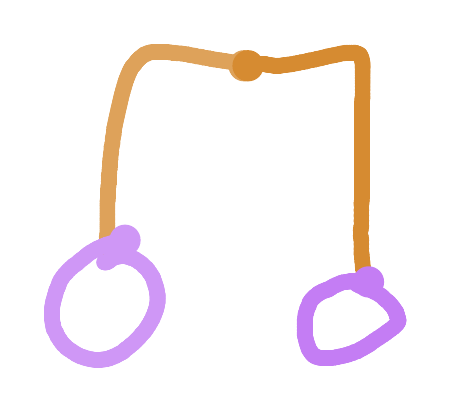}
  \includegraphics[width=\w]{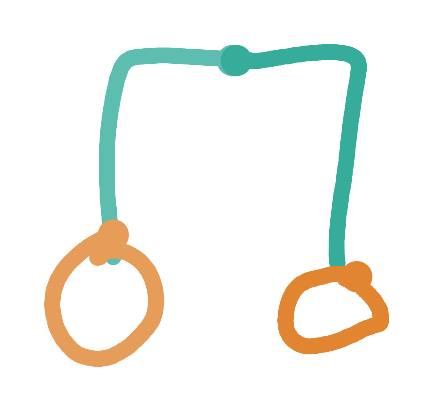}
  \includegraphics[width=\w]{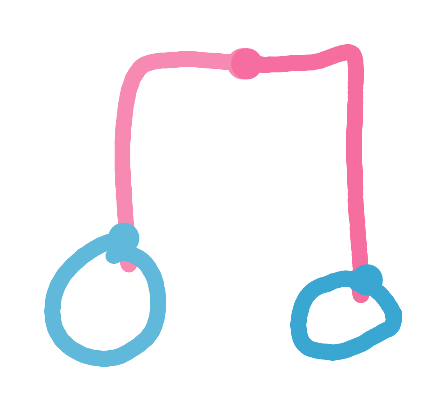}
  \includegraphics[width=\w]{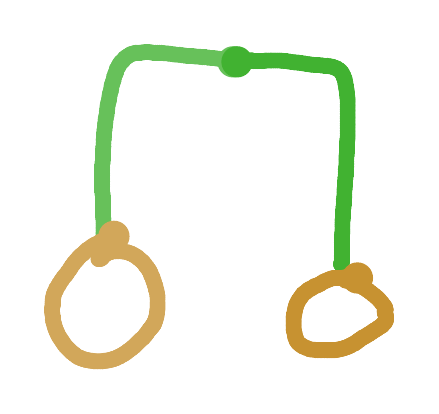}
  \includegraphics[width=\w]{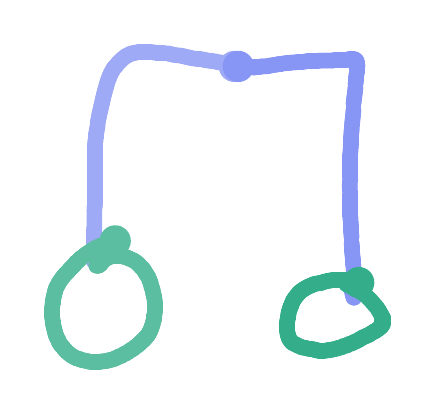}
  \includegraphics[width=\w]{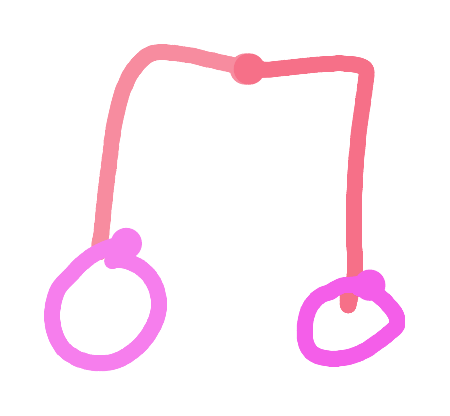}
  \includegraphics[width=\w]{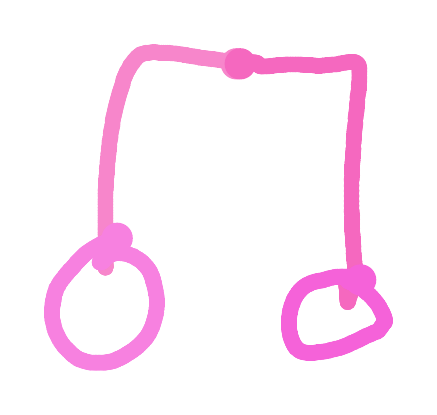}
  \includegraphics[width=\w]{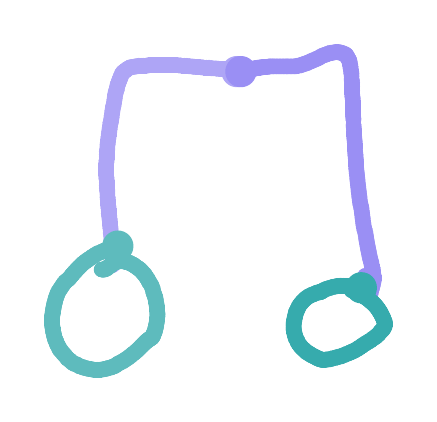}
  \includegraphics[width=\w]{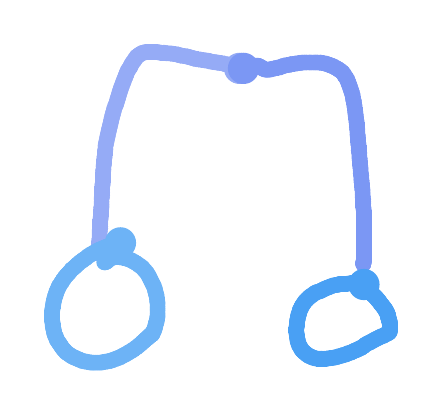}
  \\
  \includegraphics[width=\w]{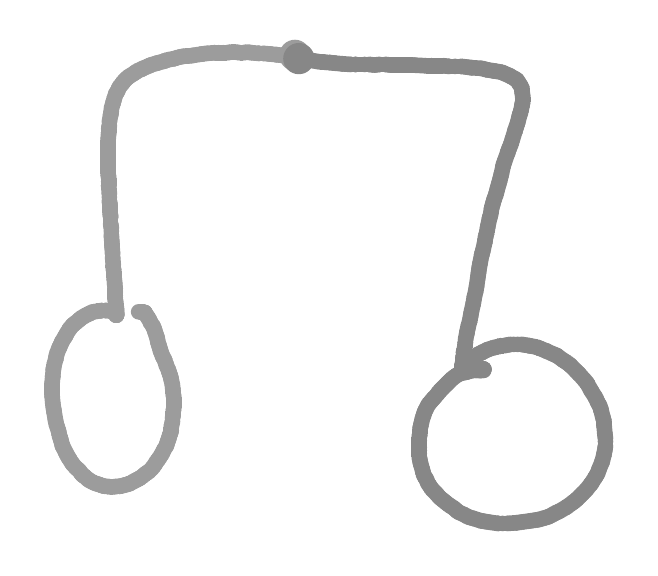}
  \includegraphics[width=\w]{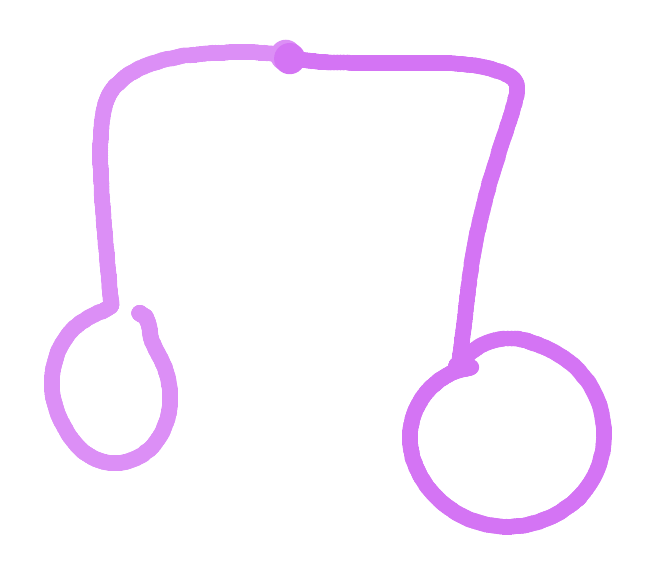}
  \includegraphics[width=\w]{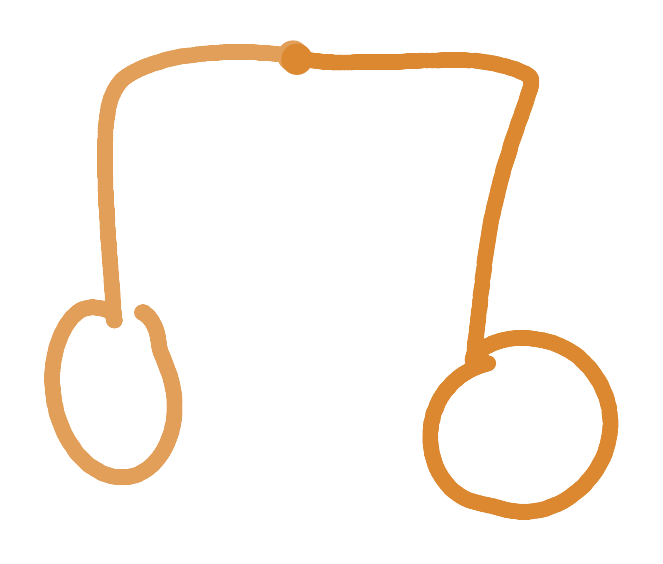}
  \includegraphics[width=\w]{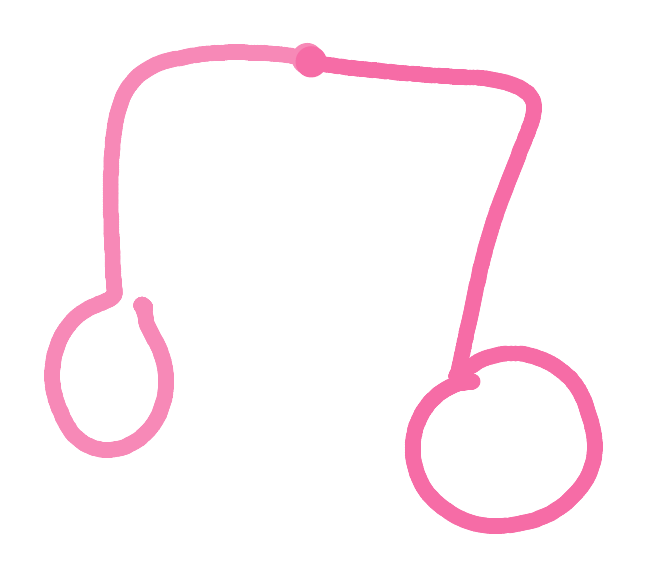}
  \includegraphics[width=\w]{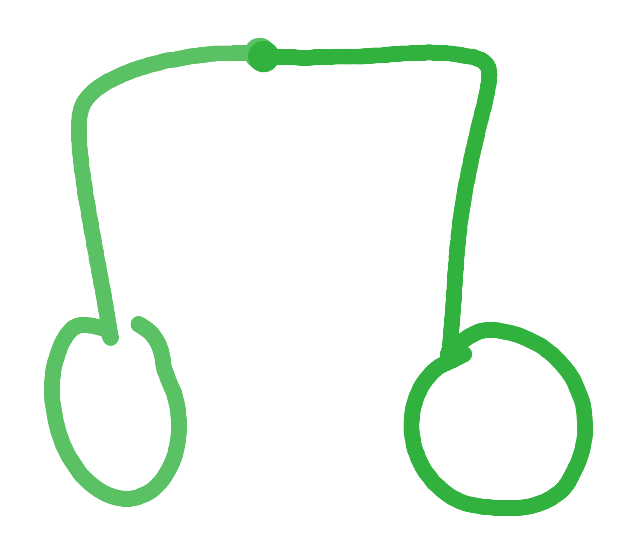}
  \includegraphics[width=\w]{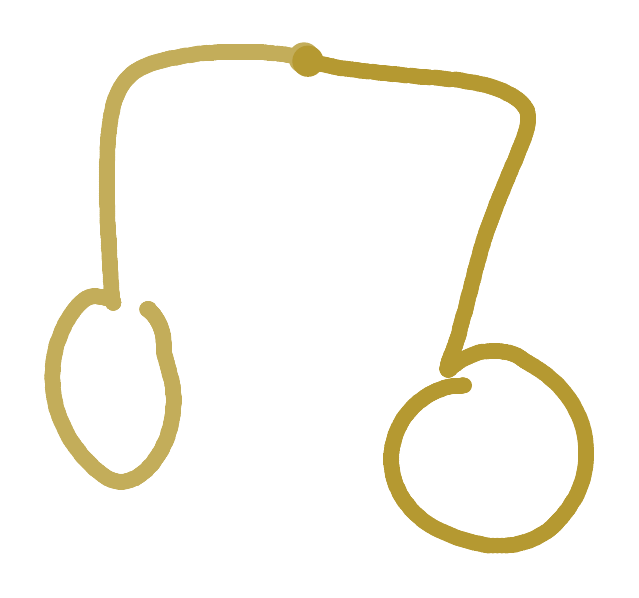}
  \includegraphics[width=\w]{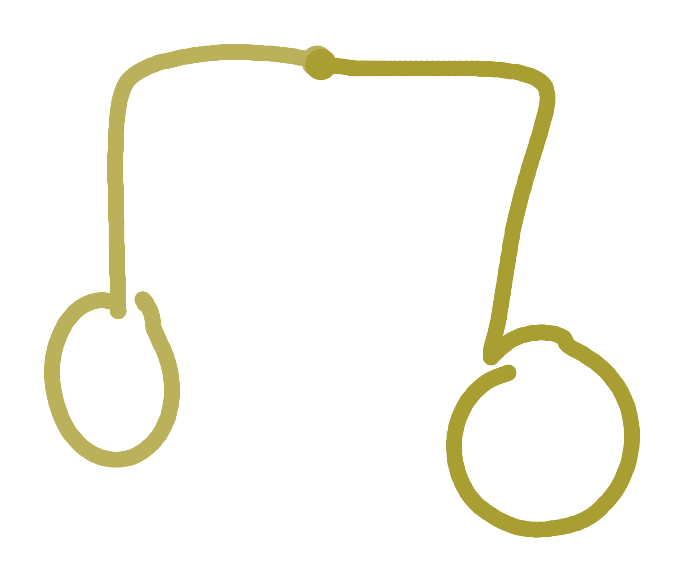}
  \includegraphics[width=\w]{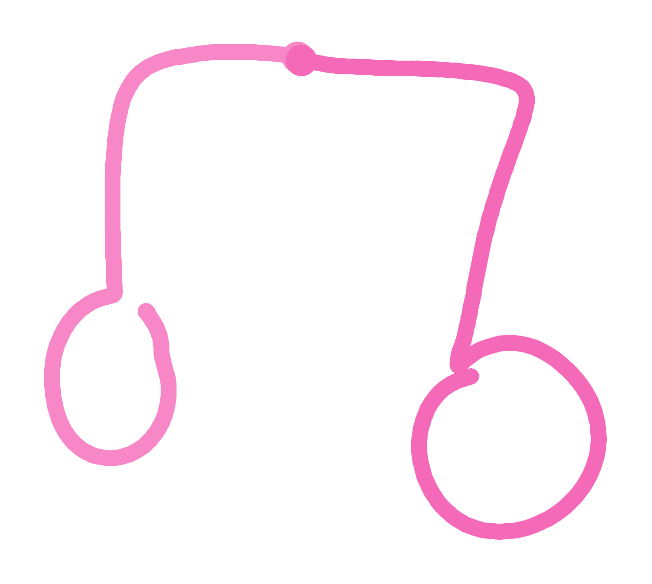}
  \includegraphics[width=\w]{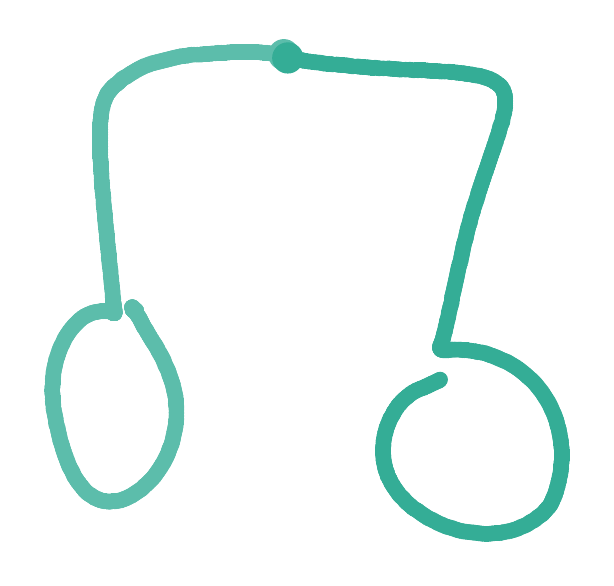}
  \includegraphics[width=\w]{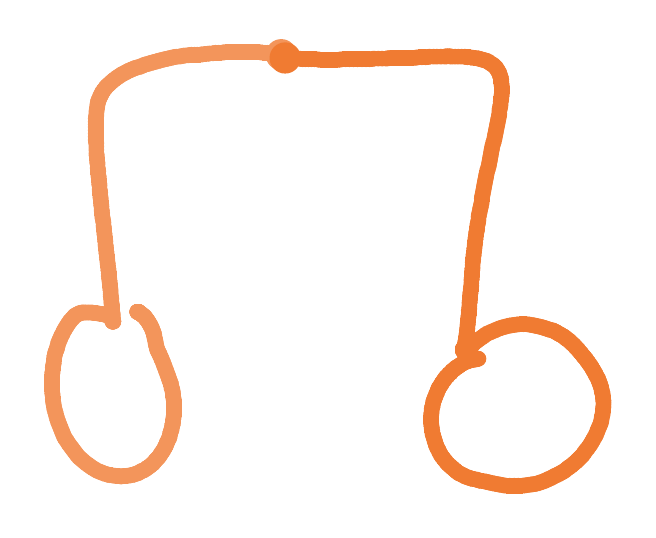}
  \includegraphics[width=\w]{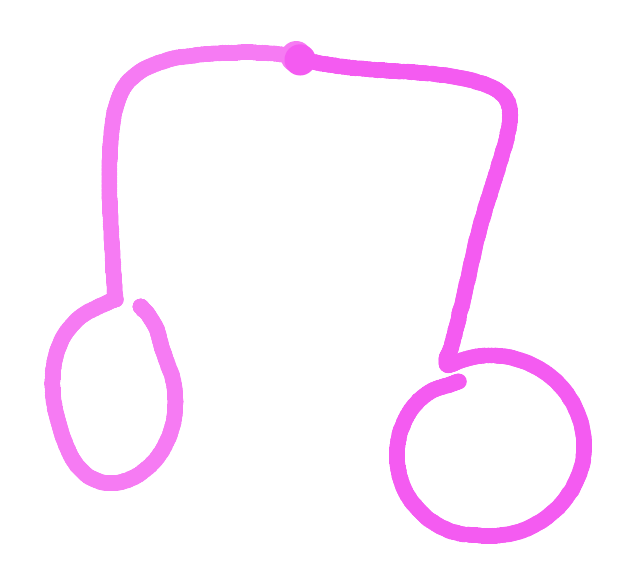}
  \\
  \includegraphics[width=\w]{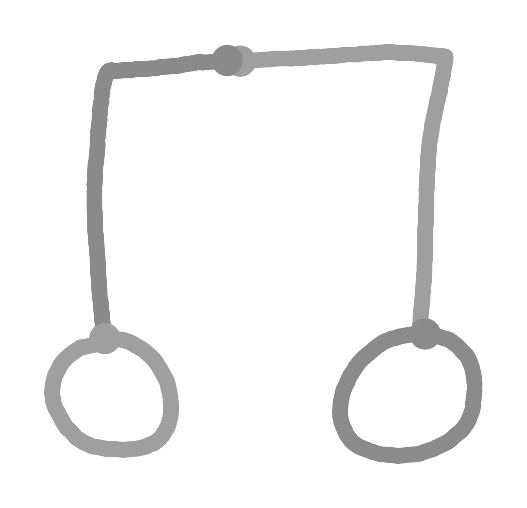}
  \includegraphics[width=\w]{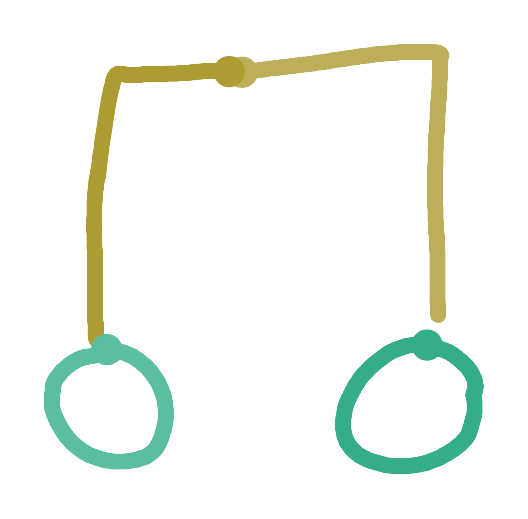}
  \includegraphics[width=\w]{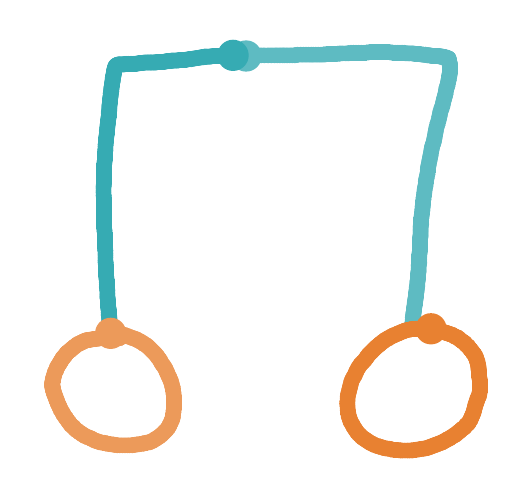}
  \includegraphics[width=\w]{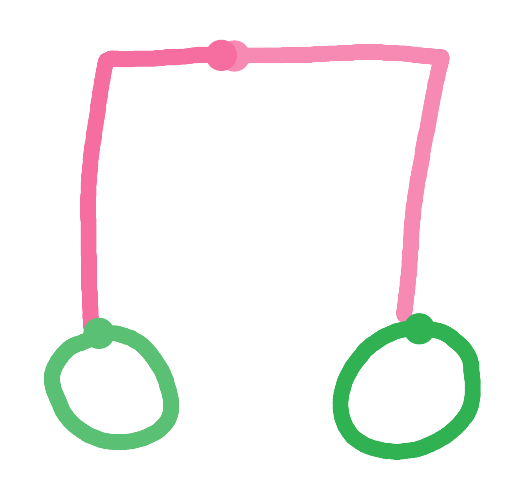}
  \includegraphics[width=\w]{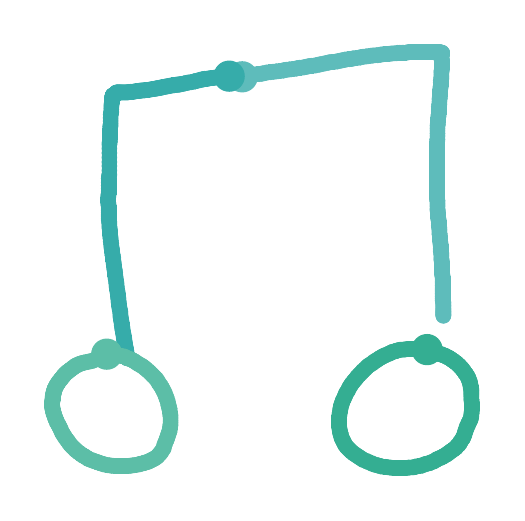}
  \includegraphics[width=\w]{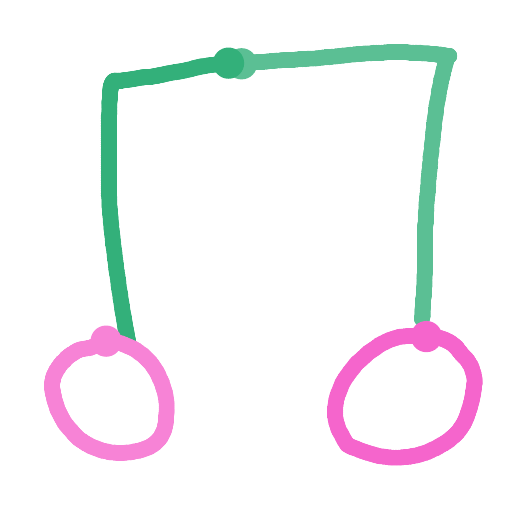}
  \includegraphics[width=\w]{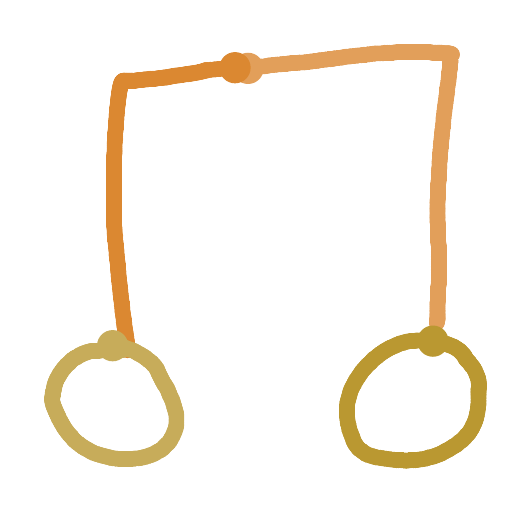}
  \includegraphics[width=\w]{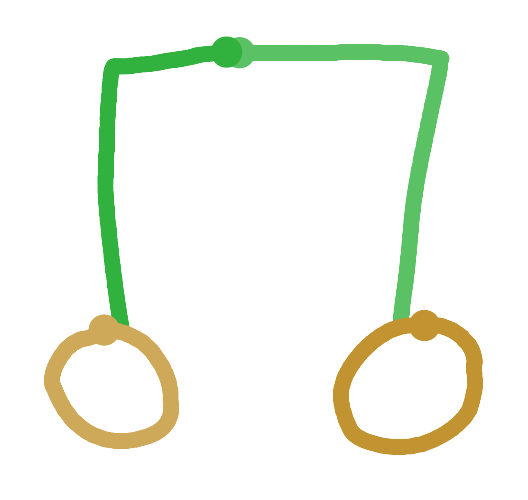}
  \includegraphics[width=\w]{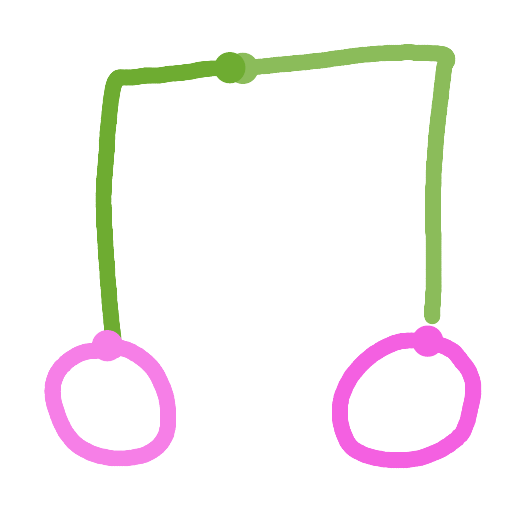}
  \includegraphics[width=\w]{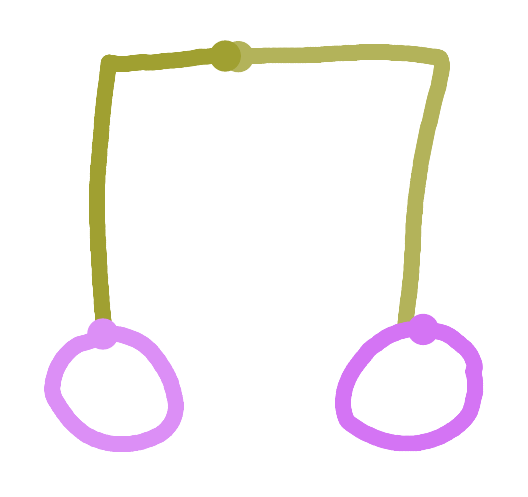}
  \includegraphics[width=\w]{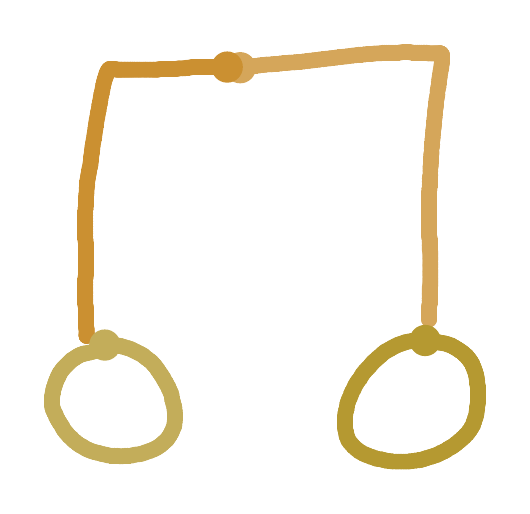}
  \\
  \includegraphics[width=\w]{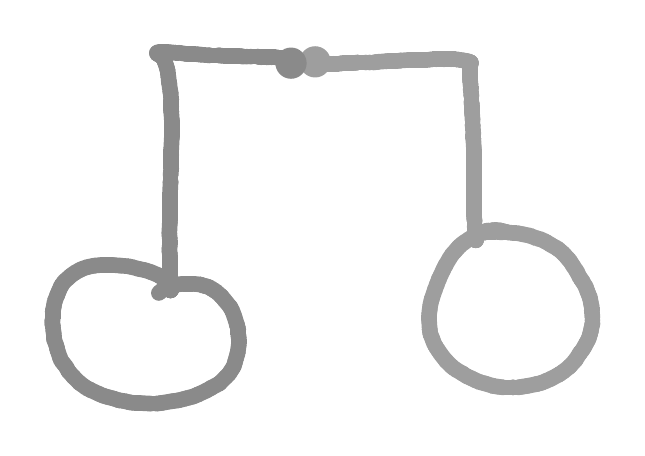}
  \includegraphics[width=\w]{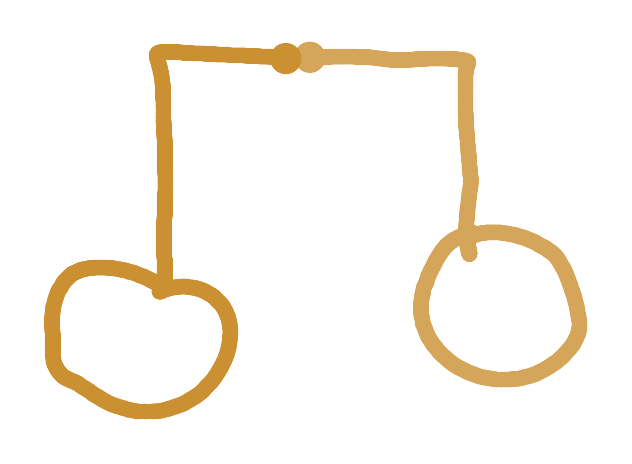}
  \includegraphics[width=\w]{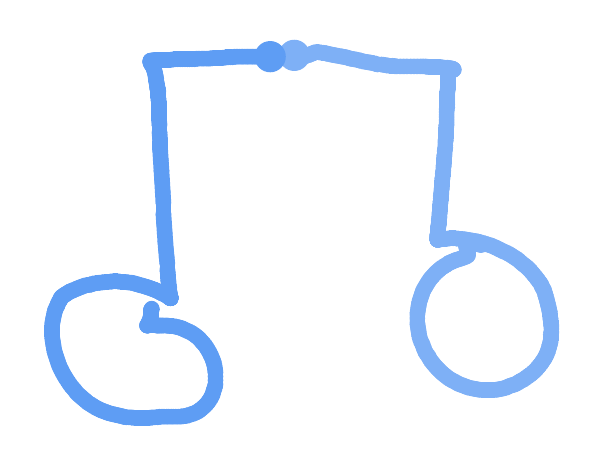}
  \includegraphics[width=\w]{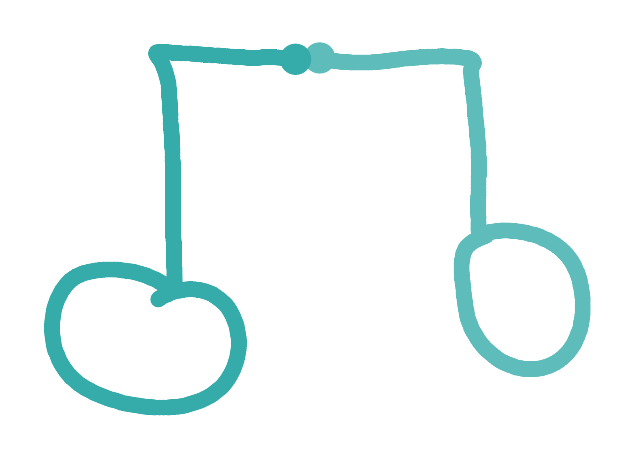}
  \includegraphics[width=\w]{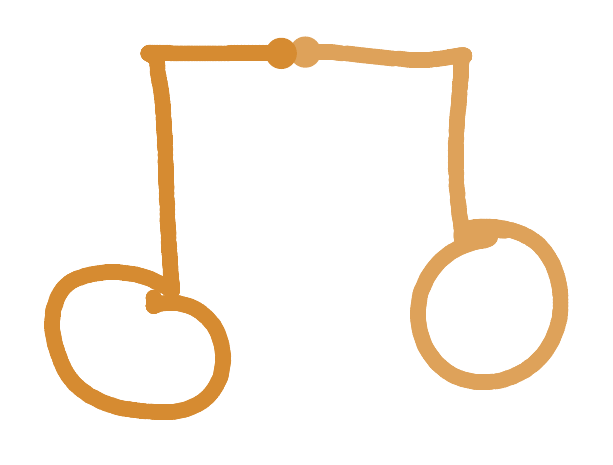}
  \includegraphics[width=\w]{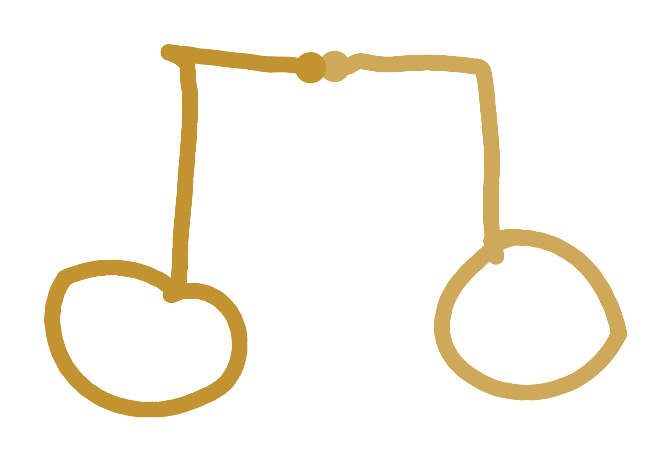}
  \includegraphics[width=\w]{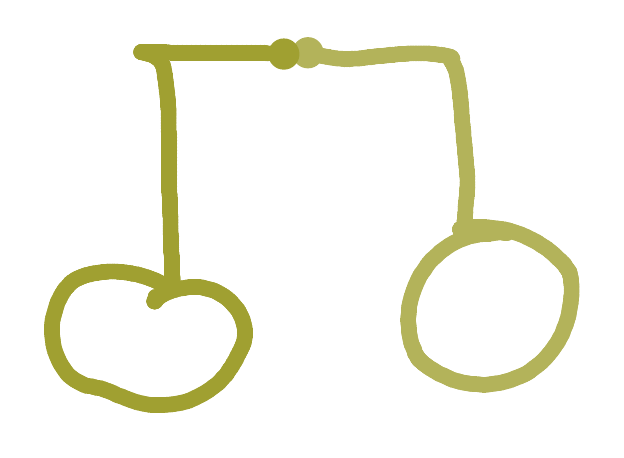}
  \includegraphics[width=\w]{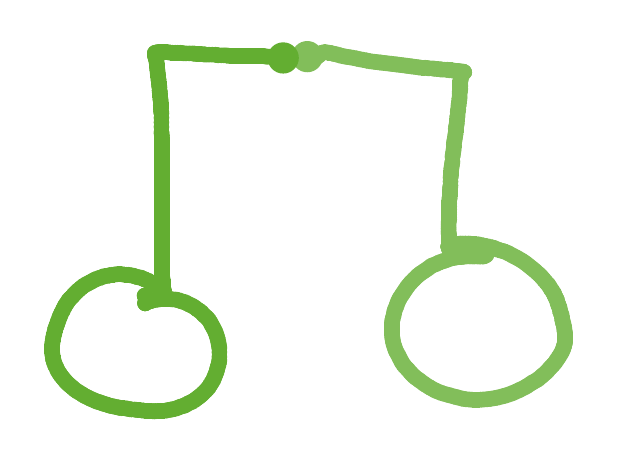}
  \includegraphics[width=\w]{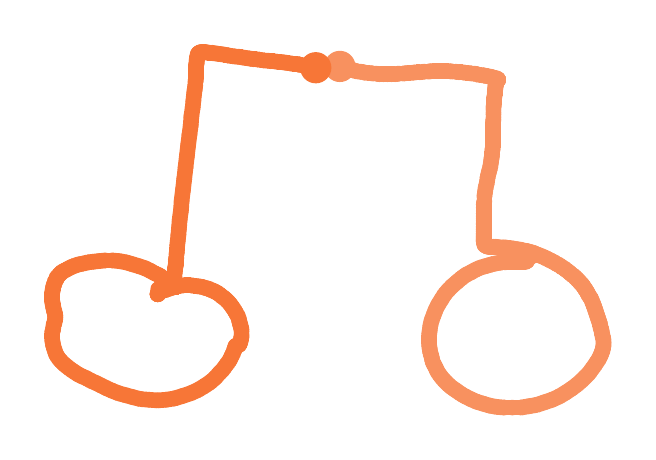}
  \includegraphics[width=\w]{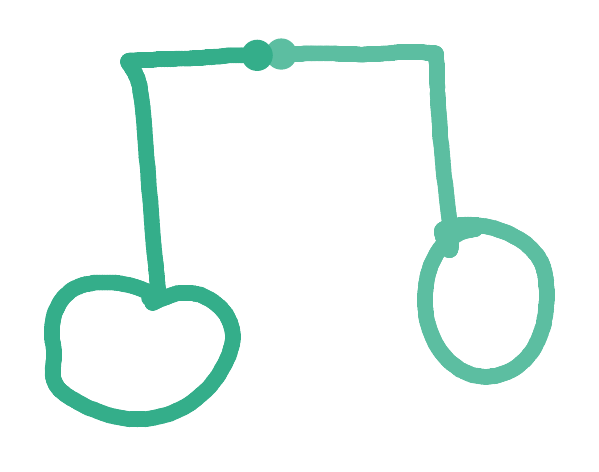}
  \includegraphics[width=\w]{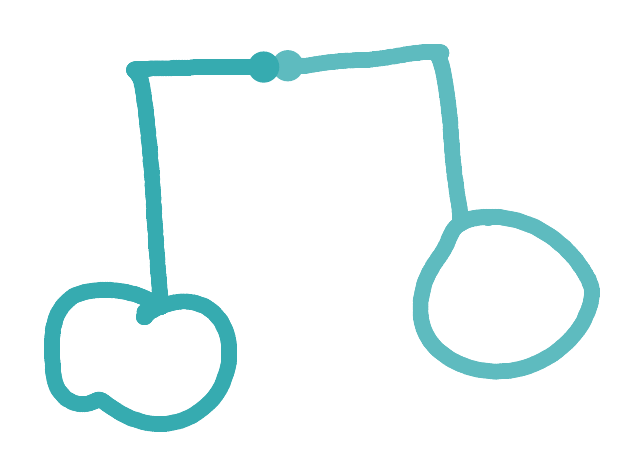}
  \caption{
    Examples of stroke gestures synthesized by \omnis.
    Note: human examples are shown in gray on the first column;
    synthetic gestures are shown using random colors.
  }
  \label{fig:synth-samples}
\end{figure}

To this end, we rely on the \slm model of the Kinematic Theory~\cite{Plamondon06}
to automatically generate a population of synthetic gestures starting from the seed $g_0$ provided by the designer.
Prior work~\cite{Leiva17_iwc} showed that stroke gestures synthesized in this manner~\cite{Leiva16_g3}
posses human-like characteristics,
and human observers can hardly discern between real and synthetic gesture shapes~\cite{Leiva17_dis}.
Let $\mathcal{G}_0 = \left\{g_1, g_2, \dots, g_N  \right\}$ denote the population of synthetic gestures derived from the gesture seed $g_0$, which
we use to approximate the actual population of all possible articulations of gesture type $g$.
Informed by results reported in previous work on successful gesture synthesis and applications~\cite{Leiva16_g3,Leiva18_keytime,Leiva18_gato}, we designed \omnis to generate a sample of $N=100$ synthetic gestures starting from the seed $g_0$.
This way, we generate a sufficient amount of variability in terms of the kinematic and geometric properties of the synthesized gestures,
reflective of the variability observed for gestures articulated by actual users as reported in~\cite{Leiva17_iwc,Leiva17_dis}.
\autoref{fig:synth-samples} exemplifies different instantiations of $\mathcal{G}_0$
for the ``car'' symbol depicted in \autoref{fig:datasets}.

To generate the distribution of some feature of interest $\phi$,
\omnis computes the value of the feature for each synthetic gesture from the set $\mathcal{G}_0$;
see \autoref{fig:teaser} for a few examples of distributions for three gesture types and features.
In our implementation, we calculate and report the mean of this synthetic distribution as our user-independent estimation of $\phi(g)$,
but we also compute other measures of location and dispersion,
such as min, max, median, standard deviation, and 95\% and 99\% confidence intervals of the mean;
see the next section for more details.
Overall, \omnis takes as input a user-provided seed gesture $g_0$,
and outputs the estimated distribution of the feature of interest $\phi$
together with a series of representative statistics.
The feature of interest $\phi$ can be specified manually using our web application by selecting it from a pre-defined list of default functions,
or can be defined programmatically via our API.

\section{Evaluation}\label{sec:evaluation}

We conducted a controlled experiment to evaluate the estimation accuracy of \omnis with respect to a variety of gesture features (\eg, path length, curvature, area size, etc.) and a wide range of stroke gesture types, including single- and multi-touch as well as single- and multi-stroke gestures.

\subsection{Gesture Datasets}
\begin{figure*}[!ht]
  \centering
  \def\w{0.32\textwidth}
  \subfloat[\label{fig:mt_strokes}\mmStrokes dataset]{\includegraphics[width=\w]{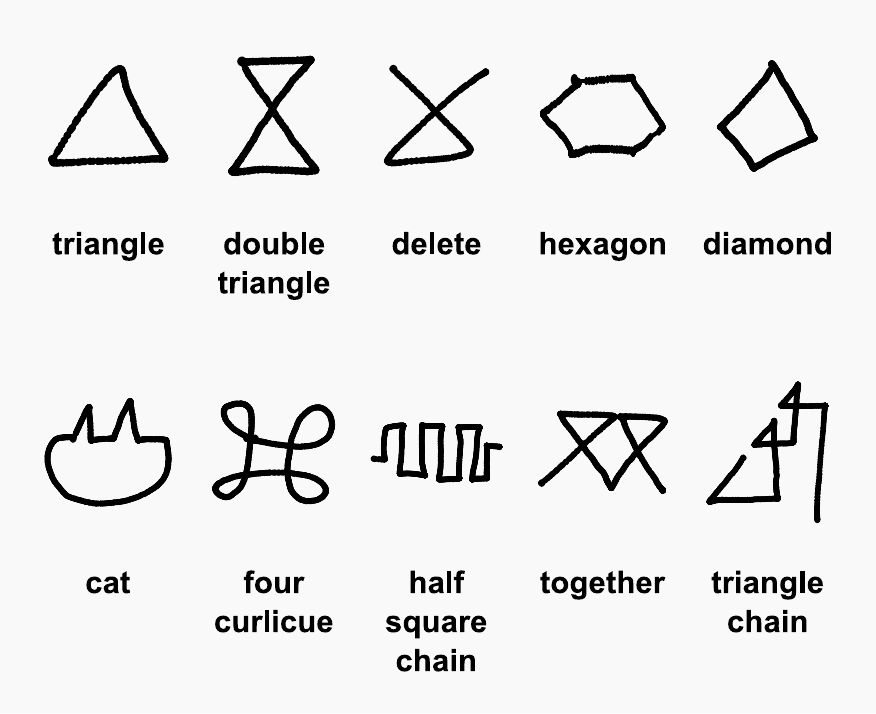}}\hfill
  \subfloat[\label{fig:mt_fingers}\mmFingers dataset]{\includegraphics[width=\w]{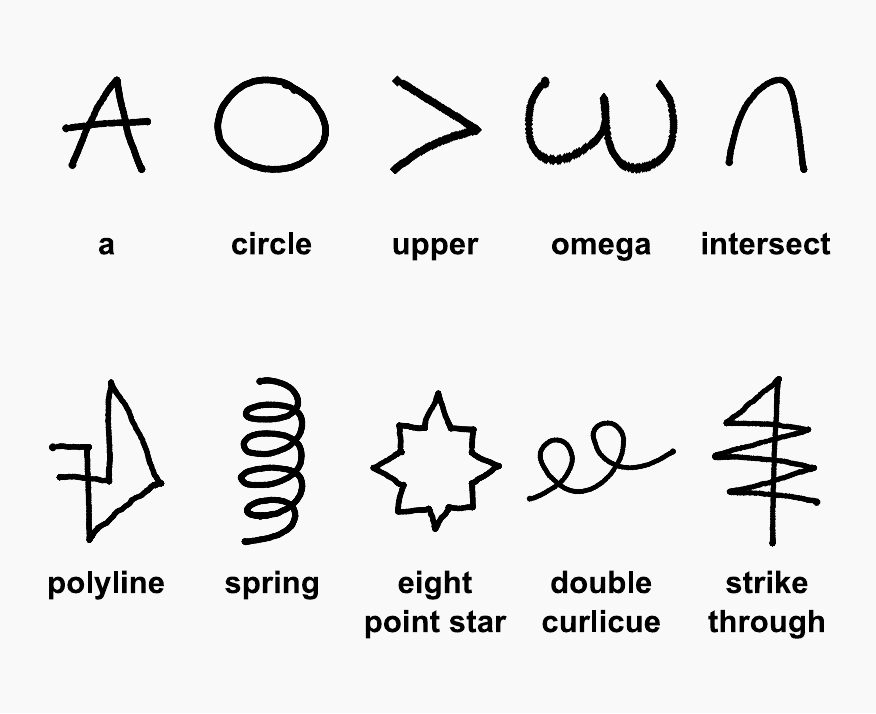}}\hfill
  \subfloat[\label{fig:mt_synchronization}\mmHands dataset]{\includegraphics[width=\w]{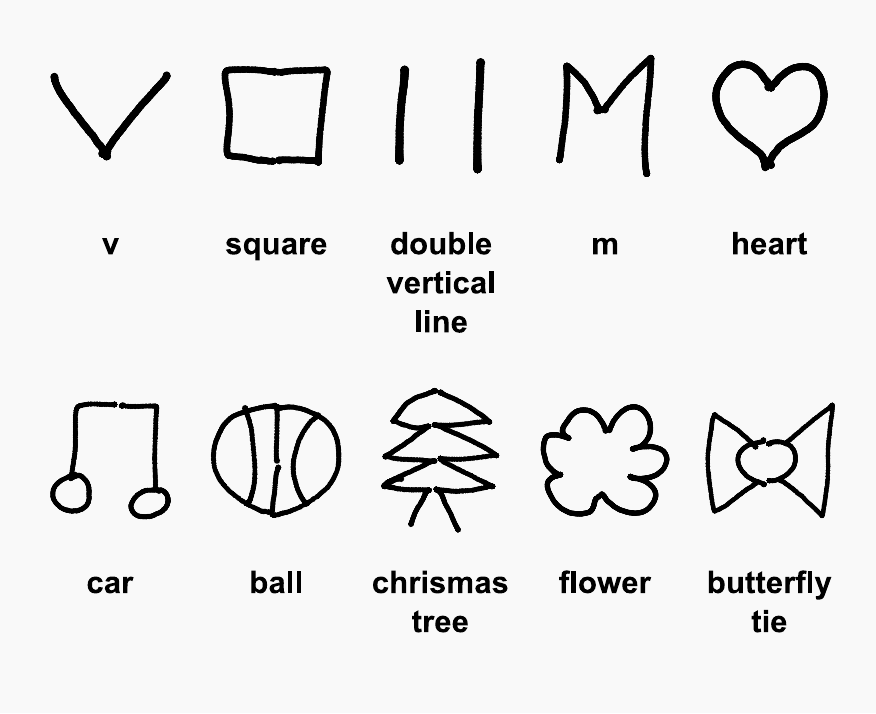}}\hfill
  \caption{
    The gesture datasets used for the evaluation of \omnis; see also \protect\autoref{tbl:datasets} for a summary of these datasets.
    Notes: each illustration from this figure is an actual sample produced by the participants from~\cite{Rekik14}.
    Top rows comprise gestures that were designed by~\cite{Rekik14} to be familiar to participants and, consequently, easy to perform. Bottom rows comprise gestures designed to be unfamiliar and, thus, more difficult to perform.
  }
  \label{fig:datasets}
\end{figure*}

We evaluated \omnis on three gesture datasets\footnote{
  \url{https://sites.google.com/site/yosrarekikresearch/projects/gesturedifficulty}
} collected by Rekik \etal~\cite{Rekik14}. The datasets comprise $30$ distinct gesture types performed repeatedly by $18$ participants on a 3M multitouch display; see \autoref{tbl:datasets}. The gestures are represented as series of 2D touch coordinates with associated timestamps and touch identifiers.
Each gesture was articulated with $5$ repetitions under various conditions:
\begin{enumerate}
  \topsep 0pt
  \itemsep 1pt
  \item \textit{Different numbers of strokes} (the \mmStrokes dataset) with three variants:
  one stroke, two strokes, and at least three strokes performed to produce a multistroke gesture.
  This dataset comprises $2,700$ samples; see \autoref{fig:mt_strokes}.

  \item \textit{Different numbers of fingers} (\mmFingers dataset) with three variants:
  one finger, two fingers, and at least three fingers touching the screen at once.
  This dataset comprises $2,700$ gesture samples; see \autoref{fig:mt_fingers}.

  \item \textit{Using one or two hands} (\mmHands dataset) with two variants:
  sequential input (one hand) and bimanual input (two hands).
  This dataset comprises $1,800$ gesture samples; see \autoref{fig:mt_synchronization} for the illustration of its gesture types.
\end{enumerate}

We specifically chose these datasets for the evaluation of \omnis
as they contain a good mixture of geometrical shapes and symbols
with a wide range of shape complexity~\cite{Isokoski01},
were specifically designed to contain both familiar and unfamiliar shapes and symbols~\cite{Rekik14},
and they cover a variety of articulations of stroke gestures,
including unistrokes, multistrokes, and multitouch gestures performed using one or both hands.
Overall, the three datasets contain a total number of $7,200$ gesture samples articulated
under various input conditions regarding the number of hands, fingers, and strokes.

\begin{table}[!ht]
  \centering
  \small
  \begin{tabular}{*2l *4r}
    \toprule
    \tth{Dataset} & \tth{Input\\conditions} & \tth{Gesture\\types} & \tth{Users} & \tth{Trials} & \tth{Num.\\samples} \\
    \midrule
    \mmStrokes & 1, 2, $\geq$3 strokes & 10 & 18 & 5 & 2,700 \\
    \mmFingers & 1, 2, $\geq$3 fingers & 10 & 18 & 5 & 2,700 \\
    \mmHands   & 1 or 2 hands          & 10 & 18 & 5 & 1,800 \\
    \midrule
    \textbf{Total} & & \textbf{30} & \textbf{18} & \textbf{5} & \textbf{7,200} \\
    \bottomrule
  \end{tabular}
  \caption{Overview of the gesture datasets from~\cite{Rekik14} used to evaluate \omnis.}
  \label{tbl:datasets}
\end{table}

\subsection{Methodology}\label{sec:method}

We employed a mixed methodology based on the principles of leave-one-out cross-validation testing in order to understand how accurately can \omnis estimate distributions of various stroke gesture features as well as statistics of those distributions, \eg, sample means and medians, for various types of stroke gestures, gesture features, but also target audiences of our technique and online tool. Thus, to make sure that we report evaluation results that are as complete as possible and both theoretically and practically relevant (\ie, the evaluation addresses both the theoretical, ideal performance of \omnis to estimate full feature distributions, but also the expected performance of \omnis in practice for estimating mean statistics from just one gesture seed provided by the designer), we evaluated \omnis in two distinct scenarios:

\begin{enumerate}[label=\textbf{EV}$_\arabic*.$, leftmargin=1cm]
	\itemsep 1pt

	\item
	Estimation of \textbf{statistics} (\eg, mean, median, trimmed means, etc.) for combinations of gesture features and stroke gesture types by using just one gesture seed $g_0$ collected from one user only. For the sake of brevity, we report estimations of mean values only. This evaluation scenario connects directly to our web implementation of \omnis, which delivers such estimations starting from one gesture example $g_0$ provided by the practitioner via the online user interface or via our API support. This evaluation reports the expected accuracy estimation performance of \omnis for such practical usage scenarios intended for our online tool.

	\item
	Estimation of \textbf{distributions} by using more gesture examples collected from one user. This scenario addresses the capacity of \omnis to compute and report estimations of the distributional forms of various combinations of gesture features and stroke gesture types, when more gesture data is available. This scenario is of practical interest to researchers addressing various aspects of gesture input, such as the variability margins of users' performance with gesture input, the types of distributions expected for various gesture features, differences between groups of end users according to specific dimensions of analysis, and so on. To this end, this scenario is centered on and evaluates the expected accuracy estimation performance of \omnis regarding gesture feature distributions.
\end{enumerate}

We also considered the following independence conditions:
\begin{description}[leftmargin=0cm]
	\itemsep 0pt

	\item[Feature-independence.] We evaluated the estimation accuracy of \omnis for a wide range of features $\phi$ relevant for stroke gesture input by surveying the available literature on gesture analysis and recognition; to be described in the next subsection. This independence condition was applied to both EV$_1$ and EV$_2$ evaluation scenarios.

  \item[Articulation-independence.] We evaluated the performance of \omnis for different articulations of the same gesture type represented by various seed gestures $g_0$, since we expected that the quality of the articulation of the gesture seed would likely have an influence on the quality of the estimations delivered by \omnis, \eg, a carefully-executed letter ``A'' may lead to better estimations vs. a sloppy and quickly-executed letter ``A.'' Therefore, to be able to catch and report such effects, we computed the \textit{best-case}, \textit{worst-case}, and \textit{average-case} performance of \omnis for the EV$_1$ evaluation scenario:

	\begin{enumerate}
		\topsep 0pt
		\itemsep 1pt

		\item
		\textit{Best-case scenario}, represented by the gesture seed $g_0$ with the estimated mean value that is closest to the groundtruth mean in terms of our evaluation measures.\footnote{The groundtruth is represented by the mean of the feature computed for all the gestures produced by all the participants, except the participant from which $g_0$ was collected.}

		\item
		\textit{Worst-case scenario}, represented by the gesture seed $g_0$ with the estimated mean value that is farthest from the groundtruth mean.

		\item
		\textit{Average-case scenario}, represented by the average of our evaluation measures across all gesture seeds $g_0$ from all participants from $\mathcal{P}$.
	\end{enumerate}

	The best-case and worst-case scenarios help us understand the lower and upper bounds of estimation accuracy attainable in practice by \omnis, highlighting the impact of the original gesture seed $g_0$ provided by the practitioner using the \omnis web application.

	\item
	\item[User-independence.] We evaluated \omnis for seed gestures $g_0$ collected from different users with two procedures specifically designed for the evaluation scenarios EV$_1$ and EV$_2$, respectively, as follows.

	For the EV$_1$ scenario regarding estimations of statistics, we computed and reported the average performance attainable by \omnis by considering gestures collected from all the participants following a leave-one-out cross-validation procedure:
	\begin{enumerate}[label=step$_\arabic*$:$\,\,$, leftmargin=1cm]
		\itemsep 0pt
		\item

		Let $p \in \mathcal{P}$ be a participant from our evaluation datasets, representative of an end user producing stroke gestures. In our experiment, the set $\mathcal{P}$ contains $18$ participants; see \autoref{tbl:datasets}.

		\item
		Let $g_0$ be a seed gesture provided by participant $p$.
		We use $g_0$ to synthesize $N=100$ artificial gestures to compute an estimation of our features of interest $\phi_i(g_0)$, according to our technique described in \autoref{sec:OmnisPraedictio}.

		\item
		We compute the groundtruth values $\Phi_i(g)$ of each feature $\phi_i$ by considering all the gestures $g$
		provided by all the other participants from the set $\mathcal{P} - \{p\}$. In our experiment, there are $17$ participants distinct from the participant from which the seed was selected.

		\item
		We compare the mean of $\phi_i(g_0)$ against the groundtruth mean of $\Phi_i(g)$, and repeat step 2 of this procedure with the next gesture $g_0$ from the current participant $p$, after which we repeat step 1 by considering the next participant from our datasets. Comparisons between estimated and groundtruth means	are performed using absolute and relative measures of accuracy, described in the next subsection.
	\end{enumerate}

	For the EV$_2$ evaluation scenario regarding feature distributions, we implemented the following variation of the previous procedure:
	\begin{enumerate}[label=step$_\arabic*$:$\,\,$, leftmargin=1cm]
		\topsep 0pt
		\itemsep 0pt

		\item
		Let $p \in \mathcal{P}$ be a participant from our evaluation datasets, representative of an end user producing stroke gestures. In our experiment, the set $\mathcal{P}$ contains $18$ participants; see \autoref{tbl:datasets}.

		\item
		Let $g_0$ be a seed gesture provided by participant $p$.
			We use $g_0$ to synthesize an artificial gesture at random,\footnote{
				This is performed automatically and without any supervision,
				to avoid cherry-picking.
			}
			and we repeat this step for each gesture of participant $p$.
			In the end, we have an estimation $\phi_i(g_0)$ of our features of interest by using data from participant $p$.

		\item
		We compute the groundtruth values $\Phi_i(g)$ of feature $\phi_i$ by considering all the human gestures $g$
			provided by all the other participants from the set $\mathcal{P} - \{p\}$. In our experiment, there are $17$ participants distinct from the participant who provided gesture examples.

		\item
		We compare the estimated distributions of features $\phi_i(g_0)$ against the groundtruth distributions of $\Phi_i(g)$, and repeat the procedure from step 1 with the next participant $p$ from our evaluation datasets.
			Comparisons between estimated and groundtruth means
			are performed using absolute and relative measures of accuracy, described in the next subsection.
	\end{enumerate}
\end{description}

As can be observed, our evaluation procedure considers various aspects of the performance of \omnis regarding (1)~the applicability of \omnis for practical purposes, where designers may wish to employ it for various types of stroke gesture features; (2)~the performance of \omnis when only one gesture example is provided by the practitioner, such as in the web application provided as a companion tool for this paper, and (3)~the theoretical performance of \omnis across different users and datasets. Next, we present our measures used to evaluate the accuracy of \omnis under these conditions.

\subsection{Evaluation Measures}
We compared the numerical estimations delivered by \omnis for various gesture features
with the feature values computed on gestures articulated by actual users
(denoted in the following tables and figures as the \emph{True} condition).
We used the following measures of \textit{relative} and \textit{absolute} accuracy:

\begin{enumerate}
  \itemsep 1pt

  \item
  \textsc{Ranking-Accuracy} evaluates the extent to which \omnis delivers the correct \emph{ranking}
  of gesture types from the perspective of our feature of interest.
  For example, if the average speed of gestures A and B are $s_\text{A}=1.5$\,px/s and $s_\text{B}=2.5$\,px/s, respectively,
  then \omnis is accurate if the estimated values also respect this relative order,
  \ie, $\hat{s}_\text{A} < \hat{s}_\text{B}$.
  To handle more than two gesture types, the ranking accuracy is evaluated
  using Spearman's rank correlation coefficient $r_s$, defined in $[-1,1]$.
  The closer $r_s$ to $1$, the more accurate \omnis is.

  \item
  \textsc{Absolute-Error} evaluates the extent to which \omnis delivers the correct \emph{magnitude}
  of the feature of interest for a given gesture type.
  For example, if the estimated bounding box area of gesture A is $\hat{b}_\text{A}=21000$\,$\text{mm}^2$,
  but the groundtruth area is $b_\text{A}=19890\,\text{mm}^2$,
  then the absolute error is $\left|\hat{b}_\text{A} - b_\text{A}\right| = 1110\,\text{mm}^2$.

\end{enumerate}

Note that \omnis computes several measures of location and dispersion for each feature,
such as the mean, median, standard deviation, variance, or confidence intervals.
However, for the sake of simplicity, we report just the estimated means in this article,
since previous work on human performance estimation has shown
that the mean is an accurate estimator of groundtruth values~\cite{Leiva18_keytime, Leiva18_gato}.
The implementation of \omnis as a web application delivers estimations for all the measures above;
see \nameref{sec:discussion} section.

\subsection{Gesture Features}

We conducted an extensive survey of the literature on gesture recognition and analysis~\cite{Rekik14, Rubine91, Blagojevic10, Vatavu11, Long99, Tu12, Vatavu13_relacc, Anthony13gi, Zhai12, Long00, Isokoski01, Shaw16, Vatavu:2013:SML, Vatavu:2018, Willems09, Kane09} in order to identify representative and useful features for stroke gesture input.
\autoref{tbl:features} provides a brief description of each feature.
For the practical purposes of our evaluation,
we considered a set of $18$ features reported by prior work as key to
evaluate user performance with gesture input (features 1--5), inform gesture set design (6--8), and recognize stroke gestures (6,9--18).

\begin{table}[!h]
  \centering
  \small
  \hspace*{-1.5cm}
  \begin{tabular}{rlccl}
    \toprule

    \tth{No.} & \tth{Feature} & \tth{Units} & \tth{Ref.} & \tth{Description} \\
    \midrule

    1 & \featsmall{production_time} & s & \cite{Cao07, Leiva18_keytime} & Total time to produce the gesture \\
    2 & \featsmall{avg_speed} & px/s & \cite{Kane11, Tu12} & Gesture path length / production time \\
    3 & \featsmall{line_similarity} & $-$ & \cite{Anthony13gi} & Distance between first and last pts / path length \\
    4 & \featsmall{aspect_ratio} & $-$ & \cite{Kane11} & Width of bounding box / height \\
    5 & \featsmall{turning_angle} & rad & \cite{Shaw16} & Sum absolute value of point angles \\

    \midrule

    6 & \featsmall{box_area} & px$^2$ & \cite{Vatavu:2013:SML} & Bounding box area  \\
    7 & \featsmall{curviness} & rad & \cite{Long00} & Sum of inter-segment angles \\
    8 & \featsmall{density} & $-$ & \cite{Long00} & Path length / distance between first and last pts \\

    \midrule

    9 & \featsmall{aspect} & rad & \cite{Long00} & abs(45$^\circ$ $-$ angle of bounding box) \\
    10 & \featsmall{path_length} & px & \cite{Anthony13gi, Rubine91} & Sum distances between adjacent pts \\
    11 & \featsmall{fl_distance} & px & \cite{Blagojevic10, Rubine91} & Distance between first and last pts \\
    12 & \featsmall{num_segments} & $-$ & \cite{Blagojevic10} & Number of fragments after corner detection \\
    13 & \featsmall{num_intersections} & $-$ & \cite{Blagojevic10} & Number of self-intersections at stroke endpts \\
    14 & \featsmall{lp_ratio} & $-$ & \cite{Blagojevic10} & Path length / convex hull perimeter \\
    15 & \featsmall{lb_ratio} & $-$ & \cite{Blagojevic10} & Path length / diagonal length of bounding box \\
    16 & \featsmall{hb_ratio} & $-$ & \cite{Blagojevic10} & Convex hull area / bounding box area \\
    17 & \featsmall{perimeter_efficiency} & $-$ & \cite{Blagojevic10} & 2$\cdot$ sqrt($\pi \cdot$ convex hull area) / hull perimeter \\
    18 & \featsmall{perimeter_to_area} & px$^{-1}$ & \cite{Blagojevic10} & Convex hull perimeter / convex hull area \\
    \bottomrule
  \end{tabular}
  \caption{
    The list of stroke gesture features evaluated in this work
    grouped into three categories of practical interest:
    features to evaluate user performance (rows 1--5),
    features to inform gesture set design (rows 6--8),
    and features for gesture recognition (rows 9--18).
    Note: the `Ref.' column indicates the papers where these features were proposed and/or employed.
  }
  \label{tbl:features}
\end{table}

Note that these feature categories are not mutually exclusive.
For example, \feat{production_time} is also a good estimator of users' perceived difficulty to articulate stroke gestures~\cite{Vatavu11, Rekik14}.
Features no. 2, 3, and 5 were found to correlate with users' consistency in gesture articulation~\cite{Anthony13gi},
whereas \feat{box_area} has been used both to estimate the user-perceived scale of stroke gestures~\cite{Vatavu:2013:SML}
and to assist partial gesture recognition~\cite{Appert:2010:SDP}.
Similarly, \feat{num_segments} was employed to distinguish text from graphics in handwritten ink~\cite{Bishop:2004:DTG}
and \feat{perimeter_efficiency} was used for sketch-based retrieval~\cite{Leung:2002}.
Finally, features 14 and 18 were used in the CALI recognizer~\cite{Fonseca:2002}.
Therefore, this set of 18 features represents a good summary of the most relevant features used in our community.

\section{Results}\label{sec:results}

We report in this section the estimation performance of \omnis
using the \textsc{Ranking-Accuracy} and \textsc{Absolute-Error} measures defined previously.
We also report the \textit{average}, \textit{best-case}, and \textit{worst-case} scenarios
using the cross-validation analysis described in \nameref{sec:method} section.

\subsection{Ranking Accuracy}

\Cref{tbl:results1} shows
Spearman correlation coefficients computed between the feature values
estimated by \omnis and groundtruth values.
On average, \omnis estimations correlate $r_s > .9$ with groundtruth
(all correlations are statistically significant at $p < .001$).
The best ranking accuracy was $r_s = .999$ for three features and the \mmHands dataset:
\feat{path_length}, \feat{num_segments}, and \feat{lb_ratio}.
Similar high accuracy levels were observed for many other features and datasets,
\eg, both \feat{production_time} and \feat{aspect_ratio} achieved $r_s = .998$
for the gestures of the \mmStrokes and \mmFingers datasets, respectively.
The lowest accuracy was $r_s = .518$ for \feat{density} computed on the gestures of the \mmHands dataset.
Taken together, these results suggest that \omnis estimations are inline with groundtruth data.

\begin{figure*}[!ht]
  \def\w{0.17\textwidth}
  \includegraphics[width=\w]{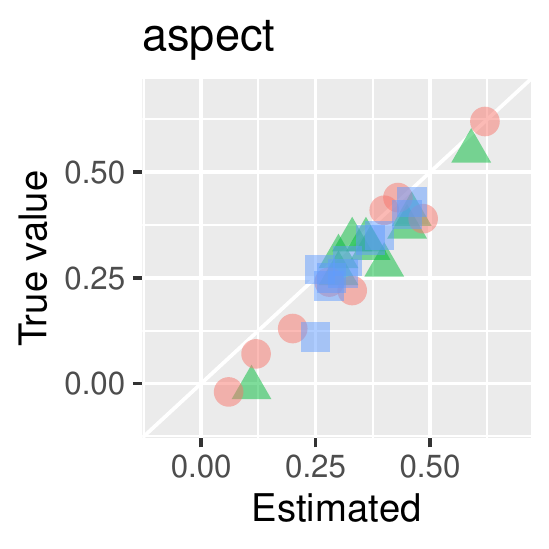}\hfill
  \includegraphics[width=\w]{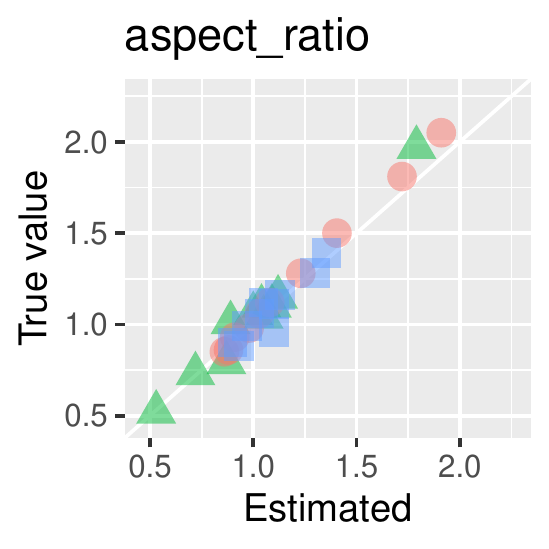}\hfill
  \includegraphics[width=\w]{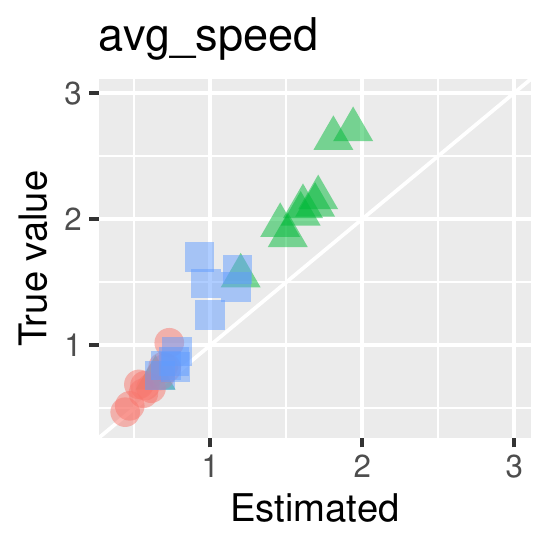}\hfill
  \includegraphics[width=\w]{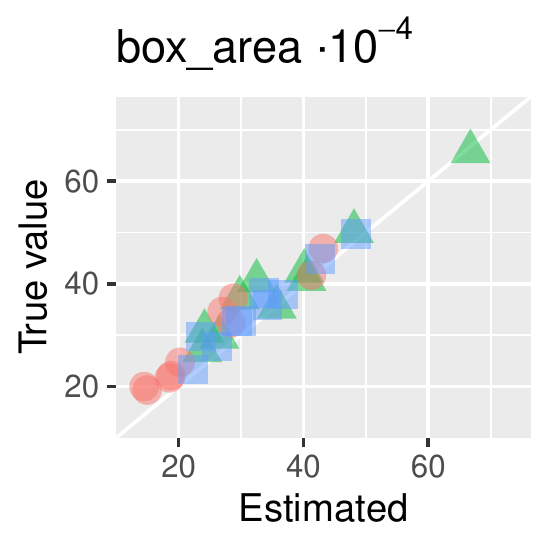}\hfill
  \includegraphics[width=\w]{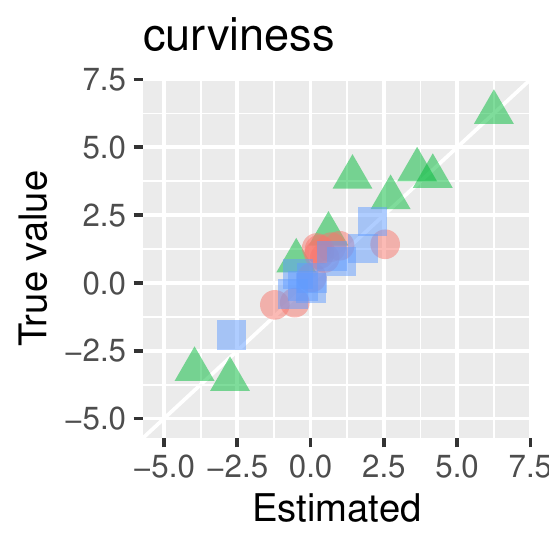}\hfill
  \\[0.5em]
  \includegraphics[width=\w]{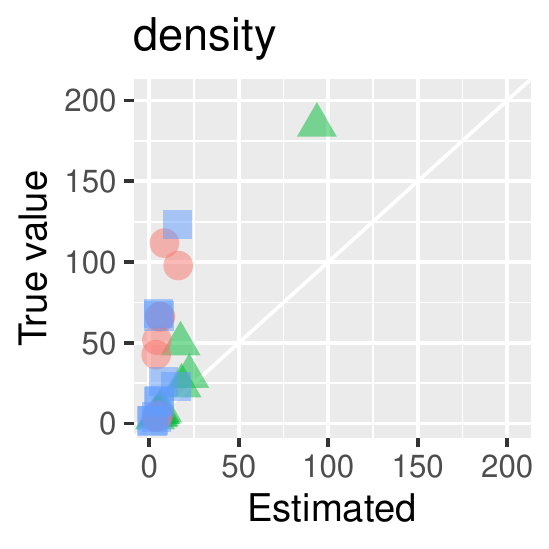}\hfill
  \includegraphics[width=\w]{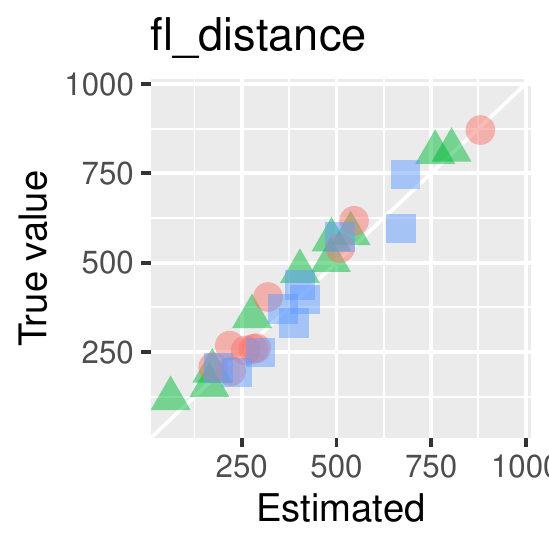}\hfill
  \includegraphics[width=\w]{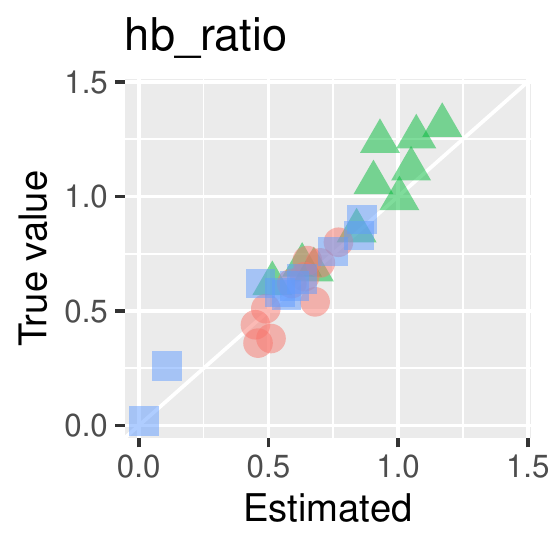}\hfill
  \includegraphics[width=\w]{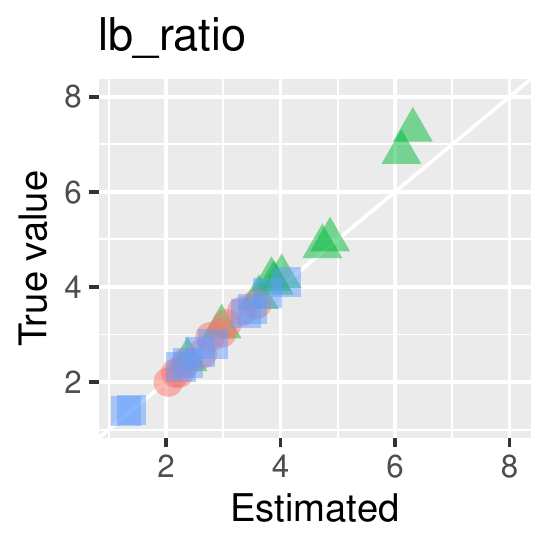}\hfill
  \includegraphics[width=\w]{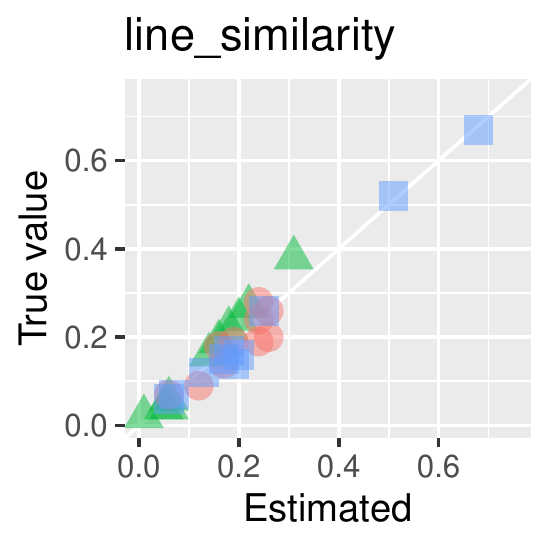}\hfill
  \\[0.5em]
  \includegraphics[width=\w]{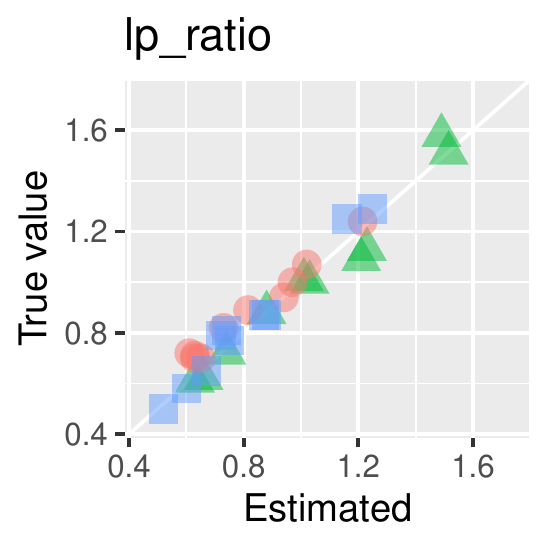}\hfill
  \includegraphics[width=\w]{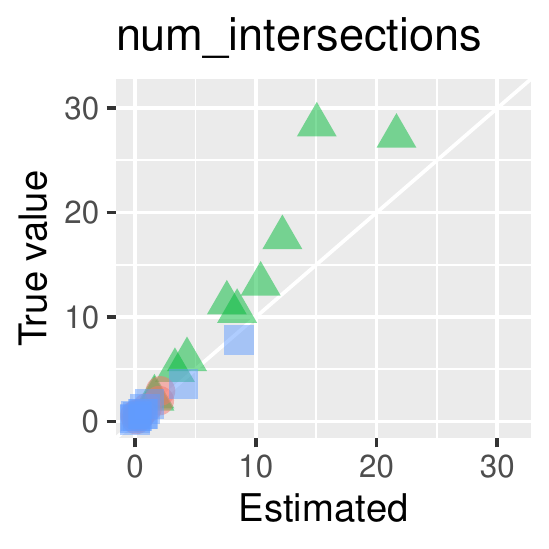}\hfill
  \includegraphics[width=\w]{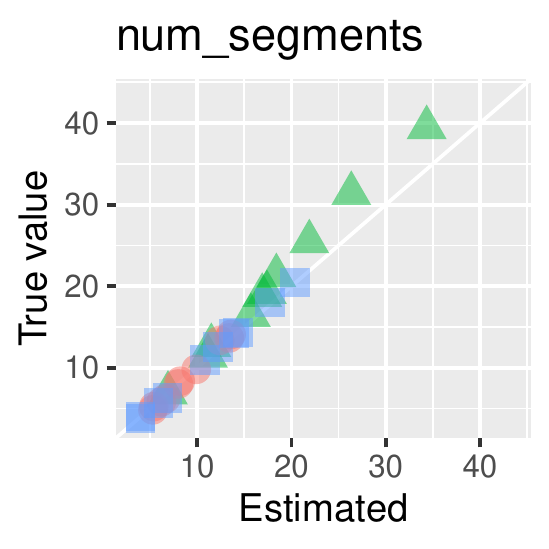}\hfill
  \includegraphics[width=\w]{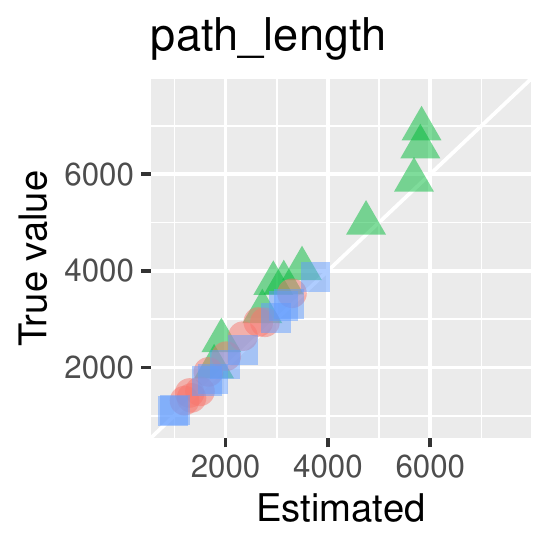}\hfill
  \includegraphics[width=\w]{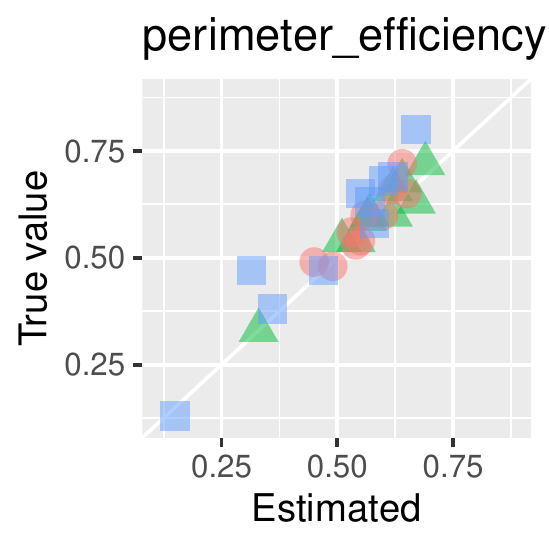}\hfill
  \\[0.5em]
  \includegraphics[width=\w]{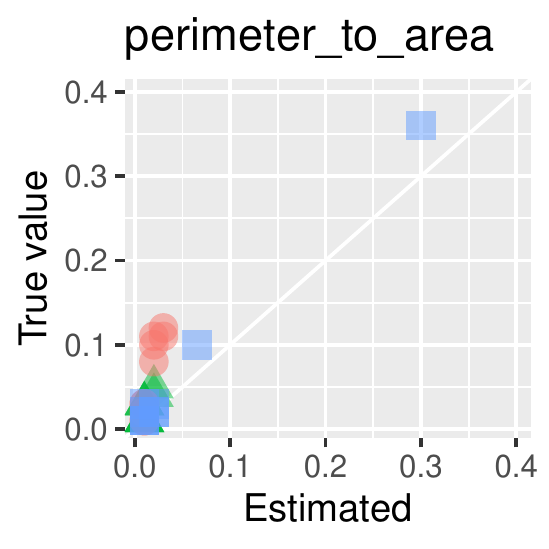}\hfill
  \includegraphics[width=\w]{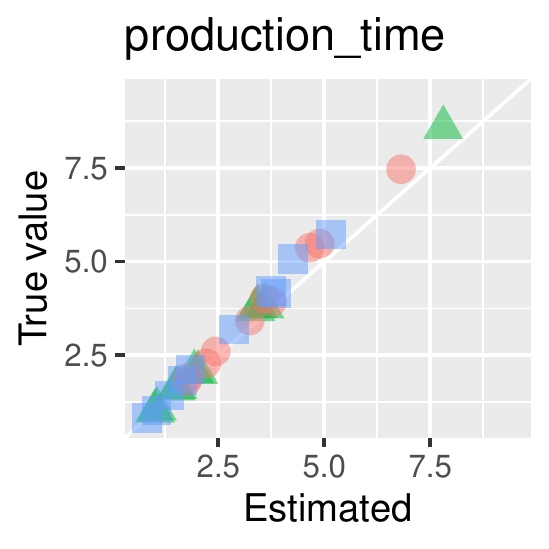}\hfill
  \includegraphics[width=\w]{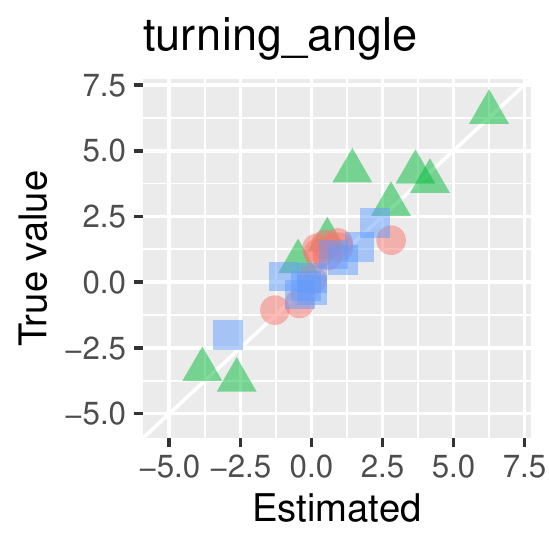}\hfill
  \phantom{\includegraphics[width=\w]{plots-min/cor-mt-turning_angle.pdf}}\hfill
  \phantom{\includegraphics[width=\w]{plots-min/cor-mt-turning_angle.pdf}}\hfill
  \\
  \caption{
    Estimated values vs. groundtruth for the feature types described in \autoref{tbl:features} aggregated by gesture type,
    corresponding to evaluation scenario EV$_1$.
    Legend:
    {\large \textcolor{ggplotGreen!60!white}{$\blacktriangle$}} \mmStrokes dataset
    {\LARGE \textcolor{ggplotRed!60!white}{$\bullet$}} \mmFingers
    {\scriptsize \textcolor{ggplotBlue!60!white}{$\blacksquare$}} \mmHands.
  }
  \label{fig:cors1}
\end{figure*}

\subsection{Estimation Errors}

Paired $t$-tests (two tails, Bonferroni-corrected) showed non-significant differences
between the estimations delivered by \omnis and groundtruth data.
Moreover, all the effect sizes were small \statp{d_\text{Mdn} < 0.3},
showing that any difference between the magnitudes of estimated and measured values
for the features from our evaluation is of small practical importance, \ie,
\omnis is very close to the actual values observable in practice for our set of features.
\Cref{fig:cors1} shows estimated \vs groundtruth feature values,
with an agglomeration of points around the diagonal.
(The diagonal represents the performance of an ideal feature estimator.)
These results confirm that the estimations delivered by \omnis
are on par with the feature values measured on gestures actually produced by users.

Regarding the best-case and worst-case scenarios (\Cref{tbl:results1}),
we can conclude that the user from which synthetic samples are generated
might have an impact, either positive or negative, on the estimations provided by \omnis.
This also suggests that it can be more beneficial to use \omnis
with several synthetic samples
in lieu of using a few trials from one user alone.
In the next section, we discuss this aspect in more detail.

\begin{table*}[!ht]
  \centering
  \scriptsize
  \setlength\tabcolsep{0.225em}
  \hspace*{-2.5cm}
  \begin{tabular}{rl*{15}r}
  \toprule
  &
  & \multicolumn{5}{c}{\tth{\mmStrokes dataset}}
  & \multicolumn{5}{c}{\tth{\mmFingers dataset}}
  & \multicolumn{5}{c}{\tth{\mmHands dataset}} \\
  \cmidrule(l){3-7} \cmidrule(l){8-12} \cmidrule(l){13-17}
  \tth{No.} & \tth{Feature}
  & \tth{True} & \tth{$\bm{r_s}$} & \tth{Error} & \tth{Best} & \tth{Worst}
  & \tth{True} & \tth{$\bm{r_s}$} & \tth{Error} & \tth{Best} & \tth{Worst}
  & \tth{True} & \tth{$\bm{r_s}$} & \tth{Error} & \tth{Best} & \tth{Worst} \\
  \midrule
1 & \featsmall{production_time}
& 2.75 & .998 & 0.83 & 0.00 & 6.67
& 3.70 & .994 & 0.99 & 0.00 & 5.16
& 2.84 & .996 & 0.83 & 0.00 & 5.25 \\
2 & \featsmall{avg_speed}
& 1.98 & .979 & 0.71 & 0.00 & 2.42
& 0.70 & .889 & 0.20 & 0.00 & 0.76
& 1.16 & .835 & 0.40 & 0.00 & 1.35 \\
3 & \featsmall{box_area} \scriptsize{$\cdot 10^{-4}$}
& 38.68 & .945 & 12.59 & 0.00 & 47.50
& 29.21 & .973 & 10.81 & 0.01 & 36.31
& 34.77 & .974 & 8.72 & 0.01 & 34.63 \\
4 & \featsmall{curviness}
& 1.20 & .949 & 3.64 & 0.00 & 18.30
& 0.71 & .660 & 1.25 & 0.00 & 4.68
& 0.37 & .904 & 1.60 & 0.00 & 5.40 \\
5 & \featsmall{density}
& 29.28 & .987 & 16.05 & 0.00 & 181.85
& 43.69 & .698 & 32.60 & 0.00 & 113.31
& 32.28 & .518 & 23.26 & 0.00 & 121.56 \\
  \midrule
6 & \featsmall{line_similarity}
& 0.15 & .985 & 0.07 & 0.00 & 0.35
& 0.18 & .775 & 0.07 & 0.00 & 0.25
& 0.23 & .995 & 0.06 & 0.00 & 0.61 \\
7 & \featsmall{aspect_ratio}
& 1.02 & .985 & 0.18 & 0.00 & 1.43
& 1.23 & .998 & 0.19 & 0.00 & 1.50
& 1.07 & .881 & 0.15 & 0.00 & 0.94 \\
8 & \featsmall{turning_angle}
& 1.15 & .928 & 3.77 & 0.01 & 19.33
& 0.74 & .588 & 1.36 & 0.00 & 5.20
& 0.34 & .891 & 1.70 & 0.00 & 5.50 \\
  \midrule
9 & \featsmall{aspect}
& 0.31 & .969 & 0.18 & 0.00 & 1.09
& 0.24 & .984 & 0.27 & 0.00 & 1.31
& 0.29 & .928 & 0.17 & 0.00 & 0.87 \\
10 & \featsmall{path_length}
& 4214.58 & .940 & 1678.31 & 0.75 & 5961.00
& 2161.48 & .995 & 441.75 & 0.48 & 1896.74
& 2352.48 & .999 & 351.52 & 0.96 & 1950.50 \\
11 & \featsmall{fl_distance}
& 456.85 & .993 & 121.56 & 0.05 & 861.49
& 378.25 & .985 & 152.59 & 0.05 & 770.13
& 410.11 & .961 & 127.09 & 0.06 & 644.89 \\
12 & \featsmall{num_segments}
& 19.49 & .952 & 7.81 & 0.01 & 30.74
& 8.89 & .998 & 1.48 & 0.00 & 8.74
& 10.97 & .999 & 1.60 & 0.00 & 10.41 \\
13 & \featsmall{num_intersections}
& 11.54 & .916 & 7.43 & 0.01 & 30.12
& 0.95 & .964 & 0.63 & 0.00 & 3.24
& 1.35 & .983 & 0.85 & 0.00 & 7.26 \\
14 & \featsmall{lp_ratio}
& 1.01 & .985 & 0.13 & 0.00 & 1.07
& 0.86 & .980 & 0.19 & 0.00 & 0.73
& 0.83 & .988 & 0.11 & 0.00 & 0.51 \\
15 & \featsmall{lb_ratio}
& 4.46 & .926 & 1.39 & 0.00 & 4.88
& 2.77 & .993 & 0.19 & 0.00 & 2.02
& 2.78 & .999 & 0.19 & 0.00 & 1.35 \\
16 & \featsmall{hb_ratio}
& 0.98 & .935 & 0.30 & 0.00 & 1.01
& 0.57 & .884 & 0.16 & 0.00 & 0.70
& 0.58 & .977 & 0.10 & 0.00 & 0.83 \\
17 & \featsmall{perimeter_efficiency}
& 0.58 & .909 & 0.14 & 0.00 & 0.61
& 0.58 & .898 & 0.16 & 0.00 & 0.59
& 0.55 & .958 & 0.13 & 0.00 & 0.62 \\
18 & \featsmall{perimeter_to_area}
& 0.02 & .776 & 0.01 & 0.00 & 0.04
& 0.06 & .965 & 0.04 & 0.00 & 0.17
& 0.05 & .997 & 0.03 & 0.00 & 0.35 \\
  \bottomrule
  \end{tabular}
  \caption{
    Feature estimation results using \omnis,
    corresponding to evaluation scenario EV$_1$.
    Spearman correlations ($r_s$) are computed between groundtruth values (columns titled ``True'')
    and \omnis estimations (shown in \autoref{fig:cors1}).
    The smaller the estimated errors, the better.
  }
  \label{tbl:results1}
\end{table*}

\subsection{Estimated Distributions of Feature Values}

\begin{figure*}[!ht]
  \centering

  \def\w{0.16\textwidth}

  \textbf{\mmFingers dataset} (\textit{different number of simultaneous fingers})\\
  \includegraphics[width=\w]{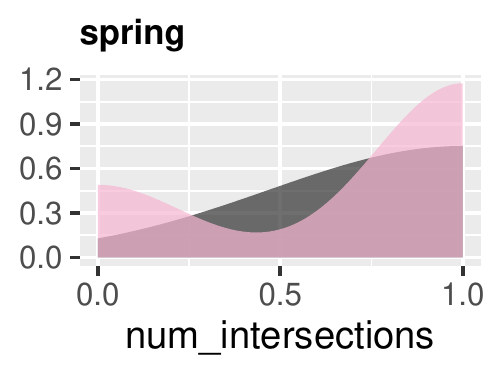}\hfill
  \includegraphics[width=\w]{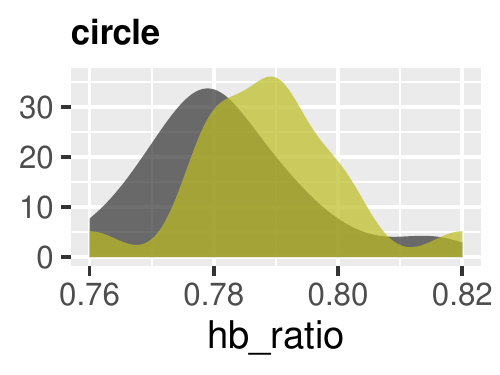}\hfill
  \includegraphics[width=\w]{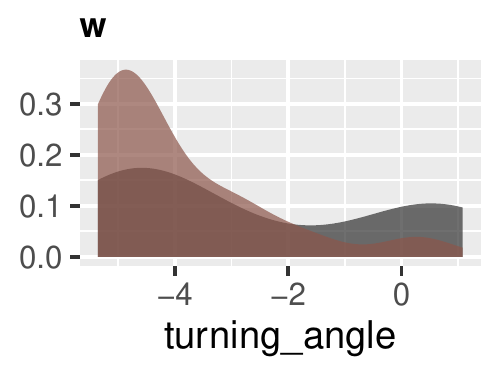}\hfill
  \includegraphics[width=\w]{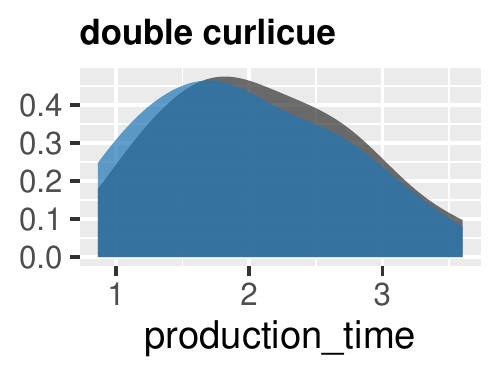}\hfill
  \includegraphics[width=\w]{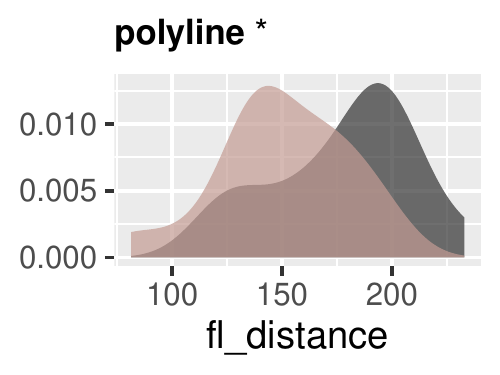}\hfill
  \\
  \includegraphics[width=\w]{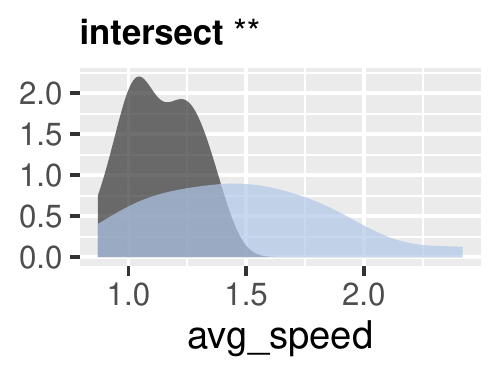}\hfill
  \includegraphics[width=\w]{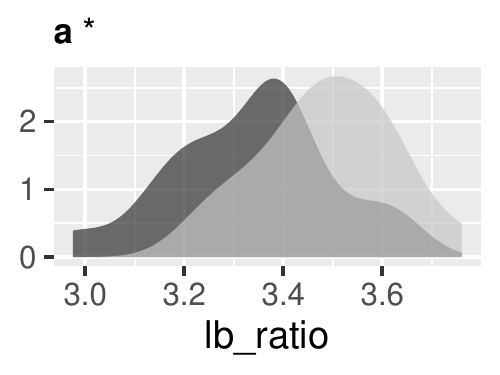}\hfill
  \includegraphics[width=\w]{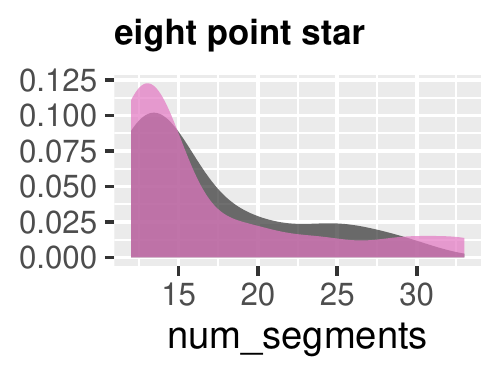}\hfill
  \includegraphics[width=\w]{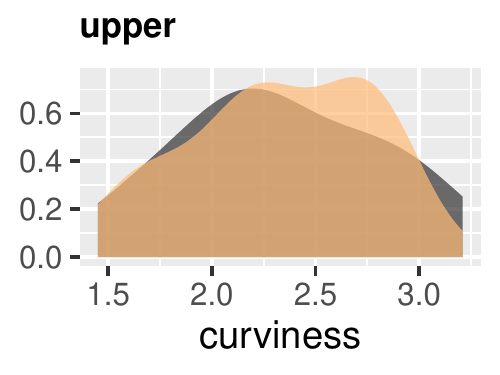}\hfill
  \includegraphics[width=\w]{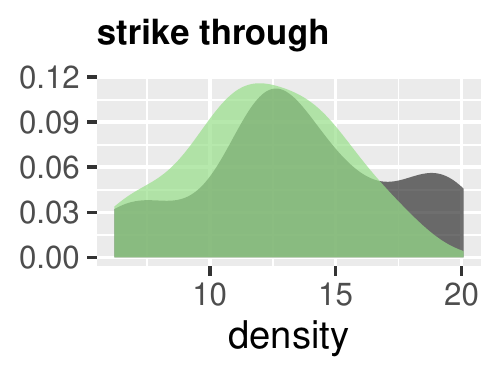}\hfill
  \\[0.5em]
  \textbf{\mmStrokes dataset} (\textit{different number of strokes})\\
  \includegraphics[width=\w]{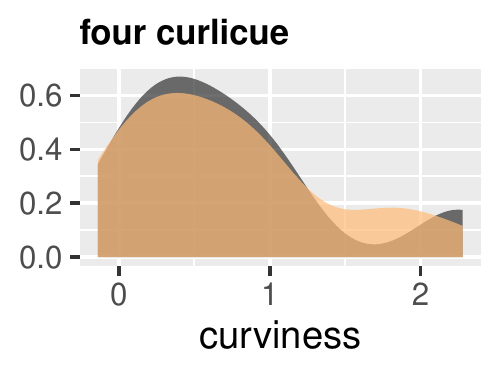}\hfill
  \includegraphics[width=\w]{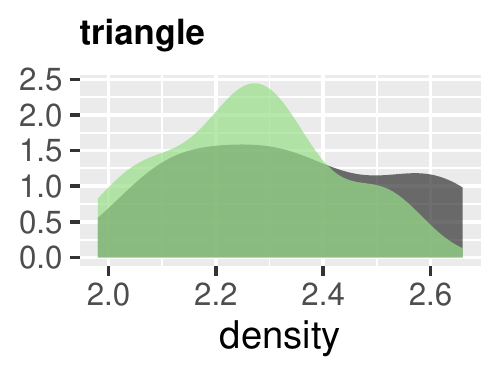}\hfill
  \includegraphics[width=\w]{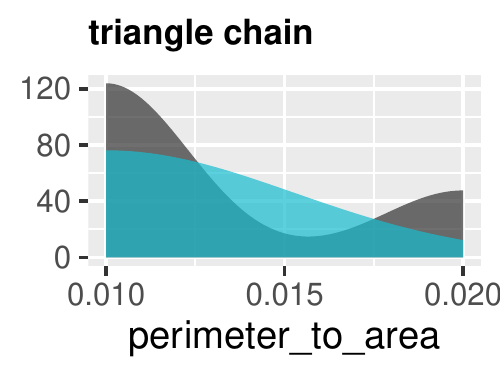}\hfill
  \includegraphics[width=\w]{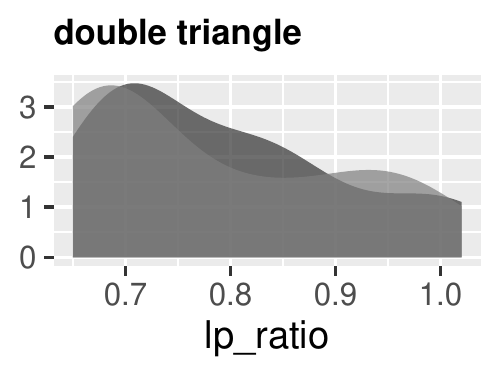}\hfill
  \includegraphics[width=\w]{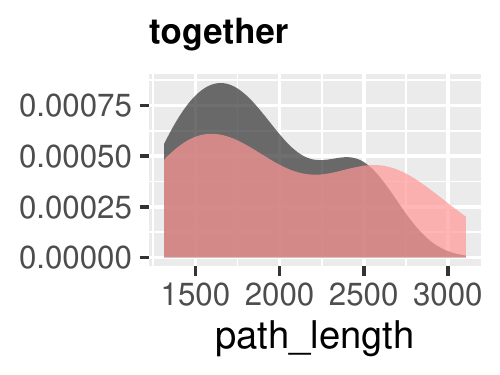}\hfill
  \\
  \includegraphics[width=\w]{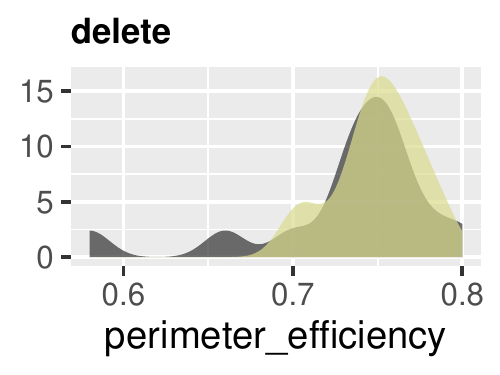}\hfill
  \includegraphics[width=\w]{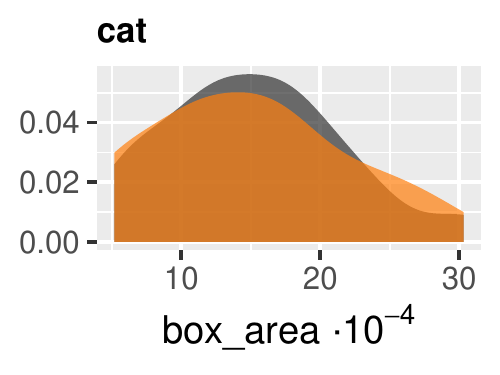}\hfill
  \includegraphics[width=\w]{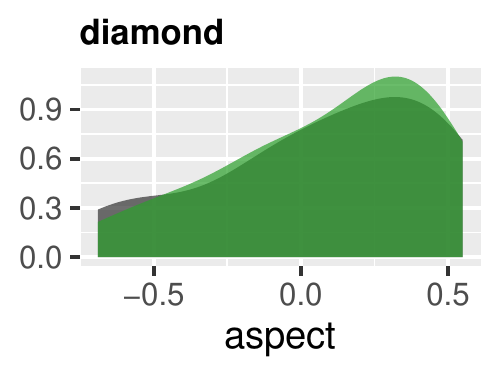}\hfill
  \includegraphics[width=\w]{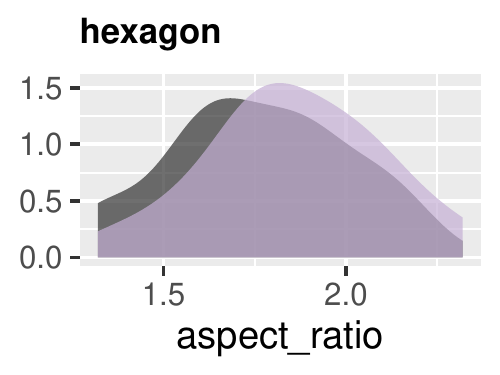}\hfill
  \includegraphics[width=\w]{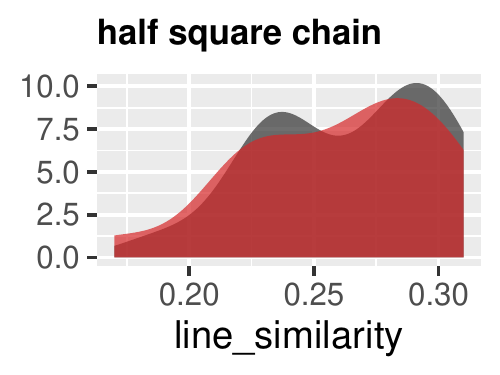}\hfill
  \\[0.5em]
  \textbf{\mmHands dataset} (\textit{sequential and bimanual articulations})\\
  \includegraphics[width=\w]{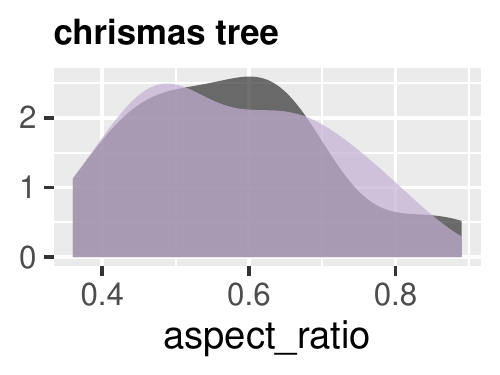}\hfill
  \includegraphics[width=\w]{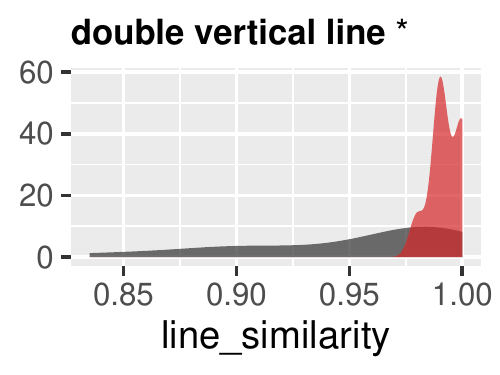}\hfill
  \includegraphics[width=\w]{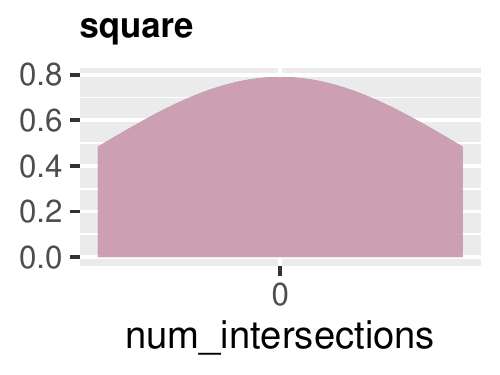}\hfill
  \includegraphics[width=\w]{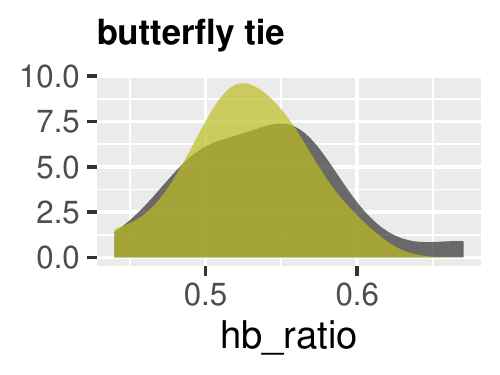}\hfill
  \includegraphics[width=\w]{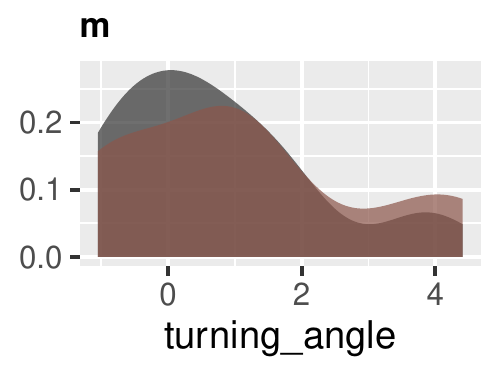}\hfill
  \\
  \includegraphics[width=\w]{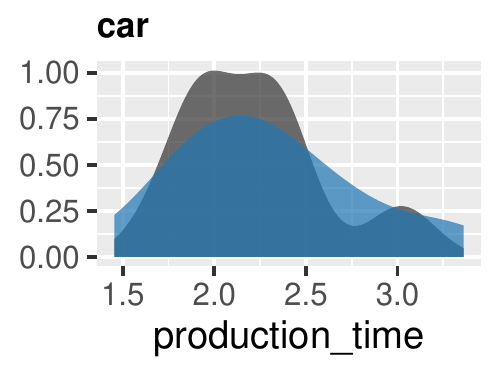}\hfill
  \includegraphics[width=\w]{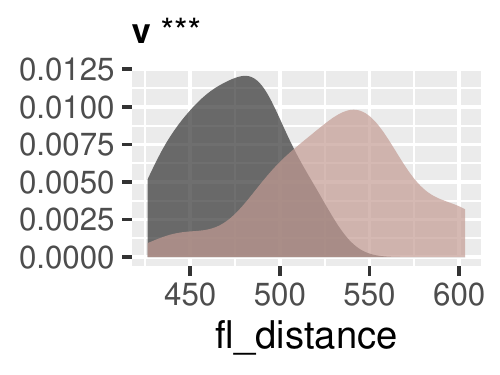}\hfill
  \includegraphics[width=\w]{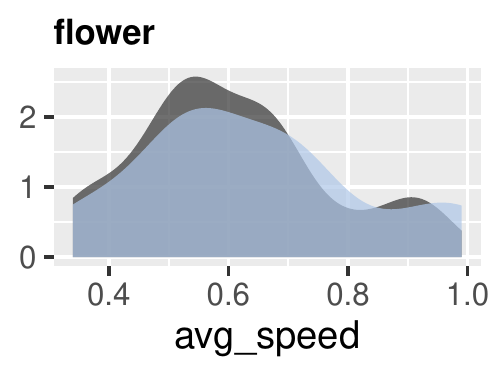}\hfill
  \includegraphics[width=\w]{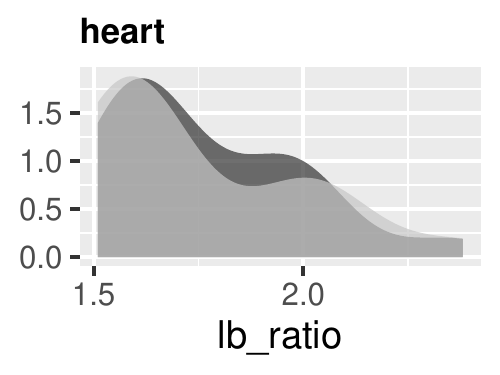}\hfill
  \includegraphics[width=\w]{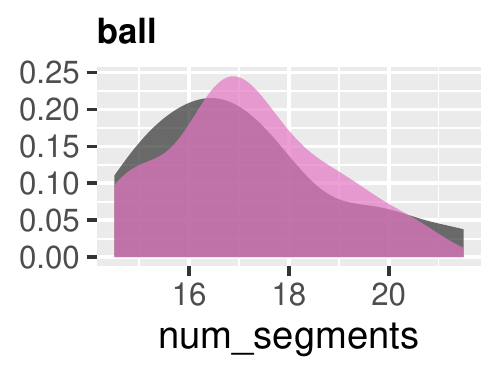}\hfill
  \\[0.5em]
  \caption{
    Feature distributions of the groundtruth data (in gray) and feature distributions estimated by \omnis (colored),
    corresponding to evaluation scenario EV$_2$.
    For the skae of brevity, we show only one feature for each gesture type.
    Where applicable, statistically significant differences between the groundtruth and estimated distributions
    are indicated using star notations, \ie,
    $^{*}~(p<.05)$, $^{**}~(p<.01)$, or $^{***}~(p<.001)$.
  }
  \label{fig:hists-summary}
\end{figure*}

To gain a deeper understanding of the estimation performance of \omnis,
we decided to plot the estimated feature distributions for each gesture.
\autoref{fig:hists-summary} illustrates some of these distributions, randomly picked from our features of interest.
Space concerns prevent us to show here all the 30 (gestures) $\times$ 18 (features) = 540 distributions,
but they are available to download from the companion website of this article.

To evaluate the extent to which the distributions estimated by \omnis (in the average case scenario) differ from the groundtruth data,
we ran Kruskal-Wallis tests\footnote{
  The Kruskal-Wallis test is a non-parametric method for testing whether samples originate from the same distribution.
  It is used for comparing two or more independent groups of equal or different sizes.
} for each combination of gesture and feature type.
We applied the Bonferroni correction to guard against over-testing the data.
Results showed that in most cases there was a significant difference between the groundtruth and the estimations delivered by \omnis.
However, while the distributions of synthetic samples were generally wider,
both the mean and median values were very close to the mean and median values of the human distributions.
Moreover, the effect sizes were small for all tests \statp{r_\text{Mdn} < 0.3},
suggesting a small practical importance of any observed difference.

\section{Discussion}\label{sec:discussion}

Our results show that \omnis can successfully estimate the location of feature distributions,
\eg, the mean or median of a given distribution,
for the majority of features and gesture types that we evaluated.
While we detected statistically significant differences in most cases
between estimations provided by \omnis and groundtruth distributions of individual gestures and features,
the corresponding effect sizes were of small practical importance.\footnote{
  Also note that, in principle, the absence of a statistical significance result does not necessarily support the absence of a true effect.
  While equivalence testing might seem like a more appropriate method to use for that purpose, we have already shown in our experiment that effect sizes are small and, therefore, equivalence testing procedures are not necessary at this point.
}
These findings show that \omnis is able to model the variation in gestures articulated by users
and estimate gesture features accurately.

\subsection{Estimating Measures of Location and Dispersion}
We showed that \omnis is a robust estimator of the mean statistic for a variety of features and stroke gesture types.
However, our empirical results regarding the similarities between the estimated and groundtruth distributions of individual gestures
hint that \omnis could be used to estimate other measures of location and dispersion,
such as the variance or the standard deviation of the feature of interest.
In support of this claim, \Cref{tbl:results-sds1} shows estimations of the standard deviations for three features,
aggregated over all the gesture types from our evaluation datasets.
For brevity, we picked only one feature from each category listed in \autoref{tbl:features}.
As can be observed, \omnis delivers similar variation compared to the one present in the groundtruth data.

The supplementary materials show the standard deviation of each feature estimated by \omnis for each dataset.
Next to our results reported so far, these additional data increase our confidence that \omnis
can be used to produce accurate estimations of potentially any stroke gesture feature.

\begin{table*}[!h]
  \centering
  \scriptsize
  \setlength\tabcolsep{0.25em}
  \hspace*{-2.5cm}
  \begin{tabular}{rl*{12}r}
  \toprule
  &
  & \multicolumn{4}{c}{\tth{\mmStrokes dataset}}
  & \multicolumn{4}{c}{\tth{\mmFingers dataset}}
  & \multicolumn{4}{c}{\tth{\mmHands dataset}} \\
  \cmidrule(l){3-6} \cmidrule(l){7-10} \cmidrule(l){11-14}
  \tth{No.} & \tth{Feature}
  & \tth{True SD} & \tth{Estimated} & \tth{Best} & \tth{Worst}
  & \tth{True SD} & \tth{Estimated} & \tth{Best} & \tth{Worst}
  & \tth{True SD} & \tth{Estimated} & \tth{Best} & \tth{Worst} \\
  \midrule
  5 & \featsmall{curviness}
  & 0.82 & 1.51 & 1.25 & 0.99
  & 3.76 & 5.23 & 4.03 & 3.27
  & 0.98 & 1.96 & 2.00 & 1.10 \\
  7 & \featsmall{aspect_ratio}
  & 0.39 & 0.40 & 0.18 & 0.45
  & 0.34 & 0.34 & 0.16 & 0.21
  & 0.15 & 0.20 & 0.23 & 0.21 \\
  10 & \featsmall{path_length}
  &  740.89 &  826.12 &  285.18 &  188.24
  & 1545.50 & 2124.92 & 1774.28 & 1220.14
  &  918.75 & 1013.39 &  524.88 &  452.40 \\
  \bottomrule
  \end{tabular}
  \caption{
    Standard deviations (SDs) for three features,
    picked at random from each of the categories listed in \autoref{tbl:features},
    corresponding to evaluation scenario EV$_1$.
    The closer the estimated SD to the true SD, the better.
  }
  \label{tbl:results-sds1}
\end{table*}

\subsection{\omnis Web Application}\label{sec:app}
\begin{figure*}[!h]
  \fbox{\includegraphics[height=2.8cm]{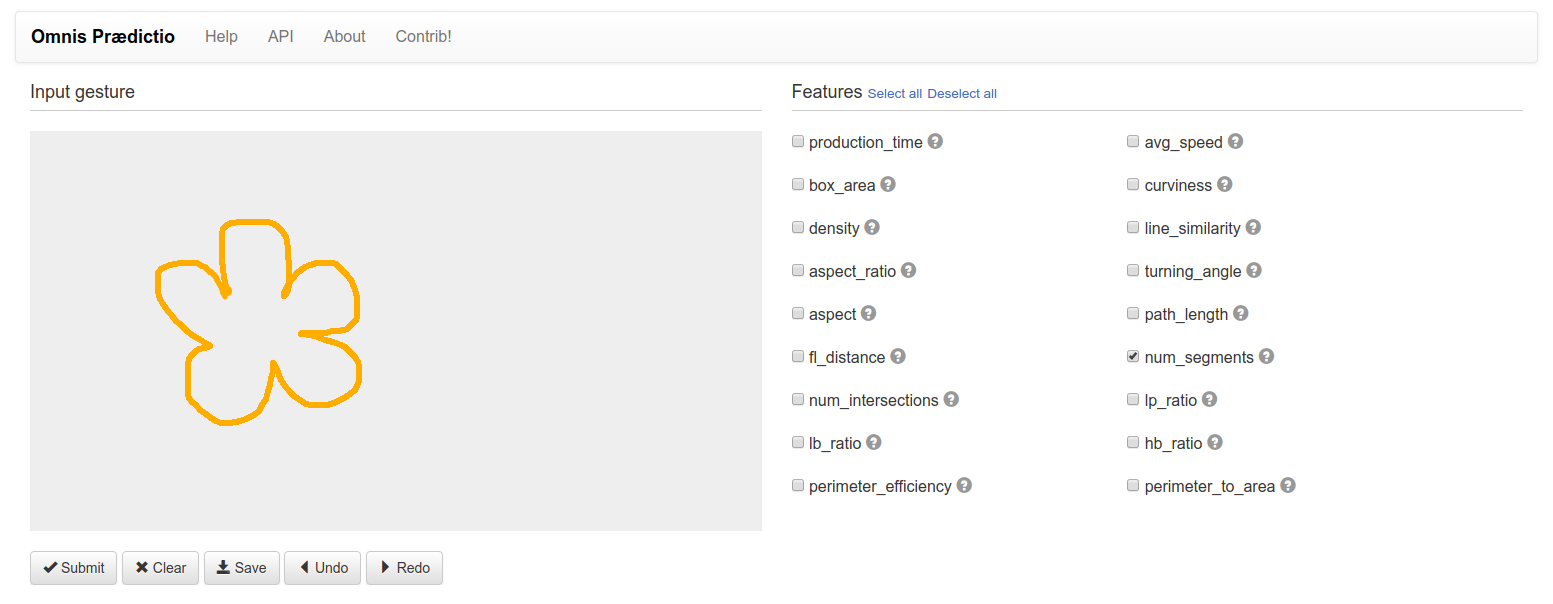}}\hfill
  \fbox{\includegraphics[height=2.8cm]{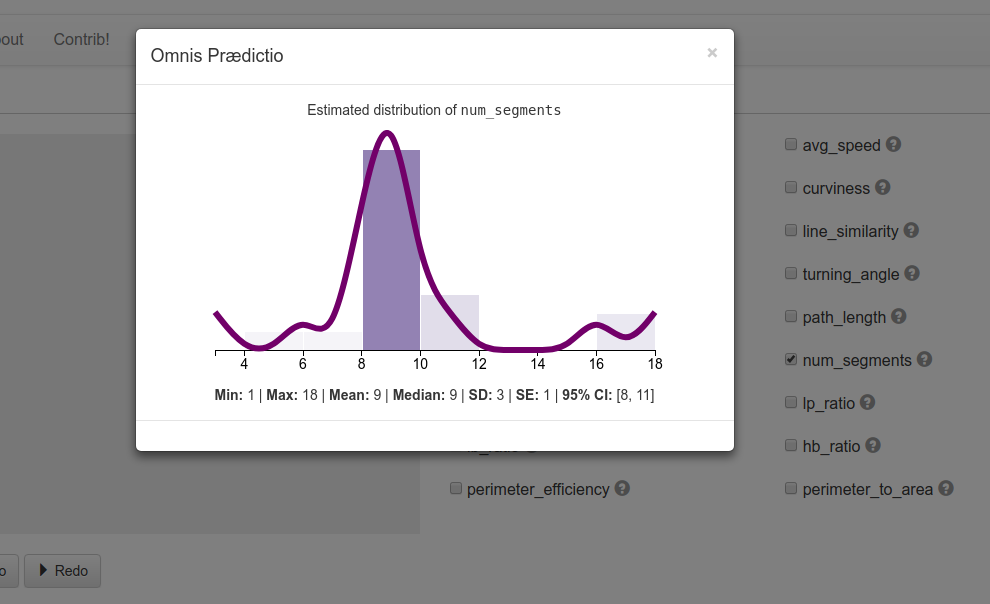}}
  \caption{
    Screenshots of the \omnis web application.
    The designer selects their desired features,
    whose values will be estimated upon submitting a gesture drawn on a web canvas.
    The application reports a wide palette of location and dispersion statistics,
    such as the mean, median, standard deviation, standard error, and confidence intervals.
  }
  \label{fig:ui}
\end{figure*}

We offer an implementation of \omnis as a web application that can be accessed at \url{https://luis.leiva.name/omnis/}
using any modern web browser running on a desktop PC, notebook, tablet, or smartphone,
which will make easier for practitioners to evaluate characteristics of stroke gestures performed on the target device itself.
The designer needs to provide just one example of a gesture by drawing it in free form on a web canvas
and indicate which features should be estimated; see \autoref{fig:ui} for a screenshot of our application.
The web canvas supports all kinds of gesture types (unistrokes, multistrokes, and multitouch gestures),
though the maximum number of simultaneous fingers touching the screen at once is device-dependent.
For example, mobile devices typically support up to ten simultaneous touch contacts.

All the 18 features listed in \autoref{tbl:features} are supported by default by the \omnis web application.
However, to make our application as reusable as possible in the form of a platform-as-a-service approach for a variety of needs of researchers, designers, developers, and educators,
we also deliver a ``custom feature definition'' module, so that practitioners can define their own gesture features.
This means that it is possible to submit a specification file using the language formalism of Python that
describes how the custom features should be computed.
In fact, the \omnis web application is a collection of several such specification files,
one for each of the three feature categories evaluated in this work; see \autoref{tbl:features}.

\subsection{\omnis Web API}\label{sec:api}
We also provide a RESTful JSON API to enable practitioners
to interface \omnis automatically from third party applications.
To this end, we designed a simple format for gesture data that developers need to follow; see \autoref{fig:api1}.
A stroke gesture is represented as an array of strokes, where a stroke is an array of points.
Each point data structure contains \feat{[x,y,time,touchId]} information.
The touch identifier is meant for multitouch gestures only
and can be omitted for unistrokes and multistrokes.
In addition, a \texttt{spec\_file} property in the input JSON enables users to specify a custom Python file
with one or more functions to apply to the individual gestures synthesized by \omnis.
The specification file must be uploaded together with the HTTP request.
An example of this use case is illustrated in \autoref{fig:api2} from the Appendix.

\subsection{Current Limitations and Workarounds}\label{sec:outlook}
As hinted in the previous sections,
the gesture example provided by the designer needs to be reconstructed with high quality by the \slm model of the Kinematic Theory~\cite{Plamondon95a,Plamondon95b}. This criterion is automatically evaluated using the signal-to-noise ratio (SNR) recommended by the Kinematic Theory. Previous work~\cite{Almaksour11, Leiva16_g3, Leiva17_iwc} suggested that SNR values below $15$\,dB denote poor execution quality and, in such cases, the gesture seed should be entered again. To address this aspect properly and prevent inaccurate estimations,
the \omnis application alerts the practitioner when the gesture seed cannot be reconstructed properly with the \slm model so that a new example could be provided. Anecdotally, in our experiments, only $73$ of the $7,200$ gestures could not be reconstructed with $\text{SNR} > 15$\,dB, which representing just a mere 1\% of all the data.

The fact that only one gesture example is required by \omnis is a convenient aspect in practice,
as it has been reported that people are reluctant to provide more than one gesture example when asked~\cite{Li10a},
but it also means that the synthesized gestures
are inherently dependent on that human sample~\cite{Leiva18_keytime, Leiva18_gato}. This fact can be interpreted as a self-limitation of \omnis
from the perspective of the variance in articulation potentially attainable with more examples of real gestures performed by actual users.
Currently, \omnis analyzes one gesture at a time,
but we plan to update its inner-workings to accept more gesture samples for its internal analysis,
enriching therefore its ability to estimate feature distributions even more accurately.
For such a scenario, designers could submit different articulations of the same gesture type,
\eg, five or ten samples collected from different users towards more accurate estimations of the feature of interest.
Intuitively, the greater the number of users that provide gesture examples for \omnis,
and the better the gesture articulation abilities of those users match the ability of the end-user population,
\omnis will deliver more accurate estimations.
Our evaluation results regarding the estimations of feature densities (second evaluation experiment)
showed how precise the performance of \omnis can get when more than one gesture sample is available.
However, since in this work we have focused primarily on the application scenario of \omnis being used with one gesture seed only (efficiency for practitioners and designers), we leave the exploration regarding the optimum size of the training set to improve \omnis for future work.
Nevertheless, we expect not too many gesture samples to be needed for \omnis to reach its peak theoretical accuracy,
since previous work already showed that gesture synthesis using the \slm principles can be successfully employed
to model the variability of large user populations.
We leave such interesting explorations for future work.

One aspect that deserves to be investigated further is that of understanding why sometimes \omnis was unable to replicate certain feature distributions. While those cases were usually rare (see Appendix and supplementary materials), we believe they might bring new insights about some potential effect of gesture shape and articulation difficulty. Moreover, in our experience, we have observed that synthetic gestures are usually more variable than their human counterparts. This, combined with the above-mentioned articulation challenges, suggests that a working theory should be devised to explain these differences.

Another aspect worthy of investigation is that of incorporating information about user demographics in the synthesis process. Currently there is no principled way to ensure an \emph{exact} representation of the user population, since, as previously mentioned, synthetic gestures are usually more variable than their human counterparts. Therefore, some deviation between feature distributions is to be expected. In addition, in order to model the articulation of new stroke gestures, the distortions introduced to the \slm model parameters are computed in this article assuming able-bodied and cognitively healthy users. This was so in order to verify our hypothesis that other gesture features besides production time can be reliably estimated in order to describe users' performance with stroke gesture input, and that synthetic gestures can be used to reproduce the distribution of those features as well. In previous work~\cite{Leiva17_chi} we were able to generate synthetic stroke gestures across user populations, i.e., modeling the articulation characteristics of a target population with data from a source population. Therefore, we believe this idea could be further explored in future work.

Finally, our current implementation allows practitioners to code in their own custom gesture features as specification files in the \omnis online application. However, because specification files are executed as part of our analysis pipeline, the \omnis application implements a strict sandbox model to prevent malicious code execution. This means that it is not possible to import Python modules, use global variables, or perform requests over the Internet. Only pure functions without side effects are allowed. Nevertheless, \omnis already features a comprehensible list of features and white-listed modules that developers can use in their own specification files and we are happy to incorporate more functionality in the future.

\subsection{Toward a Geometric Theory of Human Movement}
So far, researchers have shown that the principles and concepts of the Kinematic Theory~\cite{Plamondon95a,Plamondon95b} can be employed
to estimate gesture production times with remarkably good accuracy; see recent work by Leiva \etal~\cite{Leiva18_keytime, Leiva18_gato}.
However, such levels of performance were somewhat expected, since production time is a variable
explicitly formulated in the \slm model of the Kinematic Theory~\cite{Plamondon95a}.
The name of the theory itself shows clearly that its main purpose is to model the velocity of directed movements,
such as the elementary units that form handwriting and gesture articulation.
However, an uncertain aspect before this work
was whether the geometric properties of synthetic gestures,
as reflected by various geometric features,
are robust enough to explain real data accurately.
Our work presents the first evidence in this direction,
suggesting a possible extension of the applicability domain of the Kinematic Theory~\cite{Plamondon95a,Plamondon95b}
and its adoption in HCI~\cite{Leiva18_keytime, Leiva18_gato,Leiva17_chi,Leiva17_dis}
to explain the geometry-related aspects of gesture articulation.

According to the Kinematic Theory~\cite{Plamondon95a,Plamondon95b}, a gesture path is represented in a neural map
as a sequence of virtual targets (\autoref{fig:theory-example}) to be reached during execution with units of movement
that can be described in terms of lognormal velocity profiles.
In other words, our brains are using velocity as the control variable,
while the virtual targets are the only landmarks needed to produce the intended movement.
In this context, our empirical results represent an indirect proof of this mechanism:
if the velocity profiles are reconstructed accurately by the \slm model,
then the shape of the gesture will be accurate as well.
Using simple notions from fundamental physics,
we can see that applying the Kinematic Theory to model geometric shapes actually makes sense,
since the velocity vector is always tangent to the stroke trajectory\footnote{
  See \url{http://mathworld.wolfram.com/VelocityVector.html}
}
and, therefore, the geometric information follows from the velocity reconstruction.

Based on our empirical results and the above insights,
we propose that the Kinematic Theory is also a \emph{de facto} Geometric Theory of human movements,
to be verified by future theoretical and empirical work.
Therefore, the Kinematic Theory may be used to support conducting scientific investigations
regarding the geometric paths and shapes of stroke gestures,
not just regarding the kinematic aspects of their articulations~\cite{Leiva18_keytime, Leiva18_gato}.
We hope that future work in the form of more studies and controlled evaluations, inspired by this new perspective,
will be conducted to confirm these insights and open new ways to apply a robust motor control theory in the practice of gesture input.
Our insights are supported by several theoretical developments and empirical evidence~\cite{Plamondon95a,Almaksour11,Djioua09},
and we are excited to enable similar discoveries on topics of interest for the HCI community.
We would like to restate that efforts such as \omnis that advance the state of the art with practical,
readily usable tools are essential to help the community shape, consolidate, and advance its knowledge.

\section{Conclusion}\label{sec:conclusion}

We presented \omnis, a general technique informed by the Kinematic Theory and companion online tool,
that delivers accurate user-independent estimations of any numerical stroke gesture feature
with minimum effort required from the practitioner.
By having access to feature distributions,
various aspects of users' performance with stroke gesture input can be quantified.
\omnis is readily available to researchers, designers, developers, and educators both as an online application and as a RESTful JSON API.

Moreover, \omnis contributes to advancing our capacity as a community to model, analyze,
and understand end users' stroke gesture articulations on touchscreen devices.
Consequently, we believe that \omnis will foster more effective and efficient gesture-based user interface designs.
For example, will some new gesture recognition or machine learning approach demand a whole new set of gesture features in the future?
Or will some designer need to understand how two user groups, not studied so far, are different in terms of how they articulate stroke gestures from the perspective of some new gesture features?
\omnis will be able to readily deliver estimations for those new features, which the practitioners can define, specify, and code themselves.
From this perspective, \omnis centralizes efforts in modeling and estimating articulation features for stroke gesture input,
taking major steps towards (i)~moving from one value to a distribution
and (ii)~from focusing on a specific gesture feature (\ie, production time) to any computable feature,
including features that researchers have not invented yet,
but will be able to code and submit to the \omnis web app themselves.

\section*{Acknowledgments}
We thank the anonymous reviewers for their constructive and valuable feedback.
L.\,A. Leiva acknowledges support from the Academy of Finland (BAD project)
and the Finnish Center for Artificial Intelligence (FCAI).
R.-D. Vatavu acknowledges research conducted in the Machine Intelligence
and Information Visualization Lab (MintViz) of the MANSiD Research Center.
The infrastructure was provided by the University of Suceava
and was partially supported from the project ``Integrated center for research, development and innovation
in Advanced Materials, Nanotechnologies, and Distributed Systems for fabrication and control'', no. 671/09.04.2015,
Sectoral Operational Program for Increase of the Economic Competitiveness,
co-funded from the European Regional Development Fund.
R. Plamondon acknowledges support from NSERC-Canada grant RGPIN-2015-06409.

\section*{End Notes}
Our web application can be publicly accessed at \url{https://luis.leiva.name/omnis/}.
Due to legal restrictions, the core software for gesture synthesis cannot be made publicly available as a standalone software.
The interested people must sign a non-disclosure agreement (NDA) with the École Polytechnique de Montréal
through a collaborative project to get a license.

\balance

\clearpage

\appendix

\section{API examples}\label{apx:api}

\Cref{fig:api1,fig:api2} illustrate our API request and response examples.

\begin{figure*}
\small
\begin{minipage}[t]{0.35\linewidth}
  Request:
  \begin{lstlisting}[language=json]
{
    "measures": [
        "production_time",
        "density",
        "line_similarity"
    ],
    "strokes": [
        [
            [x1, y1, t1, p1],
            [x2, y2, t2, p1],
            ...
        ],
        [
            ...
        ]
    ],
    "spec_file": null
}
  \end{lstlisting}
\end{minipage}
\hfill
\begin{minipage}[t]{0.55\linewidth}
  Response:
  \begin{lstlisting}[language=json]
{
  "errors": null,
  "result": {
      "production_time": {
          "confidence_intervals": {
              "90
              "95
              "99
          },
          "max": 5055,
          "mean":1939,
          "median": 1863,
          "min": 501,
          "range": 4554,
          "standard_deviation": 848,
          "standard_error": 87,
          "trimmed_mean": 1862,
          "values": [1791, 2429, ...],
          "variance": 718366,
          "winsorized_mean": 1882
      },
      "density": {
          ...
      },
      "line_similarity": {
          ...
      }
  }
}
  \end{lstlisting}
\end{minipage}
\caption{
  Simple example of API request and response.
  Each point data structure contains \feat{[x,y,time,touchId]} information,
  though the touch identifier is only required for multitouch gestures.
}
\label{fig:api1}
\end{figure*}

\begin{figure*}
\small
\begin{minipage}[t]{0.35\linewidth}
  Request:
  \begin{lstlisting}[language=json]
{
  "measures": [],
  "spec_file": @some.py,
  "strokes": [
    [...],
    [...]
  ]
}
  \end{lstlisting}

  Contents of \texttt{some.py} file:
  \begin{lstlisting}[language=python]
def strokeCount(strokes):
    return len(strokes)

def my_path_len(strokes):
    lens = [len(s) for s in strokes]
    # The Numpy module is available.
    return numpy.mean(lens)
  \end{lstlisting}
\end{minipage}
\hfill
\begin{minipage}[t]{0.55\linewidth}
  Response:
  \begin{lstlisting}[language=json]
{
  "errors": null,
  "result": {
      "my_path_len": {
          "confidence_intervals": {
              "90
              "95
              "99
          },
          "max": 60.54,
          "mean":29.38,
          "median": 28.62,
          "min": 9.01,
          "range": 65.53,
          "standard_deviation": 9.47,
          "standard_error": 9.6,
          "trimmed_mean": 28.61,
          "values": [27.81, 34.19, ...],
          "variance": 8183.65,
          "winsorized_mean": 28.62
      },
      "strokeCount": {
          ...
      }
  }
}
  \end{lstlisting}
\end{minipage}
\caption{
  Advanced example of API request and response.
  The user can upload a custom feature definition file
  together with the regular HTTP JSON request.
}
\label{fig:api2}
\end{figure*}

\section{Supplementary Materials}

Many empirical results could not be accommodated in this submission, for brevity's sake.
However, we provide all the feature distributions
on the companion website at \url{https://luis.leiva.name/omnis/}.

For reviewing purposes, please see the supplementary materials submitted with this article
for an example of the kind of data that we will provide on the companion website.
\autoref{fig:hists-appendix} below is just an example of the data to be expected in such companion website.

\begin{figure*}
  \centering
  \hspace*{-3cm}
  \def\w{1.2\textwidth}
  \begin{minipage}{\w}
  \includegraphics[width=\w]{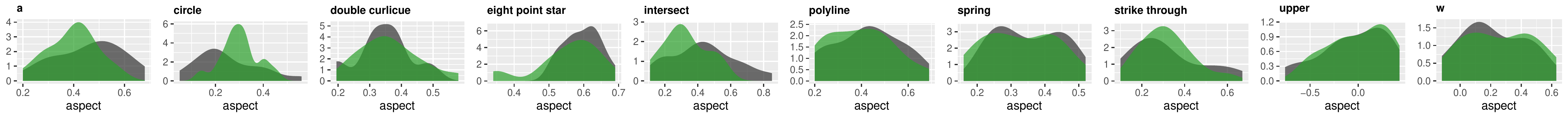}
  \includegraphics[width=\w]{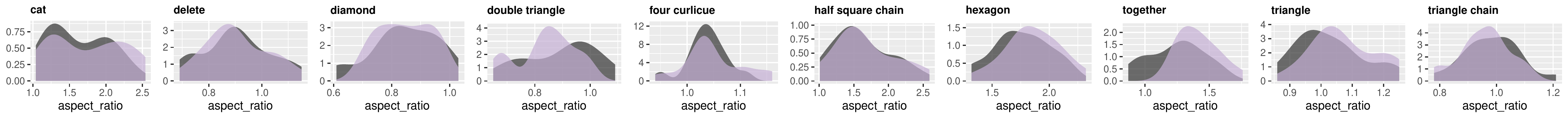}
  \includegraphics[width=\w]{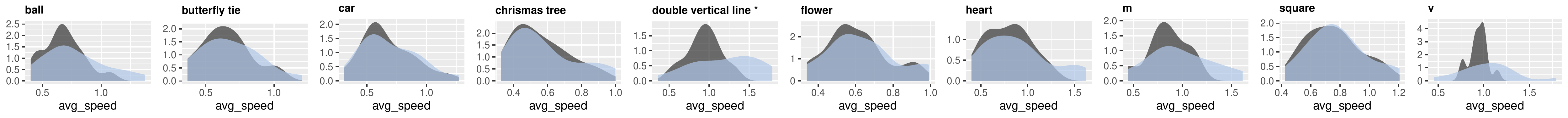}
  \includegraphics[width=\w]{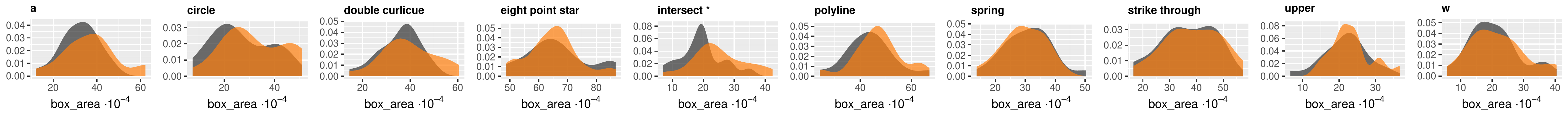}
  \includegraphics[width=\w]{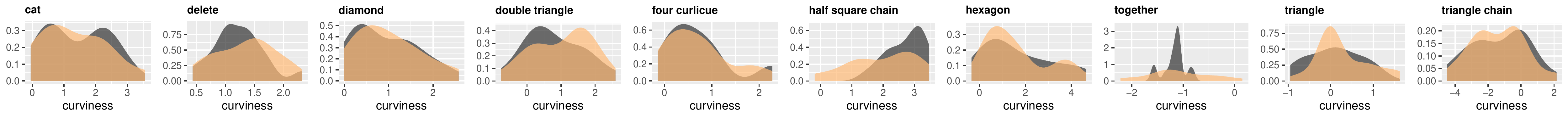}
  \includegraphics[width=\w]{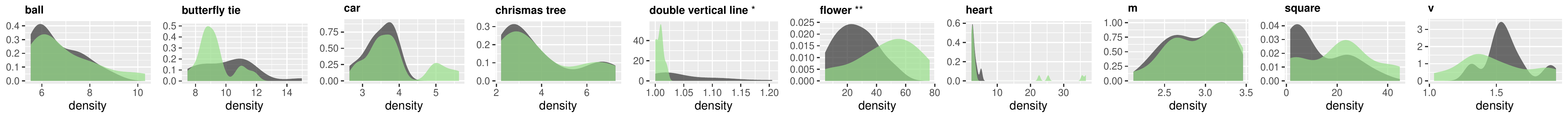}
  \includegraphics[width=\w]{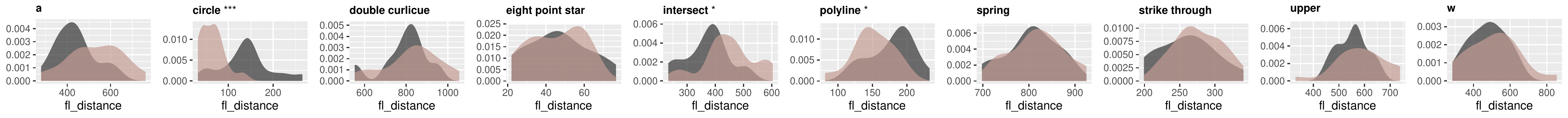}
  \includegraphics[width=\w]{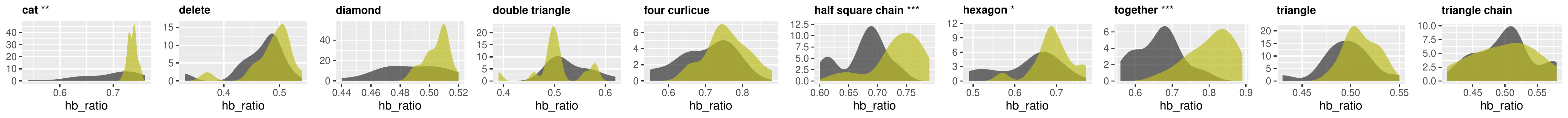}
  \includegraphics[width=\w]{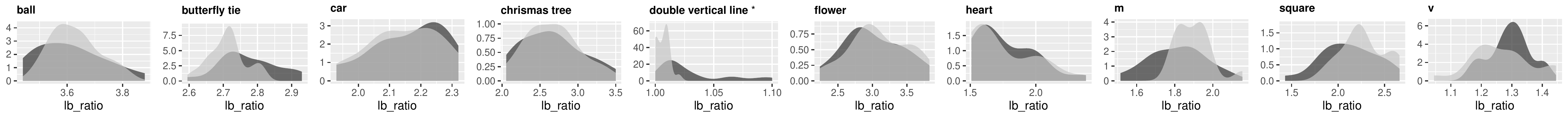}
  \includegraphics[width=\w]{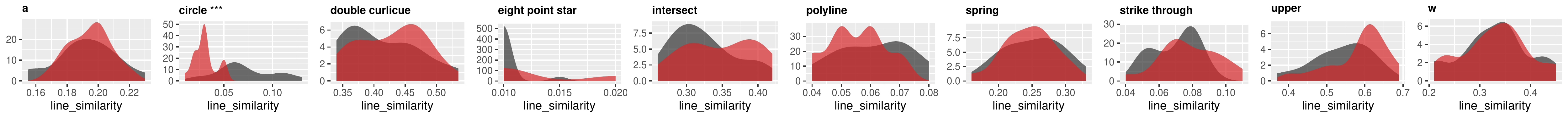}
  \end{minipage}
  \caption{
    A subset of the sample distributions of each feature,
    corresponding to evaluation scenario EV$_2$.
    Please refer to \protect\autoref{fig:hists-summary} and our companion website for more information.
  }
  \label{fig:hists-appendix}
\end{figure*}

\begin{figure*}
  \centering
  \hspace*{-3cm}
  \def\w{1.2\textwidth}
  \begin{minipage}{\w}
  \includegraphics[width=\w]{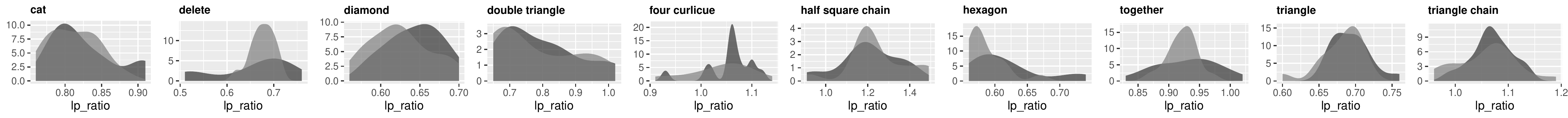}
  \includegraphics[width=\w]{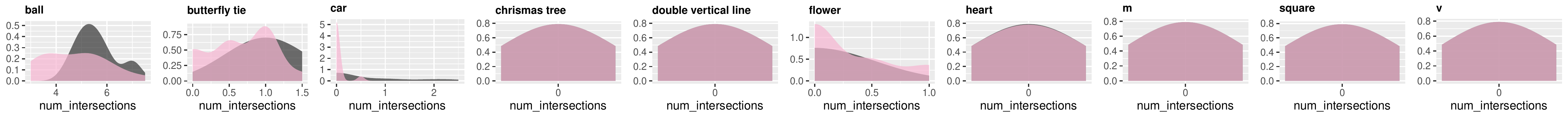}
  \includegraphics[width=\w]{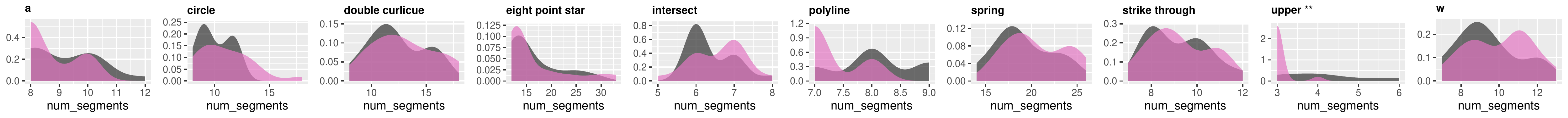}
  \includegraphics[width=\w]{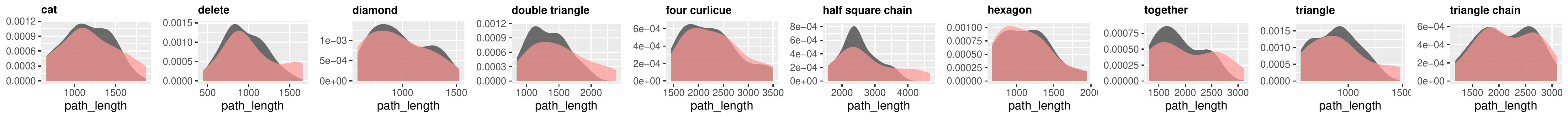}
  \includegraphics[width=\w]{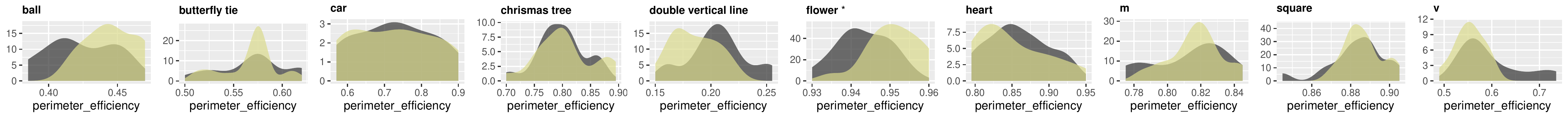}
  \includegraphics[width=\w]{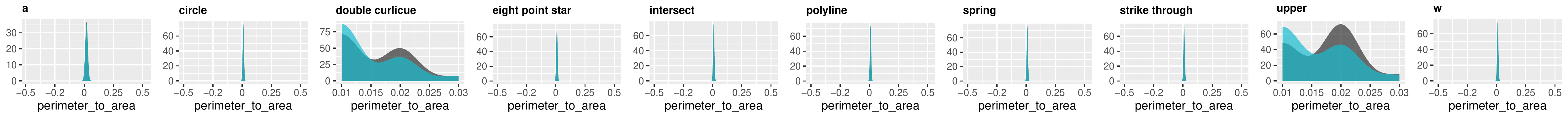}
  \includegraphics[width=\w]{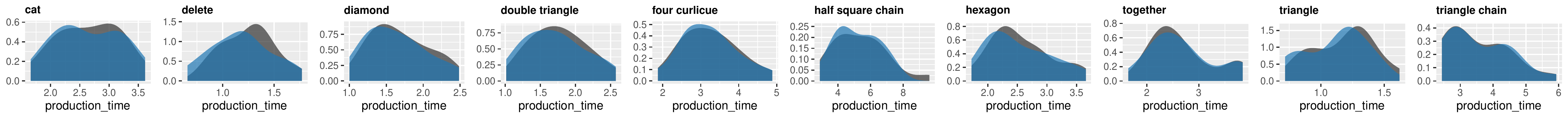}
  \includegraphics[width=\w]{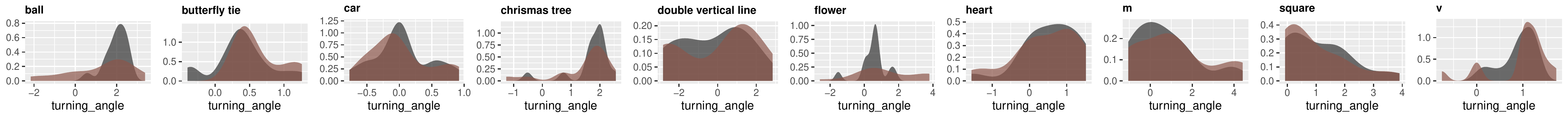}
  \end{minipage}
  \caption{
    Cont. of \protect\autoref{fig:hists-appendix}.
    Please refer to \protect\autoref{fig:hists-summary} and our companion website for more information.
  }
  \label{fig:hists-appendix-cont}
\end{figure*}


\small
\begin{thebibliography}{118}
\expandafter\ifx\csname natexlab\endcsname\relax\def\natexlab#1{#1}\fi
\expandafter\ifx\csname url\endcsname\relax
  \def\url#1{\texttt{#1}}\fi
\expandafter\ifx\csname urlprefix\endcsname\relax\def\urlprefix{URL: }\fi

\bibitem[{Accot and Zhai(1997)}]{Accot97}
Accot, J., Zhai, S., 1997. Beyond {Fitts'} law: Models for trajectory-based
  {HCI} tasks. In: Proceedings of SIGCHI Conf. on Human Factors in Computing
  Systems (CHI). pp. 295--302.

\bibitem[{Almaksour et~al.(2011)Almaksour, Anquetil, Plamondon, and
  O'Reilly}]{Almaksour11}
Almaksour, A., Anquetil, E., Plamondon, R., O'Reilly, C., 2011. Synthetic
  handwritten gesture generation using {Sigma-Lognormal} model for evolving
  handwriting classifiers. In: Proceedings of Biennial Conf. of the Intl.
  Graphonomics Society (IGS). pp. 98--101.

\bibitem[{Anderson and Bischof(2013)}]{Anderson:2013:LPG}
Anderson, F., Bischof, W.~F., 2013. Learning and performance with gesture
  guides. In: Proceedings of SIGCHI Conf. on Human Factors in Computing Systems
  (CHI). pp. 1109--1118.

\bibitem[{Anthony et~al.(2013)Anthony, Vatavu, and Wobbrock}]{Anthony13gi}
Anthony, L., Vatavu, R.-D., Wobbrock, J.~O., 2013. Understanding the
  consistency of users' pen and finger stroke gesture articulation. In:
  Proceedings of Graphics Interface (GI). pp. 87--94.

\bibitem[{Anthony and Wobbrock(2010)}]{Anthony10}
Anthony, L., Wobbrock, J.~O., 2010. A lightweight multistroke recognizer for
  user interface prototypes. In: Proceedings of Graphics Interface (GI). pp.
  245--252.

\bibitem[{Anthony and Wobbrock(2012)}]{Anthony12}
Anthony, L., Wobbrock, J.~O., 2012. \${N}-protractor: a fast and accurate
  multistroke recognizer. In: Proceedings of Graphics Interface (GI). pp.
  117--120.

\bibitem[{Appert and Bau(2010)}]{Appert:2010:SDP}
Appert, C., Bau, O., 2010. Scale detection for a priori gesture recognition.
  In: Proceedings of SIGCHI Conf. on Human Factors in Computing Systems (CHI).
  pp. 879--882.

\bibitem[{Appert and Zhai(2009)}]{Appert09}
Appert, C., Zhai, S., 2009. Using strokes as command shortcuts: Cognitive
  benefits and toolkit support. In: Proceedings of SIGCHI Conf. on Human
  Factors in Computing Systems (CHI). pp. 2289--2298.

\bibitem[{Azenkot et~al.(2012)Azenkot, Rector, Ladner, and
  Wobbrock}]{Azenkot12}
Azenkot, S., Rector, K., Ladner, R., Wobbrock, J., 2012. {PassChords}: Secure
  multi-touch authentication for blind people. In: Proceedings of Intl. ACM
  SIGACCESS Conference on Computers \& Accessibility (ASSETS). pp. 159--166.

\bibitem[{Bailly et~al.(2012)Bailly, M\"{u}ller, and Lecolinet}]{Bailly12}
Bailly, G., M\"{u}ller, J., Lecolinet, E., 2012. Design and evaluation of
  finger-count interaction: Combining multitouch gestures and menus. Int. J.
  Hum.-Comput. Stud. 70~(10), 673--689.

\bibitem[{Bi et~al.(2013)Bi, Li, and Zhai}]{Bi13}
Bi, X., Li, Y., Zhai, S., 2013. Ffitts law: Modeling finger touch with {Fitts'}
  law. In: Proceedings of SIGCHI Conf. on Human Factors in Computing Systems
  (CHI). pp. 1363--1372.

\bibitem[{Bishop et~al.(2004)Bishop, Svensen, and Hinton}]{Bishop:2004:DTG}
Bishop, C.~M., Svensen, M., Hinton, G.~E., 2004. Distinguishing text from
  graphics in on-line handwritten ink. In: Proceedings of Intl. Workshop on
  Frontiers in Handwriting Recognition (IWFHR). pp. 142--147.

\bibitem[{Blagojevic et~al.(2010)Blagojevic, Chang, and Plimmer}]{Blagojevic10}
Blagojevic, R., Chang, S. H.-H., Plimmer, B., 2010. The power of automatic
  feature selection: Rubine on steroids. In: Proceedings of Sketch-Based
  Interfaces and Modeling Symposium (SBIM). pp. 79--86.

\bibitem[{Bragdon and Ko(2011)}]{Bragdon:2011:GSA}
Bragdon, A., Ko, H.-S., 2011. Gesture select:: Acquiring remote targets on
  large displays without pointing. In: Proceedings of SIGCHI Conf. on Human
  Factors in Computing Systems (CHI). pp. 187--196.

\bibitem[{Bullock and Grossberg(1988)}]{Bullock88}
Bullock, D., Grossberg, S., 1988. The {VITE} model: a neural command circuit
  for generating arm and articulator trajectories. In: Dynamic Patterns in
  Complex Systems. World Scientific, pp. 305--326.

\bibitem[{Burgbacher and Hinrichs(2015)}]{Burgbacher:2015:MHP}
Burgbacher, U., Hinrichs, K., 2015. Modeling human performance of stroke-based
  text entry. In: Proceedings of ACM Conf. on Human-computer interaction with
  mobile devices and services (MobileHCI). pp. 137--143.

\bibitem[{Cao and Zhai(2007)}]{Cao07}
Cao, X., Zhai, S., 2007. Modeling human performance of pen stroke gestures. In:
  Proceedings of SIGCHI Conf. on Human Factors in Computing Systems (CHI). pp.
  1495--1504.

\bibitem[{Carcangiu and Spano(2018)}]{Carcangiu:2018:GGA}
Carcangiu, A., Spano, L.~D., 2018. G-gene: A gene alignment method for online
  partial stroke gestures recognition. Proceedings of the ACM on Human-Computer
  Interaction 2~(EICS), 13:1--13:17.

\bibitem[{Card et~al.(1980)Card, Moran, and Newell}]{Card80}
Card, S.~K., Moran, T.~P., Newell, A., 1980. The keystroke-level model for user
  performance time with interactive systems. Commun. ACM 23~(1), 396--410.

\bibitem[{Castellucci and MacKenzie(2008)}]{Castellucci08}
Castellucci, S.~J., MacKenzie, I.~S., 2008. Graffiti vs. {Unistrokes}: An
  empirical comparison. In: Proceedings of SIGCHI Conf. on Human Factors in
  Computing Systems (CHI). pp. 305--308.

\bibitem[{Chen et~al.(2014)Chen, Grossman, and
  Fitzmaurice}]{Chen:2014:STE:2642918.2647354}
Chen, X.~A., Grossman, T., Fitzmaurice, G., 2014. Swipeboard: A text entry
  technique for ultra-small interfaces that supports novice to expert
  transitions. In: Proceedings of Annual ACM Symp. on User Interface Software
  and Technology (UIST). pp. 615--620.

\bibitem[{Cordiez(2008)}]{LovelyCharts}
Cordiez, J., 2008. Lovely charts for {iPad on Vimeo}. Accessed on May 2019.
\urlprefix\url{https://vimeo.com/32450225}

\bibitem[{Djioua and Plamondon(2009)}]{Djioua09}
Djioua, M., Plamondon, R., 2009. Studying the variability of handwriting
  patterns using the {Kinematic Theory}. Hum. Mov. Sci. 28~(5), 588--601.

\bibitem[{Djioua and Plamondon(2010)}]{Djioua10}
Djioua, M., Plamondon, R., 2010. The limit profile of a rapid movement
  velocity. Hum. Mov. Sci. 29~(1), 48--61.

\bibitem[{DroidByDesign(2014)}]{oftSeen}
DroidByDesign, 2014. {OftSeen} gestures. Accessed on May 2019.
\urlprefix\url{https://play.google.com/store/apps/details?id=com.proofbydesign.oftSeenGestures}

\bibitem[{Fitts(1954)}]{Fitts54}
Fitts, P.~M., 1954. The information capacity of the human motor system in
  controlling the amplitude of movement. J. Exp. Psychol. 47~(6), 381--391.

\bibitem[{Flash and Hogan(1985)}]{Flash85}
Flash, T., Hogan, N., 1985. The coordination of arm movements: an
  experimentally confirmed mathematical model. J. Neurosci. 5~(7), 1688--1703.

\bibitem[{Fonseca et~al.(2002)Fonseca, Pimentel, and Jorge}]{Fonseca:2002}
Fonseca, M.~J., Pimentel, C., Jorge, J.~A., 2002. Cali: An online scribble
  recognizer for calligraphic interfaces. In: AAAI Spring Symposium on Sketch
  Understanding. pp. 51--58.

\bibitem[{Funk et~al.(2014)Funk, Sahami, Henze, and Schmidt}]{Funk:2014:UTW}
Funk, M., Sahami, A., Henze, N., Schmidt, A., 2014. Using a touch-sensitive
  wristband for text entry on smart watches. In: Proceedings of Extended
  Abstracts on Human Factors in Computing Systems (CHI EA). pp. 2305--2310.

\bibitem[{Galbally et~al.(2012)Galbally, Plamondon, Fierrez, and
  Ortega-García}]{Galbally12b}
Galbally, J., Plamondon, R., Fierrez, J., Ortega-García, J., 2012. Synthetic
  on-line signature generation. {Part II}: Experimental validation. Pattern
  Recogn. 45~(7), 2622--2632.

\bibitem[{Ghomi et~al.(2013)Ghomi, Huot, Bau, Beaudouin-Lafon, and
  Mackay}]{Ghomi13}
Ghomi, E., Huot, S., Bau, O., Beaudouin-Lafon, M., Mackay, W.~E., 2013.
  Arp\`{e}ge: Learning multitouch chord gestures vocabularies. In: Proceedings
  of the ACM Intl. Conf. on Interactive Tabletops and Surfaces (ITS). pp.
  209--218.

\bibitem[{Gordon et~al.(2016)Gordon, Ouyang, and
  Zhai}]{Gordon:2016:WTG:2858036.2858242}
Gordon, M., Ouyang, T., Zhai, S., 2016. {WatchWriter}: Tap and gesture typing
  on a smartwatch miniature keyboard with statistical decoding. In: Proceedings
  of SIGCHI Conf. on Human Factors in Computing Systems (CHI). pp. 3817--3821.

\bibitem[{Grossman et~al.(2015)Grossman, Chen, and Fitzmaurice}]{Grossman15}
Grossman, T., Chen, X.~A., Fitzmaurice, G., 2015. Typing on glasses: Adapting
  text entry to smart eyewear. In: Proceedings of ACM Conf. on Human-computer
  interaction with mobile devices and services (MobileHCI). pp. 144--152.

\bibitem[{Han et~al.(2017)Han, Ahn, Park, and Lee}]{Han:2017:DTG}
Han, J., Ahn, S., Park, K., Lee, G., 2017. Designing touch gestures using the
  space around the smartwatch as continuous input space. In: Proceedings of the
  ACM Intl. Conf. on Interactive Surfaces and Spaces (ISS). pp. 210--219.

\bibitem[{Harrison et~al.(2011)Harrison, Schwarz, and Hudson}]{Harrison11}
Harrison, C., Schwarz, J., Hudson, S.~E., 2011. Tapsense: Enhancing finger
  interaction on touch surfaces. In: Proceedings of Annual ACM Symp. on User
  Interface Software and Technology (UIST). pp. 627--636.

\bibitem[{Hidtechs(2014)}]{Quickify}
Hidtechs, 2014. Quickify - gesture shortcuts. Accessed on May 2019.
\urlprefix\url{https://play.google.com/store/apps/details?id=com.hidtechs.mylauncher.activities}

\bibitem[{Hinrichs and Carpendale(2011)}]{Hinrichs11}
Hinrichs, U., Carpendale, S., 2011. Gestures in the wild: Studying multi-touch
  gesture sequences on interactive tabletop exhibits. In: Proceedings of SIGCHI
  Conf. on Human Factors in Computing Systems (CHI). pp. 3023--3032.

\bibitem[{Honan(2013)}]{AppleNewton}
Honan, M., 2013. Remembering the apple newton's prophetic failure and lasting
  impact. Accessed on May 2019.
\urlprefix\url{https://www.wired.com/2013/08/remembering-the-apple-newtons-prophetic-failure-and-lasting-ideals/}

\bibitem[{Isokoski(2001)}]{Isokoski01}
Isokoski, P., 2001. Model for unistroke writing time. In: Proceedings of SIGCHI
  Conf. on Human Factors in Computing Systems (CHI). pp. 357--364.

\bibitem[{Kane et~al.(2008)Kane, Bigham, and Wobbrock}]{Kane08}
Kane, S.~K., Bigham, J.~P., Wobbrock, J.~O., 2008. Slide rule: Making mobile
  touch screens accessible to blind people using multi-touch interaction
  techniques. In: Proceedings of Intl. ACM SIGACCESS Conference on Computers \&
  Accessibility (ASSETS). pp. 73--80.

\bibitem[{Kane et~al.(2009)Kane, Jayant, Wobbrock, and Ladner}]{Kane09}
Kane, S.~K., Jayant, C., Wobbrock, J.~O., Ladner, R.~E., 2009. Freedom to roam:
  A study of mobile device adoption and accessibility for people with visual
  and motor disabilities. In: Proceedings of Intl. ACM SIGACCESS Conference on
  Computers \& Accessibility (ASSETS). pp. 115--122.

\bibitem[{Kane et~al.(2011)Kane, Wobbrock, and Ladner}]{Kane11}
Kane, S.~K., Wobbrock, J.~O., Ladner, R.~E., 2011. Usable gestures for blind
  people: Understanding preference and performance. In: Proceedings of SIGCHI
  Conf. on Human Factors in Computing Systems (CHI). pp. 413--422.

\bibitem[{Kin et~al.(2012{\natexlab{a}})Kin, Hartmann, DeRose, and
  Agrawala}]{Kin12b}
Kin, K., Hartmann, B., DeRose, T., Agrawala, M., 2012{\natexlab{a}}. Proton++:
  A customizable declarative multitouch framework. In: Proceedings of Annual
  ACM Symp. on User Interface Software and Technology (UIST). pp. 477--486.

\bibitem[{Kin et~al.(2012{\natexlab{b}})Kin, Hartmann, DeRose, and
  Agrawala}]{Kin12a}
Kin, K., Hartmann, B., DeRose, T., Agrawala, M., 2012{\natexlab{b}}. Proton:
  Multitouch gestures as regular expressions. In: Proceedings of SIGCHI Conf.
  on Human Factors in Computing Systems (CHI). pp. 2885--2894.

\bibitem[{Kristensson and Zhai(2004)}]{Kristensson04}
Kristensson, P.~O., Zhai, S., 2004. {SHARK}$^2$: A large vocabulary shorthand
  writing system for pen-based computers. In: Proceedings of Annual ACM Symp.
  on User Interface Software and Technology (UIST). pp. 43--52.

\bibitem[{Kristensson and Zhai(2007)}]{Kristensson:2007:CSW}
Kristensson, P.~O., Zhai, S., 2007. Command strokes with and without preview:
  Using pen gestures on keyboard for command selection. In: Proceedings of
  SIGCHI Conf. on Human Factors in Computing Systems (CHI). pp. 1137--1146.

\bibitem[{Leiva(2017)}]{Leiva17_dis}
Leiva, L.~A., 2017. Large-scale user perception of synthetic stroke gestures.
  In: Proceedings of the ACM Conf. on Conference on Designing Interactive
  Systems (DIS). pp. 1135--1140.

\bibitem[{Leiva et~al.(2014)Leiva, Alabau, Romero, Toselli, and
  Vidal}]{Leiva14_iwc}
Leiva, L.~A., Alabau, V., Romero, V., Toselli, A.~H., Vidal, E., 2014.
  Context-aware gestures for mixed-initiative text editing {UI}s. Interact.
  Comput. 27~(1), 675--696.

\bibitem[{Leiva et~al.(2016)Leiva, Martín-Albo, and Plamondon}]{Leiva16_g3}
Leiva, L.~A., Martín-Albo, D., Plamondon, R., 2016. {Gestures à Go Go}:
  Authoring synthetic human-like stroke gestures using the kinematic theory of
  rapid movements. ACM T. Intel. Syst. Tec. 7~(2), 15:1--15:29.

\bibitem[{Leiva et~al.(2017{\natexlab{a}})Leiva, Martín-Albo, and
  Plamondon}]{Leiva17_iwc}
Leiva, L.~A., Martín-Albo, D., Plamondon, R., 2017{\natexlab{a}}. The
  {Kinematic Theory} produces human-like stroke gestures. Interact. Comput.
  29~(4).

\bibitem[{Leiva et~al.(2018{\natexlab{a}})Leiva, Martín-Albo, Plamondon, and
  Vatavu}]{Leiva18_keytime}
Leiva, L.~A., Martín-Albo, D., Plamondon, R., Vatavu, R.-D.,
  2018{\natexlab{a}}. Keytime: Super-accurate prediction of stroke gesture
  production times. In: Proceedings of SIGCHI Conf. on Human Factors in
  Computing Systems (CHI). pp. 239:1--239:12.

\bibitem[{Leiva et~al.(2017{\natexlab{b}})Leiva, Martín-Albo, and
  Vatavu}]{Leiva17_chi}
Leiva, L.~A., Martín-Albo, D., Vatavu, R.-D., 2017{\natexlab{b}}. Synthesizing
  stroke gestures across user populations: A case for users with visual
  impairments. In: Proceedings of SIGCHI Conf. on Human Factors in Computing
  Systems (CHI). pp. 4182--4193.

\bibitem[{Leiva et~al.(2018{\natexlab{b}})Leiva, Martín-Albo, and
  Vatavu}]{Leiva18_gato}
Leiva, L.~A., Martín-Albo, D., Vatavu, R.-D., 2018{\natexlab{b}}. Gato:
  Predicting human performance with multistroke and multitouch gesture input.
  In: Proceedings of ACM Conf. on Human-computer interaction with mobile
  devices and services (MobileHCI). pp. 32:1--32:11.

\bibitem[{Leung and Chen(2002)}]{Leung:2002}
Leung, W.~H., Chen, T., 2002. User-independent retrieval of free-form
  hand-drawn sketches. In: Proceedings of Intl. Conf. on Acoustics, Speech, and
  Signal Processing (ICASSP). pp. 2029--2032.

\bibitem[{Li(2010{\natexlab{a}})}]{Li10b}
Li, Y., 2010{\natexlab{a}}. {Gesture Search}: A tool for fast mobile data
  access. In: Proceedings of Annual ACM Symp. on User Interface Software and
  Technology (UIST). pp. 87--96.

\bibitem[{Li(2010{\natexlab{b}})}]{Li10a}
Li, Y., 2010{\natexlab{b}}. Protractor: a fast and accurate gesture recognizer.
  In: Proceedings of SIGCHI Conf. on Human Factors in Computing Systems (CHI).
  pp. 2169--2172.

\bibitem[{Liu et~al.(2017)Liu, Clark, and Lindqvist}]{Liu:2017:USG}
Liu, C., Clark, G.~D., Lindqvist, J., 2017. Where usability and security go
  hand-in-hand: Robust gesture-based authentication for mobile systems. In:
  Proceedings of SIGCHI Conf. on Human Factors in Computing Systems (CHI). pp.
  374--386.

\bibitem[{{Long Jr.} et~al.(1999){Long Jr.}, Landay, and Rowe}]{Long99}
{Long Jr.}, A.~C., Landay, J.~A., Rowe, L.~A., 1999. Implications for a gesture
  design tool. In: Proceedings of SIGCHI Conf. on Human Factors in Computing
  Systems (CHI). pp. 40--47.

\bibitem[{{Long Jr.} et~al.(2000){Long Jr.}, Landay, Rowe, and
  Michiels}]{Long00}
{Long Jr.}, A.~C., Landay, J.~A., Rowe, L.~A., Michiels, J., 2000. Visual
  similarity of pen gestures. In: Proceedings of SIGCHI Conf. on Human Factors
  in Computing Systems (CHI). pp. 360--367.

\bibitem[{Lu and Li(2015)}]{Lu:2015:GEA}
Lu, H., Li, Y., 2015. Gesture on: Enabling always-on touch gestures for fast
  mobile access from the device standby mode. In: Proceedings of SIGCHI Conf.
  on Human Factors in Computing Systems (CHI). pp. 3355--3364.

\bibitem[{Luo and Vogel(2015)}]{Luo15}
Luo, Y., Vogel, D., 2015. Pin-and-cross: A unimanual multitouch technique
  combining static touches with crossing selection. In: Proceedings of Annual
  ACM Symp. on User Interface Software and Technology (UIST). pp. 323--332.

\bibitem[{Lü and Li(2012)}]{Lu12}
Lü, H., Li, Y., 2012. {Gesture Coder}: A tool for programming multi-touch
  gestures by demonstration. In: Proceedings of SIGCHI Conf. on Human Factors
  in Computing Systems (CHI). pp. 2875--2884.

\bibitem[{Lü and Li(2013)}]{Lu13}
Lü, H., Li, Y., 2013. {Gesture Studio}: Authoring multi-touch interactions
  through demonstration and declaration. In: Proceedings of SIGCHI Conf. on
  Human Factors in Computing Systems (CHI). pp. 257--266.

\bibitem[{Martín-Albo and Leiva(2016)}]{MartinAlbo16_g3}
Martín-Albo, D., Leiva, L.~A., 2016. {G3}: bootstrapping stroke gestures
  design with synthetic samples and built-in recognizers. In: Proceedings of
  ACM Conf. on Human-computer interaction with mobile devices and services
  (MobileHCI). pp. 633--637.

\bibitem[{Martín-Albo et~al.(2015)Martín-Albo, Plamondon, and
  Vidal}]{MartinAlbo15}
Martín-Albo, D., Plamondon, R., Vidal, E., 2015. Improving sigma-lognormal
  parameter extraction. In: Proceedings of Intl. Conf. on Document Analysis and
  Recognition (ICDAR). pp. 286--290.

\bibitem[{Matulic et~al.(2017)Matulic, Vogel, and Dachselt}]{Matulic17}
Matulic, F., Vogel, D., Dachselt, R., 2017. Hand contact shape recognition for
  posture-based tabletop widgets and interaction. In: Proceedings of the ACM
  Intl. Conf. on Interactive Surfaces and Spaces (ISS). pp. 3--11.

\bibitem[{Mauney(2010)}]{Mauney2010}
Mauney, D., 2010. What gestures do people actually use?
  https://www.lukew.com/ff/entry.asp?1197.

\bibitem[{Microsoft(2003)}]{WindowsTablet}
Microsoft, 2003. {Microsoft Tablet PC}. Accessed on May 2019.
\urlprefix\url{https://docs.microsoft.com/en-us/previous-versions/ms840465(v=msdn.10)}

\bibitem[{Mobotap(2011)}]{Dolphin}
Mobotap, 2011. Dolphin browser for {Android}. Accessed on May 2019.
\urlprefix\url{https://dolphin.com}

\bibitem[{Morris et~al.(2010)Morris, Wobbrock, and Wilson}]{Morris10}
Morris, M.~R., Wobbrock, J.~O., Wilson, A.~D., 2010. Understanding users'
  preferences for surface gestures. In: Proceedings of Graphics Interface (GI).
  pp. 261--268.

\bibitem[{M\"{u}ller et~al.(2017)M\"{u}ller, Oulasvirta, and
  Murray-Smith}]{Muller17}
M\"{u}ller, J., Oulasvirta, A., Murray-Smith, R., 2017. Control theoretic
  models of pointing. ACM Trans. Comput.-Hum. Interact. 24~(4), 27:1--27:36.

\bibitem[{Nacenta et~al.(2013)Nacenta, Kamber, Qiang, and
  Kristensson}]{Nacenta13}
Nacenta, M.~A., Kamber, Y., Qiang, Y., Kristensson, P.~O., 2013. Memorability
  of pre-designed and user-defined gesture sets. In: Proceedings of SIGCHI
  Conf. on Human Factors in Computing Systems (CHI). pp. 1099--1108.

\bibitem[{Plamondon(1995{\natexlab{a}})}]{Plamondon95a}
Plamondon, R., 1995{\natexlab{a}}. A kinematic theory of rapid human movements.
  {Part I}: Movement representation and control. Biol. Cybern. 72~(4),
  295--307.

\bibitem[{Plamondon(1995{\natexlab{b}})}]{Plamondon95b}
Plamondon, R., 1995{\natexlab{b}}. A kinematic theory of rapid human movements.
  {Part II}: Movement time and control. Biol. Cybern. 72~(4), 309--320.

\bibitem[{Plamondon et~al.(1993)Plamondon, Alimi, Yergeau, and
  Leclerc}]{Plamondon93}
Plamondon, R., Alimi, A.~M., Yergeau, P., Leclerc, F., 1993. Modelling velocity
  profiles of rapid movements: a comparative study. Biol. Cybern. 69~(1),
  119--128.

\bibitem[{Plamondon and Djioua(2006)}]{Plamondon06}
Plamondon, R., Djioua, M., 2006. A multi-level representation paradigm for
  handwriting stroke generation. Hum. Mov. Sci. 25~(4--5), 586--607.

\bibitem[{Plamondon et~al.(2014)Plamondon, O'Reilly, Galbally, Almaksour, and
  Anquetil}]{Plamondon14}
Plamondon, R., O'Reilly, C., Galbally, J., Almaksour, A., Anquetil, E., 2014.
  Recent developments in the study of rapid human movements with the {Kinematic
  Theory}: Applications to handwriting and signature synthesis. Pattern Recogn.
  Lett. 35, 225--235.

\bibitem[{Poppinga et~al.(2014)Poppinga, Shirazi, Henze, Heuten, and
  Boll}]{Poppinga14}
Poppinga, B., Shirazi, A.~S., Henze, N., Heuten, W., Boll, S., 2014.
  Understanding shortcut gestures on mobile touch devices. In: Proceedings of
  ACM Conf. on Human-computer interaction with mobile devices and services
  (MobileHCI). pp. 173--182.

\bibitem[{{POW Studios}(2008)}]{POWStudios}
{POW Studios}, 2008. Mr. {Spiff}'s revenge. Accessed on May 2019.
\urlprefix\url{https://youtu.be/nqeIuNTbT-8}

\bibitem[{Quinn and Zhai(2018)}]{Quinn16}
Quinn, P., Zhai, S., 2018. Modeling gesture-typing movements. Hum.-Comput.
  Interact. 33~(2), 234--280.

\bibitem[{Rekik et~al.(2014)Rekik, Vatavu, and Grisoni}]{Rekik14}
Rekik, Y., Vatavu, R.-D., Grisoni, L., 2014. Understanding users' perceived
  difficulty of multi-touch gesture articulation. In: Proceedings of Intl.
  Conf. on Multimodal Interaction (ICMI). pp. 232--239.

\bibitem[{Rick(2010)}]{Rick:2010:POV}
Rick, J., 2010. Performance optimizations of virtual keyboards for stroke-based
  text entry on a touch-based tabletop. In: Proceedings of Annual ACM Symp. on
  User Interface Software and Technology (UIST). pp. 77--86.

\bibitem[{Rubine(1991)}]{Rubine91}
Rubine, D., 1991. Specifying gestures by example. In: Proceedings of annual
  Conf. on Computer graphics and interactive techniques (SIGGRAPH). pp.
  329--337.

\bibitem[{Schneegass and Voit(2016)}]{Schneegass:2016:GUT}
Schneegass, S., Voit, A., 2016. Gesturesleeve: Using touch sensitive fabrics
  for gestural input on the forearm for controlling smartwatches. In:
  Proceedings of the ACM Intl. Symp. on Wearable Computers (ISWC). pp.
  108--115.

\bibitem[{Shaw and Anthony(2016)}]{Shaw16}
Shaw, A., Anthony, L., 2016. Analyzing the articulation features of children's
  touchscreen gestures. In: Proceedings of Intl. Conf. on Multimodal
  Interaction (ICMI). pp. 333--340.

\bibitem[{Tanuwidjaja et~al.(2014)Tanuwidjaja, Huynh, Koa, Nguyen, Shao,
  Torbett, Emmenegger, and Weibel}]{Tanuwidjaja:2014:CWA:2632048.2632091}
Tanuwidjaja, E., Huynh, D., Koa, K., Nguyen, C., Shao, C., Torbett, P.,
  Emmenegger, C., Weibel, N., 2014. Chroma: A wearable augmented-reality
  solution for color blindness. In: Proceedings of Intl. Conf. on Ubiquitous
  Computing (UbiComp). pp. 799--810.

\bibitem[{Taranta~II and {LaViola Jr.}(2015)}]{Taranta:2015:PPB}
Taranta~II, E.~M., {LaViola Jr.}, J.~J., 2015. Penny pincher: A blazing fast,
  highly accurate \$-family recognizer. In: Proceedings of Graphics Interface
  (GI). pp. 195--202.

\bibitem[{Taranta~II et~al.(2016{\natexlab{a}})Taranta~II, Maghoumi, Pittman,
  and {LaViola Jr.}}]{Taranta16}
Taranta~II, E.~M., Maghoumi, M., Pittman, C.~R., {LaViola Jr.}, J.~J.,
  2016{\natexlab{a}}. A rapid prototyping approach to synthetic data generation
  for improved {2D} gesture recognition. In: Proceedings of Annual ACM Symp. on
  User Interface Software and Technology (UIST). pp. 873--885.

\bibitem[{Taranta~II et~al.(2017)Taranta~II, Samiei, Maghoumi, Khaloo, Pittman,
  and {LaViola Jr.}}]{Taranta17}
Taranta~II, E.~M., Samiei, A., Maghoumi, M., Khaloo, P., Pittman, C.~R.,
  {LaViola Jr.}, J.~J., 2017. Jackknife: A reliable recognizer with few samples
  and many modalities. In: Proceedings of SIGCHI Conf. on Human Factors in
  Computing Systems (CHI). pp. 5850--5861.

\bibitem[{Taranta~II et~al.(2016{\natexlab{b}})Taranta~II, Vargas, and {LaViola
  Jr.}}]{Taranta:2016:SAG}
Taranta~II, E.~M., Vargas, A.~N., {LaViola Jr.}, J.~J., 2016{\natexlab{b}}.
  Streamlined and accurate gesture recognition with penny pincher. Comput.
  Graph. 55~(C), 130--142.

\bibitem[{{TCB Networks}(2002)}]{StrokeIt}
{TCB Networks}, 2002. Stroke it - mouse gestures for windows. Accessed on May
  2019.
\urlprefix\url{https://www.tcbmi.com/strokeit/}

\bibitem[{Thomassen et~al.(1983)Thomassen, Keuss, and van Galen}]{Thomassen83}
Thomassen, A.~J., Keuss, P.~J., van Galen, G.~P., 1983. Motor aspects of
  handwriting. Acta Psychol. 54~(1--3).

\bibitem[{Tu et~al.(2012)Tu, Ren, and Zhai}]{Tu12}
Tu, H., Ren, X., Zhai, S., 2012. A comparative evaluation of finger and pen
  stroke gestures. In: Proceedings of SIGCHI Conf. on Human Factors in
  Computing Systems (CHI). pp. 1287--1296.

\bibitem[{Ungurean et~al.(2018{\natexlab{a}})Ungurean, Vatavu, Leiva, and
  Mart\'{\i}n-Albo}]{Ungurean18mobilehci}
Ungurean, O.-C., Vatavu, R.-D., Leiva, L.~A., Mart\'{\i}n-Albo, D.,
  2018{\natexlab{a}}. Predicting stroke gesture input performance for users
  with motor impairments. In: Proceedings of ACM Conf. on Human-computer
  interaction with mobile devices and services (MobileHCI). pp. 23--30.

\bibitem[{Ungurean et~al.(2018{\natexlab{b}})Ungurean, Vatavu, Leiva, and
  Plamondon}]{Ungurean18chi}
Ungurean, O.-C., Vatavu, R.-D., Leiva, L.~A., Plamondon, R.,
  2018{\natexlab{b}}. Gesture input for users with motor impairments on
  touchscreens: Empirical results based on the kinematic theory. In:
  Proceedings of Extended Abstracts on Human Factors in Computing Systems (CHI
  EA). pp. LBW537:1--LBW537:6.

\bibitem[{Vanderdonckt et~al.(2018)Vanderdonckt, Roselli, and
  P{\'e}rez-Medina}]{Vanderdonckt:2018:FAS}
Vanderdonckt, J., Roselli, P., P{\'e}rez-Medina, J.~L., 2018. !ftl, an
  articulation-invariant stroke gesture recognizer with controllable position,
  scale, and rotation invariances. In: Proceedings of Intl. Conf. on Multimodal
  Interaction (ICMI). pp. 125--134.

\bibitem[{Vatavu et~al.(2012)Vatavu, Anthony, and Wobbrock}]{Vatavu12_p}
Vatavu, R.-D., Anthony, L., Wobbrock, J.~O., 2012. Gestures as point clouds: a
  \${P} recognizer for user interface prototypes. In: Proceedings of Intl.
  Conf. on Multimodal Interaction (ICMI). pp. 273--280.

\bibitem[{Vatavu et~al.(2013{\natexlab{a}})Vatavu, Anthony, and
  Wobbrock}]{Vatavu13_relacc}
Vatavu, R.-D., Anthony, L., Wobbrock, J.~O., 2013{\natexlab{a}}. Relative
  accuracy measures for stroke gestures. In: Proceedings of Intl. Conf. on
  Multimodal Interaction (ICMI). pp. 279--286.

\bibitem[{Vatavu et~al.(2014)Vatavu, Anthony, and Wobbrock}]{Vatavu14ghost}
Vatavu, R.-D., Anthony, L., Wobbrock, J.~O., 2014. Gesture heatmaps:
  Understanding gesture performance with colorful visualizations. In:
  Proceedings of Intl. Conf. on Multimodal Interaction (ICMI). pp. 172--179.

\bibitem[{Vatavu et~al.(2018{\natexlab{a}})Vatavu, Anthony, and
  Wobbrock}]{Vatavu:2018:QSA}
Vatavu, R.-D., Anthony, L., Wobbrock, J.~O., 2018{\natexlab{a}}. {\$Q}: A
  super-quick, articulation-invariant stroke-gesture recognizer for
  low-resource devices. In: Proceedings of ACM Conf. on Human-computer
  interaction with mobile devices and services (MobileHCI). pp. 23:1--23:12.

\bibitem[{Vatavu et~al.(2013{\natexlab{b}})Vatavu, Casiez, and
  Grisoni}]{Vatavu:2013:SML}
Vatavu, R.-D., Casiez, G., Grisoni, L., 2013{\natexlab{b}}. Small, medium, or
  large?: Estimating the user-perceived scale of stroke gestures. In:
  Proceedings of SIGCHI Conf. on Human Factors in Computing Systems (CHI). pp.
  277--280.

\bibitem[{Vatavu et~al.(2018{\natexlab{b}})Vatavu, Gheran, and
  Schipor}]{Vatavu:2018}
Vatavu, R.-D., Gheran, B.-F., Schipor, M.~D., 2018{\natexlab{b}}. The impact of
  low vision on touch-gesture articulation on mobile devices. IEEE Pervasive
  Computing 17~(1), 27--37.

\bibitem[{Vatavu et~al.(2011)Vatavu, Vogel, Casiez, and Grisoni}]{Vatavu11}
Vatavu, R.-D., Vogel, D., Casiez, G., Grisoni, L., 2011. Estimating the
  perceived difficulty of pen gestures. In: Proceedings of IFIP Intl. Conf. on
  Human-computer Interaction (INTERACT). pp. 89--106.

\bibitem[{Viviani and Flash(1995)}]{Viviani95}
Viviani, P., Flash, T., 1995. Minimum-jerk, two-thirds power law, and
  isochrony: converging approaches to movement planning. J. Exp. Psychol.
  21~(1), 32--53.

\bibitem[{Viviani and Terzuolo(1982)}]{Viviani82}
Viviani, P., Terzuolo, C., 1982. Trajectory determines movement dynamics.
  Neuroscience 7~(2), 431--437.

\bibitem[{{WB Games and Niantic}(2019)}]{HPWU}
{WB Games and Niantic}, 2019. Harry potter: Wizards unite. Accessed on Jun
  2019.
\urlprefix\url{https://wizardsunitehub.info/spells/}

\bibitem[{Willems et~al.(2009)Willems, Niels, van Gerven, and
  Vuurpijl}]{Willems09}
Willems, D., Niels, R., van Gerven, M., Vuurpijl, L., 2009. Iconic and
  multi-stroke gesture recognition. Pattern Recogn. 42~(12), 3303--3312.

\bibitem[{Wobbrock(2018)}]{WobbrockDollarFamilyImpact}
Wobbrock, J.~O., 2018. Impact of \$-family. Accessed on May 2019.
\urlprefix\url{http://depts.washington.edu/madlab/proj/dollar/impact.html}

\bibitem[{Wobbrock et~al.(2008)Wobbrock, Cutrell, Harada, and
  MacKenzie}]{Wobbrock08}
Wobbrock, J.~O., Cutrell, E., Harada, S., MacKenzie, I.~S., 2008. An error
  model for pointing based on fitts' law. In: Proceedings of SIGCHI Conf. on
  Human Factors in Computing Systems (CHI). pp. 1613--1622.

\bibitem[{Wobbrock et~al.(2009)Wobbrock, Morris, and Wilson}]{Wobbrock09}
Wobbrock, J.~O., Morris, M.~R., Wilson, A.~D., 2009. User-defined gestures for
  surface computing. In: Proceedings of SIGCHI Conf. on Human Factors in
  Computing Systems (CHI). pp. 1083--1092.

\bibitem[{Wobbrock et~al.(2007)Wobbrock, Wilson, and Li}]{Wobbrock07}
Wobbrock, J.~O., Wilson, A.~D., Li, Y., 2007. Gestures without libraries,
  toolkits or training: A \$1 recognizer for user interface prototypes. In:
  Proceedings of Annual ACM Symp. on User Interface Software and Technology
  (UIST). pp. 159--168.

\bibitem[{Yang et~al.(2016)Yang, Clark, Lindqvist, and
  Oulasvirta}]{Yang:2016:FGA}
Yang, Y., Clark, G.~D., Lindqvist, J., Oulasvirta, A., 2016. Free-form gesture
  authentication in the wild. In: Proceedings of SIGCHI Conf. on Human Factors
  in Computing Systems (CHI). pp. 3722--3735.

\bibitem[{Yu et~al.(2016)Yu, Sun, Zhong, Li, Zhao, and
  Shi}]{Yu:2016:OHI:2858036.2858542}
Yu, C., Sun, K., Zhong, M., Li, X., Zhao, P., Shi, Y., 2016. One-dimensional
  handwriting: Inputting letters and words on smart glasses. In: Proceedings of
  SIGCHI Conf. on Human Factors in Computing Systems (CHI). pp. 71--82.

\bibitem[{Zhai et~al.(2012)Zhai, Kristensson, Appert, Anderson, and
  Cao}]{Zhai12}
Zhai, S., Kristensson, P.~O., Appert, C., Anderson, T.~H., Cao, X., February
  2012. Foundational issues in touch-surface stroke gesture design: An
  integrative review. Foundations and Trends in Human-Computer Interaction
  5~(2), 97--205.
\urlprefix\url{http://dx.doi.org/10.1561/1100000012}

\bibitem[{Zhang(2017)}]{Zhang:2017:IAL}
Zhang, C., 2017. Improving app look up speed on mobile via user-defined touch
  gesture. In: Proceedings of SIGCHI Conf. on Human Factors in Computing
  Systems (CHI). pp. 196--201.

\bibitem[{Zhang et~al.(2016)Zhang, Jiang, and Tian}]{Zhang:2016:AMA}
Zhang, C., Jiang, N., Tian, F., 2016. Accessing mobile apps with user defined
  gesture shortcuts: An exploratory study. In: Proceedings of the ACM Intl.
  Conf. on Interactive Surfaces and Spaces (ISS). pp. 385--390.

\bibitem[{Zhao et~al.(2015)Zhao, Szpiro, and
  Azenkot}]{Zhao:2015:FCH:2700648.2809865}
Zhao, Y., Szpiro, S., Azenkot, S., 2015. {ForeSee}: A customizable head-mounted
  vision enhancement system for people with low vision. In: Proceedings of
  Intl. ACM SIGACCESS Conference on Computers \& Accessibility (ASSETS). pp.
  239--249.

\bibitem[{Zheng et~al.(2018)Zheng, Bi, Li, Li, and Zhai}]{Zheng:2018:MGM}
Zheng, J., Bi, X., Li, K., Li, Y., Zhai, S., 2018. M3 gesture menu: Design and
  experimental analyses of marking menus for touchscreen mobile interaction.
  In: Proceedings of SIGCHI Conf. on Human Factors in Computing Systems (CHI).
  pp. 249:1--249:14.

\end{thebibliography}

\normalsize
\end{document}